\documentstyle[12pt,axodraw,epsf]{article}
\catcode`@=11
\newcount\@tempcntc
\def\@citex[#1]#2{\if@filesw\immediate\write\@auxout{\string\citation{#2}}\fi
  \@tempcnta\z@\@tempcntb\m@ne\def\@citea{}\@cite{\@for\@citeb:=#2\do
    {\@ifundefined
       {b@\@citeb}{\@citeo\@tempcntb\m@ne\@citea\def\@citea{,}{\bf ?}\@warning
       {Citation `\@citeb' on page \thepage \space undefined}}%
    {\setbox\z@\hbox{\global\@tempcntc0\csname b@\@citeb\endcsname\relax}%
     \ifnum\@tempcntc=\z@ \@citeo\@tempcntb\m@ne
       \@citea\def\@citea{,}\hbox{\csname b@\@citeb\endcsname}%
     \else
      \advance\@tempcntb\@ne
      \ifnum\@tempcntb=\@tempcntc
      \else\advance\@tempcntb\m@ne\@citeo
      \@tempcnta\@tempcntc\@tempcntb\@tempcntc\fi\fi}}\@citeo}{#1}}
\def\@citeo{\ifnum\@tempcnta>\@tempcntb\else\@citea\def\@citea{,}%
  \ifnum\@tempcnta=\@tempcntb\the\@tempcnta\else
   {\advance\@tempcnta\@ne\ifnum\@tempcnta=\@tempcntb \else \def\@citea{--}\fi
    \advance\@tempcnta\m@ne\the\@tempcnta\@citea\the\@tempcntb}\fi\fi}
\catcode`@=12

\def\theequation{\arabic{section}.\arabic{equation}}

\begin{document}

\begin{flushright}
CERN-TH/97-293\\
MPI/PhT/97-67\\
hep-ph/9710426\\
October 1997
\end{flushright}

\begin{center}
{\Large {\bf Gauge- and Renormalization-Group- Invariant}}\\[0.4cm]
{\Large {\bf Formulation of the Higgs-Boson Resonance}}\\[2.4cm]
{\large Joannis Papavassiliou}$^a$ 
                           {\large and Apostolos Pilaftsis}$^b$\\[0.4cm]
$^a${\em Theory Division, CERN, CH-1211 Geneva 23, Switzerland}\\[0.3cm]
$^b${\em Max-Planck-Institut f\"ur Physik, F\"ohringer Ring 6, 80805 
Munich, Germany}
\end{center}

\vskip0.7cm     \centerline{\bf   ABSTRACT}  \noindent
A gauge- and renormalization-group-  invariant approach implemented by
the pinch technique is formulated  for resonant transitions  involving
the Higgs boson.  The lineshape of the Higgs boson is shown to consist
of two  distinct     and  physically   meaningful  contributions:    a
process-independent resonant part and a process-dependent non-resonant
background,  which are separately  gauge independent,  invariant under
the renormalization group, satisfy  naive, tree-level Ward identities,
and  respect the optical  and equivalence  theorem individually.   The
former process-independent quantity serves as the natural extension of
the concept of  the effective charge to the  case of the Higgs scalar,
and constitutes a  common ingredient of every Born-improved amplitude.
The difference in the phenomenological predictions obtained within our
approach and those found with other methods is briefly discussed.

\newpage

\section{Introduction}
\indent

The production of the Higgs boson, the only as yet unobserved building
block  of the  Standard  Model (SM), and   the  detailed study of  its
properties will be of central interest for several years to come.  The
Higgs boson is  intimately connected to the prevailing field-theoretic
mechanism  for endowing gauge  bosons, leptons, and quarks with masses
\cite{Higgs}.  Since understanding  the origin  of mass constitutes  a
major challenge for all models aspiring to describe physics beyond the
SM, accurate  experimental  information about    the Higgs sector   is
indispensable  for determining both   their theoretical  relevance and
their phenomenological viability.

Within the  SM, the  mass   $M_{H}$ of  the  Higgs   boson is a   free
parameter.  The experimental lower bound on the SM Higgs boson through
direct searches at the CERN Large  Electron Positron collider (LEP) is
$M_H > 65.2$ GeV \cite{Grivaz}, whereas the theoretical upper bound is
about 700 GeV \cite{HiggsBoundsTheory,HiggsBoundsLattice}.  Since  the
SM  observables  depend logarithmically  on $M_{H}$ \cite{VeltmanLog},
the high precision electroweak data, even  though they favour slightly
a ``light'' Higgs boson of about 150  GeV, they can only impose rather
loose bounds   on    $M_H$.  In  particular,  from  the   LEP data  on
$\sin^2\theta^{\rm  lept}_{\rm eff}$,  the electroweak observable most
sensitive to $M_H$, the upper bound $M_H < 550$ GeV is obtained at the
1.64$\sigma$ level \cite{Precision}, whereas  a tighter upper bound of
$M_H  < 443$  GeV at the   1.64$\sigma$  has been advocated  after the
inclusion  of two-loop  top  quark  effects   in the  calculations  of
$\sin^2\theta^{\rm lept}_{\rm eff}$ \cite{PG}.

A Higgs  boson with mass  of about 100 GeV  can  be discovered at LEP2
\cite{HiggsRep}, through   the   Bjorken process,  or Higgs-strahlung,
$e^{+}e^{-}\to ZH$ \cite{BjorkenProcess}.    If the Higgs boson  turns
out to be  heavier, its discovery   will become again possible  at the
CERN    Large Hadron Collider  (LHC).   In  that case, the Higgs-boson
production will proceed through a variety of sub-processes.  In all of
the above scenarios, depending on the value of $M_{H}$ and the specific
kinematic conditions, the Higgs-boson  production may be resonant.  In
that   case, exactly  as has  happened  in  the  case  of the resonant
$Z$-boson  production \cite{ZRES}, the well-known theoretical problems
associated with  the self-consistent  treatment of resonant transition
amplitudes    are bound  to    resurface,  but   with  the  additional
phenomenological complication that, in contrast to the $Z$-boson case,
bosonic and     fermionic    channels   give   numerically  comparable
contributions to the Higgs-boson decay rate.

{}From the theoretical point of view, the self-consistent treatment of
the Higgs  boson resonance  in the context   of the  SM has  attracted
significant attention,  due   to     a variety  of   open    questions
\cite{Dicusetal}.  In the vicinity  of resonances transition amplitude
become    singular and must be   regulated  by a  Breit-Wigner type of
propagator.  The most obvious signal that  a method more sophisticated
than  a standard resummation  of   conventional self-energy graphs  is
needed  in the case  of  non-Abelian  gauge  theories, comes  from the
simple  calculational fact that  the  bosonic radiative corrections to
the self-energies of  vector  ($\gamma$, $W$,  $Z$) or  scalar (Higgs)
bosons  induce a non-trivial dependence  on the gauge-fixing parameter
(GFP), used to define the tree-level  bosonic propagators appearing in
the quantum loops.    This is  to   be  contrasted to  the   fermionic
radiative corrections, which, even in the context of non-Abelian gauge
theories behave as in quantum  electrodynamics (QED), {\em i.e.}, they
are  GFP  independent.     In  addition,   formal    field-theoretical
considerations as well  as direct calculations  show that, contrary to
the QED  case, the non-Abelian  Green's functions do not satisfy their
naive,  tree-level  Ward  identities  (WI's), after  bosonic  one-loop
corrections   are included.  A   careful   analysis  shows that   this
fundamental difference   between Abelian and  non-Abelian theories has
far-reaching    consequences;    the  naive generalization     of  the
Breit-Wigner method to the   latter case gives rise   to Born-improved
amplitudes, which  do not faithfully  capture the underlying dynamics.
Most noticeably,  due to    violation  of the optical  theorem   (OT),
unphysical thresholds  and artificial resonances appear, which distort
the Higgs boson lineshape.  In addition, the high energy properties of
such amplitudes are altered, and are in  direct contradiction with the
equivalence theorem (ET) \cite{EqTh,CG}.

Recently however,  a formalism  based   on  the pinch  technique  (PT)
\cite{PT,JMC} has been developed  in a series of papers \cite{PP1,PP2}
which  bypasses  all the aforementioned   difficulties, and provides a
self-consistent  framework  for dealing  with unstable   particles and
resonant transition  amplitudes in the context of  non-Abelian gauge
theories.   In \cite{PP1} the  general methodology has been presented,
whereas  in \cite{PP2}  the   crucial   physical requirements for    a
physically meaningful resummation have been  discussed in detail.   In
addition, it was shown that the  resummation algorithm based on the PT
satisfies    all those requirements;  in  fact,   to  the  best of our
knowledge,  it   is the  only   algorithm  known to  date   which can
accomplish  this.  Several applications  of the above formalism may be
found in the literature  \cite{Applications}.  In this paper we employ
the  above formalism in  order   to develop a  systematic  approach to
resonant transition  amplitudes involving  the SM  Higgs   boson.  The
theoretical highlights of  our  study have  been presented in  a short
communication   \cite{HiggsPRL}.  In this  longer  paper we address in
detail the most important  calculational aspects of this analysis, and
discuss extensively the multitude of physical issues involved.

The paper is  organized as follows:  In Section  2, we  use  the PT to
compute  the GFP-independent Higgs-boson   self-energy at the one-loop
level,  within the context of   three different characteristic gauges,
namely the Feynman-t'Hooft gauge, the general renormalizable $R_{\xi}$
gauges,  and the covariant  background field gauges (BFG's).  Explicit
expressions  are reported, and  the  pathologies associated with gauge
dependences in the conventional formulation are discussed.  In Section
3, we employ  arguments of unitarity and  analyticity and show how the
effective  Higgs-boson self-energy  of   the previous section   may be
obtained  from  tree-level amplitudes involving  the  Higgs  boson; in
fact, it satisfies {\em individually} the OT, {\em both} for fermionic
as well as bosonic contributions.  In addition, we apply our formalism
to  the process  $t\bar{t}\to  H^*\to t\bar{t}$, and  discuss how  our
predictions differ  from those obtained by  other methods.  In Section
4, we review the notion of the effective charge in QED and discuss how
this  concept  may be  extended   to the case     of gluon in  quantum
chromodynamics (QCD), based on  properties of the PT gluon self-energy
under   the  renormalization    group  \cite{PT}.    Furthermore,   we
demonstrate explicitly that    in pure scalar   theories,  {\em e.g.},
$(\phi^3)_6$ in  six space-time dimensions,  the scalar  particle does
not admit the  construction of a renormalization-group-invariant (RGI)
quantity which could serve as the analog  of the QED effective charge. 
However,  if the  scalar theory is   endowed  with a global  symmetry,
which,  in  turn, is  broken spontaneously,  we  find  that   a scalar
effective ``charge'' may still be formed. This latter example provides
useful insight  and  sets   up  the stage  for  addressing  the   more
complicated case  of the full SM.   The next two  sections contain the
main theoretical  thrust of our paper: In  Section 5, after discussing
how the {\em process-independent} PT self-energies for the $W$ and $Z$
bosons  of the  SM  can  give rise to   RGI  quantities which  may  be
identified as  effective     charges \cite{PRW}, we   show  that   the
construction of  a process-independent and  RGI quantity involving the
Higgs-boson propagator is  indeed possible.  The above construction of
a Higgs-boson effective charge becomes  only possible by virtue of the
naive,    tree-level  WI's  satisfied  by    the   GFP-independent  PT
sub-amplitudes.  In  Section 6, we show  with an explicit example that
the PT sub-amplitudes satisfy the ET {\em individually}, and that with
the help of the same PT WI's, this  fact remains true even {\em after}
resummation.   In Section 7,   we  present our  conclusions.  Finally,
lengthy analytic  expressions   pertaining to   the $HWW$  and   $HZZ$
vertices are relegated to the Appendices.

\setcounter{equation}{0}
\section {One loop calculations in the Pinch Technique} 
\indent

In this section we show how the application of the PT gives rise to an
effective self-energy for the Higgs boson, which is independent of the
GFP, and displays a high-energy behaviour which is consistent with the
ET.  For  definiteness  we  focus  on the gauge   invariant  subset of
Feynman diagrams  containing  two $W$ bosons  (and their corresponding
would-be Goldstone bosons and  ghosts) in a typical  S-matrix element,
{\em e.g.}, $t\bar{t}\to H^*\to t\bar{t}$, shown in Fig.\ 1.  We carry
out  this calculation in three  representative gauges, {\em i.e.}, the
renormalizable Feynman-'t Hooft gauge, the general $R_\xi$ gauges, and
the background field method (BFM) in the covariant $R_{\xi_Q}$ gauges.
We  discuss  the  relevant   technical  points  and  present  explicit
intermediate and final results.

\subsection{The Feynman--'t Hooft gauge}

First we present the calculation for the special GFP choice $\xi=1$ in
the renormalizable $R_\xi$ gauges. This particular  choice is known to
simplify  computations;  of course, as we   will see explicitly in the
next subsection, the same final answer emerges for any other choice of
$\xi$, after the PT algorithm has been carried out.

We first calculate the diagrams contributing to the conventional Higgs
boson self-energy  (Figs.\ 1(a)--(d)).   A straightforward calculation
yields (we omit contributions from tadpole and seagull graphs):
\begin{equation}
  \label{DFEY}
\Pi^{HH}(q^2)\ =\ \frac{\alpha_w}{4\pi}
\Big( -q^2\, +\, 3 M_W^2\, +\, \frac{M_H^4}{4M_W^2}\, \Big)\, 
B_0(q^2,M^2_W,M^2_W)\, ,
\end{equation}
where $\alpha_w=g^2_w/(4\pi)$ and 
\begin{equation}
  \label{B0}
B_0(p^2,m^2_1,m^2_2)\ =\ (2\pi\mu)^{4-d}\, \int\, \frac{d^dk}{i\pi^2}\,
\frac{1}{(k^2 - m_1^2)[(k+p)^2 - m_2^2 ]}
\end{equation}
is the Veltman--'t Hooft function \cite{HV} defined in $d=4-2\epsilon$
dimensions, using the conventions   of Ref.\ \cite{BAK}.   {}From  the
integrand of Eq.\  (\ref{B0}), it is clear that $B_0(p^2,m^2_1,m^2_2)$
develops absorptive (imaginary) parts,  when $\sqrt{p^2} \ge m_1+m_2$. 
The mass parameters $m_1,\ m_2$ may  represent either physical masses,
such as that of the $W$ and/or $Z$ bosons, or masses of the respective
unphysical would-be Goldstone bosons and ghosts.

According to the PT \cite{PT}, we must now extract the propagator-like
pieces concealed  inside vertex and  box diagrams. Such  pieces emerge
every  time longitudinal momenta   coming from propagators or vertices
trigger elementary WI's of the form
\begin{equation}
  \label{EWI}
\not\! k  P_L = (\not\! k + \not\! p -m_b) P_L - P_R (\not\! p -m_t)
+m_b P_L - m_t P_R\, ,
\end{equation}
where $P_{R(L)} = [1  + (-) \gamma_5]/2$  is the  chirality projection
operator.  The first term in Eq.\ (\ref{EWI}) pinches out the internal
propagator   of the  $b$  quark,  whereas  the   second  one dies when
contracted with the spinor of the external on-shell  $t$ quark.

In the Feynman-'t Hooft gauge,  the only graphs that  can give rise to
propagator-like contributions are the vertex graphs of Figs.\ 1(f) and
1(g), denoted by ${\cal V}_{GW}$ and ${\cal V}_{WG}$, and their mirror
images,    Figs.\   1(j)  and 1(k).    Setting     $p'=p-q$, we define
$S^{\nu}_{R}$ and $S^{\nu}_{L}$ as follows:
\begin{eqnarray}
  \label{SR}
S^\nu_R &=& \frac{g_w^2}{2}\, \bar{v}(p')\, (m_t P_L-m_b P_R)\, 
\frac{1}{\not\! k+\not\! p -m_b}\,  \gamma^\nu  P_L\,  u(p)\, ,\\
  \label{SL}
S^\mu_L &=& \frac{g_w^2}{2}\, \bar{v}(p')\, \gamma^\mu P_L\, 
\frac{1}{\not\! k + \not\! p -m_b}\, (m_t P_R - m_b P_L)\,  u(p)\, .
\end{eqnarray}
The action  of the longitudinal   momenta from the  vertices on  these
expressions gives
\begin{eqnarray}
  \label{KSLR}
k_\nu\, S^\nu_R &=& \frac{g_w^2}{2}\, \bar{v}(p')\, m_t P_L\, u(p)\, + 
\dots\, , \nonumber \\
{(k+q)}_{\mu}\, S^\mu_L &=& \frac{g_w^2}{2}\, \bar{v}(p')\, m_t P_R\, u(p)\,
+ \dots\, ,
\end{eqnarray}
where the  ellipses mean omission of vertex  like  pieces, {\em e.g.},
pieces still containing the  $b$ quark tree-level propagator, $(\not\! 
k+\not\! p - m_b)^{-1}$. With the help of Eq.\ (\ref{KSLR}), we find
\begin{eqnarray}
  {\cal V}_{WG}+{\cal V}_{GW} &=& -\, \Big[ (2k+q)_\nu
  S^\nu_R + (2k+q)_\mu S^\mu_L\, \Big]\, B_0(q^2,M^2_W,M^2_W)
  \nonumber\\ &=& -\, \frac{\alpha_w}{8\pi}m_t B_0(q^2,M^2_W,M^2_W) \,
  \bar{v}(p')\, u(p)\, + \dots
\end{eqnarray}
We notice that the only propagator-like piece  couples to the external
$t\bar{t}$  pair   exactly as a  Higgs   boson.  In addition,  no term
proportional  to $\gamma_{5}$  has   survived; had such  a  term  been
present,  it ought  to be  alloted  to the  effective  $HG^0$ one-loop
mixing self-energy, thus breaking  the CP invariance of the underlying
theory \cite{APRL}. This exercise demonstrates   explicitly how the  PT
preserves the  discrete   symmetries of  the  classical  action  after
quantization.

Finally,   the pinch contribution  ${\cal  V}_{GW}$ to the Higgs-boson
self-energy stemming   from the vertex  (mirror  image graphs  give an
extra factor of 2) is
\begin{equation}
  \label{VFEY}
{\cal V}_{GW}\ =\ -\, \frac{\alpha_w}{4\pi} (q^{2}-M_{H}^{2})\, 
B_0(q^2,M^2_W,M^2_W)\, .
\end{equation}
Adding the contribution  from  Eq.\ (\ref{VFEY}) to the   conventional
result   of  Eq.\ (\ref{DFEY}), we   finally  arrive  at the following
expression    for   the  PT     one-loop   Higgs   boson   self-energy
$\widehat{\Pi}^{HH}(q^2)$,
\begin{equation}
  \label{HPT}
\widehat{\Pi}^{HH}_{(WW)}(q^2)\ =\ \frac{\alpha_w}{16\pi}\frac{M_H^4}{M_W^2}
\Big[\, 1+4\frac{M_W^2}{M_H^2}- 4\frac{M_W^2}{M_H^4}
(2q^2-3M_W^2)\, \Big]\, B_0(q^2,M^2_W,M^2_W)\, .
\end{equation}

\subsection{$R_\xi$ renormalizable gauges}

After   this introductory calculation,  we  turn to  the general case,
where the  GFP  $\xi$ is kept  arbitrary.   In these  gauges, the free
$W$-boson propagator is given by
\begin{eqnarray}
  \label{RxiProp}
\Delta^{(\xi_w)}_{\mu\nu}(q) &=& \Big[ -g_{\mu\nu}+ 
(1-\xi_w)\, \frac{q_\mu q_\nu}{q^2 - \xi_w M^2_W}\, \Big]\,
\frac{1}{q^2 - M^2_W}
\nonumber\\
&=& U_{\mu\nu}(q)\, -\, \frac{q_\mu q_\nu}{M^2_W}D^{(\xi_w)}(q^2)
\, ,
\end{eqnarray}
where 
\begin{equation}
  \label{RxiGoldProp}
D^{(\xi_w)}(q^2)=\frac{1}{q^2-\xi_w M^{2}_W} 
\end{equation}
is the propagator of the would-be Goldstone boson and ghost fields.  A
straightforward calculation for the conventional one-loop Higgs
self-energy yields 
\begin{eqnarray}
  \label{DRKSI}
\Pi^{HH}(q^2) &=& \frac{\alpha_w}{4 \pi}\, \Big[ \Big(\,
\frac{(q^2)^2}{4 M_W^2} - q^2 + 3M_W^2\Big)B_0(q^2,M^2_W,M^2_W)\nonumber\\ 
&& +\, \frac{M_H^4-(q^2)^2}{4M_W^2}\, B_0(q^2,\xi_wM^2_W,\xi_wM^2_W)\,
\Big]\, ,\quad
\end{eqnarray}
where tadpole  and  seagull terms  have  again   been omitted.    Some
comments are now in order regarding Eq.\ (\ref{DRKSI}):
\begin{itemize}

\item[(a)] The term proportional  to $(q^2)^2$ is  absent only for the
  special choice $\xi_w=1$, in which case $B_0$ factorizes out.

\item[(b)]  The term proportional to   $(q^2)^2$  is ultraviolet  (UV)
  finite, {\em   i.e.},  it  does   not  depend on  the  UV  regulator
  $1/\epsilon$ for any value of $\xi_w$.  Of course, this is expected,
  since we are working within a renormalizable gauge.

\item[(c)] Even   though  terms  proportional  to $B_0(q^2,M^2_W,\xi_w
  M^2_W)$ appear in  intermediate calculations of individual diagrams,
  they finally cancel  in the sum.  So, there  are no terms with mixed
  poles; we  only  have thresholds   at $q^2=4M_W^2$  and  $q^2=4\xi_w
  M_W^2$.    This  result can  be traced  back  to  the fact  that the
  tree-level   $HW^+_\mu   W^-_\nu$    coupling   is  proportional  to
  $g_{\mu\nu}$ and hence, any contraction between the longitudinal and
  transverse parts  of the two  $W$-boson propagators in the loop will
  vanish.   The  transverse   part  of the   $W$-boson propagator   is
  associated with  the physical  pole    at $q^2=M^2_W$, whereas   the
  longitudinal  one possesses an unphysical singularity  at $q^2 = \xi
  M^2_W$.  Since only   terms  arising  from the contraction   between
  transverse-transverse  and longitudinal-longitudinal  parts  of  the
  $W$-boson  progators  can survive, the   absence  of mixed  poles is
  expected. However,  this last feature may   change in higher orders,
  since new  momentum-dependent form-factors for the  vertex $HW^+_\mu
  W^-_\nu$  are  radiatively induced, which could   give rise to mixed
  poles.

\item[(d)] Setting $\xi_w=1$ in the  expression of Eq.\ (\ref{DRKSI}),
  we recover the result of Eq.\ (\ref{DFEY}).

\end{itemize}

Next we collect  the pinch contributions  which are kinematically akin
to a  Higgs  boson self-energy.   Due to   the additional longitudinal
momenta   proportional    to   $1-\xi_w$,    we  receive   extra pinch
contributions, from the vertex- as well as the box-diagrams.  The only
technically  subtle point in  this context is that the propagator-like
parts  related to  the Higgs  boson  arise  from {\em  two} successive
contractions of the longitudinal momenta on the elementary vertex: the
first  momentum pinches,  giving  rise to  propagator-like terms whose
coupling to the external quarks is proportional to $\gamma_{\mu} P_L$;
clearly   this   coupling is not  Higgs-boson-like.    In  addition, a
vertex-like   term proportional  to $m_t$  survives.  After the second
longitudinal momentum is contracted with that latter vertex-like term,
it  removes   the internal  fermion propagator  and   gives  rise to a
propagator-like contribution, which  couples to  the external fermions
proportionally to $m_{t}$.  To see that  mechanism in detail, consider
the typical quantity ${\cal T}_{t}^{\mu\nu}$,  appearing in the graphs
in question, defined as
\begin{equation}
  \label{Vmunu}
{\cal T}_{t}^{\mu\nu}\ =\ 
\frac{g_w^2}{2}\, \bar{v}(p')\, \gamma^\mu P_L\, 
\frac{1}{\not\! k+\not\! p -m_b}\,  \gamma^\nu P_L\,  u(p)\, . 
\end{equation}
The action of the first longitudinal momentum $k_\nu$ gives
\begin{equation}
  \label{DOUBLEK}
k_\nu\, V^{\mu\nu}\ =\ \frac{g_w^2}{2}\, \bar{v}(p')\, \gamma^\mu P_L\,\, 
u(p)\, +\, S^{\mu}_{L}\, .
\end{equation}
When the  second momentum $k_{\mu}$ acts on  the first term on the RHS
of Eq.\ (\ref{DOUBLEK}), by virtue  of Eq.\ (\ref{KSLR}) gives rise to
a  propagator-like term proportional  to $m_{t}$, as  is also shown in
Fig.\  2, according to equation  Eq.\ (\ref{KSLR}).  As for the second
term on the RHS of Eq.\ (\ref{DOUBLEK}), after it gets contracted with
the second momentum  $k_{\mu}$, it will  be judiciously alloted to the
various remaining   effective  self-energies, such  as $\gamma\gamma$,
$ZZ$, $\gamma Z$,  $Z\gamma$,  {\em   etc.}, according to   the  rules
established in \cite{JP90}.  These latter terms  are  not displayed in
Fig.\ 2.

We now gather all relevant pinch contributions from the box diagrams: 
\begin{eqnarray}
  \label{BWW}
{\cal B}_{WW}(q^2) &=& \frac{\alpha_w}{16\pi M_W^2}\, 
\Big[ B_0(q^2,M^2_W,M^2_W)
- 2B_0(q^2,M^2_W,\xi_wM^2_W) + B_0(q^2,\xi_wM^2_W,\xi_wM^2_W) \Big]\nonumber\\
&& \times\, (q^2-M_H^2)^2\, ,\\
  \label{BGW}
{\cal B}_{GW}(q^2) &=& \frac{\alpha_w}{8\pi M_W^2}\, 
\Big[ B_0(q^2,M^2_W,\xi_w M^2_W)
- B_0(q^2,\xi_wM^2_W,\xi_wM^2_W) \Big]\, (q^2-M_H^2)^2 .
\end{eqnarray}
The net pinch contribution  to  the effective Higgs boson  self-energy
originating from the box graphs may be summarized by
\begin{equation}
  \label{BTOT}
{\cal B}(q^2)\ =\ \,\frac{\alpha_w}{16\pi M_W^2}\, \Big[\, 
B_0(q^2,M^2_W,M^2_W) - B_0(q^2,\xi_wM^2_W,\xi_wM^2_W)\, \Big]\, 
(q^2-M_H^2)^2\, .
\end{equation}
Again, the terms  proportional  to $B_0(q^2,M^2_W,\xi_wM^2_W)$ cancel.
In  addition, for $\xi_w  = 1$,  the above  expression vanishes as  it
should,  since in the  Feynman gauge there  are no pinch contributions
coming from boxes.

Similarly, the individual pinch  contributions from vertex graphs  are
listed below
\begin{eqnarray}
  \label{VWW}
{\cal V}_{WW}(q^2) &=& -\frac{\alpha_w}{4\pi}
\Big[ \Big( 1+\frac{q^2}{2M_W^2}\Big) 
B_0(q^2,M^2_W,M^2_W)\, - \Big(1-\xi_w+\frac{q^2}{M_W^2}\Big)
B_0(q^2,M^2_W,\xi_wM^2_W)\nonumber\\
&&+\Big(\, \frac{q^2}{2M_W^2}-\xi_w\Big)B_0(q^2,\xi_wM^2_W,\xi_wM^2_W)\Big]
(q^2-M_H^2)\, ,\\
  \label{VGW}
{\cal V}_{GW}(q^2) &=& -\frac{\alpha_w}{4\pi} \Big[ \Big(
1-\xi_w+\frac{q^2}{M_W^2}\Big) B_0(q^2,M^2_W,\xi_wM^2_W)\nonumber\\ 
&&+\, \Big(\xi_w-\frac{q^2}{M_W^2}\, \Big) B_0(q^2,\xi_w M^2_W,
\xi_wM^2_W)\Big]\, (q^2-M_H^2)\, ,
\end{eqnarray}
which gives in the sum 
\begin{equation}
  \label{VTOT}
{\cal V}(q^2)\ =\ -\frac{\alpha_w}{4\pi}
\Big[\Big( 1+\frac{q^2}{2M_W^2}\, \Big)
B_0(q^2,M^2_W,M^2_W) - \frac{q^2}{2M_W^2}B_0(q^2,\xi_wM^2_W,\xi_wM^2_W) 
\Big](q^2-M_H^2)\, .\quad
\end{equation}
Again, the terms proportional to $B_0(q^2,M^2_W,\xi_wM^2_W)$ cancel in
the  final  result, and  the  analytic expression of Eq.\ (\ref{VFEY})
emerges for $\xi_w=1$.

Adding Eq.\ (\ref{BTOT}) and Eq.\  (\ref{VTOT}), we  find that in  the
linear renormalizable gauges the total pinch contribution to effective
Higgs boson self-energy is given by
\begin{eqnarray}
  \label{PRKSI}
\Pi^{HH,P}(q^2) &=& - \frac{\alpha_w}{4\pi}
\Big[ \Big( 1\, +\, \frac{q^2+M_H^2}{4M_W^2}\, \Big)B_0(q^2,M^2_W,M^2_W)
\nonumber\\
&& -\, \frac{q^2+ M_H^2}{4M_W^2}\, B_0(q^2,\xi_w M^2_W,\xi_w M^2_W)\Big]
\, (q^2-M_H^2)\, .
\end{eqnarray}
Adding Eq.\ (\ref{PRKSI}) to the   conventional result given in   Eq.\
(\ref{DRKSI}),  we     see   that    all   terms    proportional    to
$B_0(q^2,\xi_wM^2_W,\xi_wM^2_W)$,  which  are the only terms depending
on  $\xi_w$, cancel, and   we find again  the PT  result given in Eq.\
(\ref{HPT}).

\subsection{The covariant background field gauge}

We   shall   consider   the BFM  applied    to    the covariant gauges
\cite{BFM,DDW,Ken}; a   detailed   discussion  of  the   BFM    in the
non-covariant gauges may be found in \cite{GPT}.  The calculation here
is  particularly illuminating, because it  shows  that the results are
plagued with pathologies away from $\xi_Q=1$ \cite{PP2}.

Using the   Feynman rules  of  the  covariant  background field  gauge
\cite{DDW}, we obtain for the Higgs-boson self-energy in an arbitrary
$\xi_Q$ gauge
\begin{eqnarray}
  \label{DBFG}
\Pi^{\widehat{H}\widehat{H}}(q^2) &=& \frac{\alpha_w}{4\pi}
\Big\{
\Big(\, \frac{(q^2)^2}{4M_W^2} - q^2 + 3 M_W^2 \Big)
B_0(q^2,M^2_W,M^2_W)\nonumber\\
&&+ \Big[ \frac{M_H^4-(q^2)^2}{4M_W^2} - \xi_Q (q^2-M_H^2)\Big]\, 
B_0(q^2,\xi_Q M^2_W,\xi_Q M^2_W)\, \Big \}\, .
\end{eqnarray}

Some important comments must be made:
\begin{itemize}

\item[(a)] Setting $\xi_Q=1$ in  the expression of Eq.\ (\ref{DBFG}) ,
  we  recover the full PT   answer of Eq.\  (\ref{HPT}), in accordance
  with earlier observations \cite{DDW,Ken}
  
\item[(b)] We see  that for $\xi_Q\neq 1$  the $(q^2)^2$ term survives
  and    is          proportional          to   the         difference
  $B_0(q^2,M^2_W,M^2_W)-B_0(q^2,\xi_Q M^2_W,\xi_Q M^2_W)$.     For any
  finite value of $\xi_Q$  this  term vanishes for sufficiently  large
  $q^2$,   {\em  i.e.}, $q^2\gg   M_W^2$ and  $q^2\gg   \xi_Q 
  M_W^2$.\footnote{A   noticeable exception   is  the   unitary  gauge
    ($\xi_Q\to\infty$), in  which such a  term  survives and, in fact,
    gives  rise   to a   divergent,  non-renormalizable contribution.}
  Therefore,  the quantity in   Eq.\ (\ref{DBFG})  displays  good high
  energy behaviour in compliance with  unitarity.  Notice however that
  the onset of this good behaviour depends crucially  on the choice of
  $\xi_Q$.  Since $\xi_Q$ is a free parameter and may  be chosen to be
  arbitrarily  large, but finite, the restoration  of unitarity may be
  arbitrarily delayed as  well. This fact poses no  problem as long as
  one is  restricted to the  computation  of physical  amplitudes at a
  finite order    in  perturbation  theory.  However,  if    the above
  self-energy was  to    be resummed in   order  to  regulate resonant
  transition  amplitudes,  it would  lead  to an   artificial delay of
  unitarity  restoration.   Specific  quantitative examples   of  such
  artifacts will be presented in Section 7.
  
\item[(c)] In addition  to the problem  described above, which becomes
  significant for large values  of $\xi_Q$, a serious pathology occurs
  for any value of $\xi_Q\neq 1$,  namely the appearance of unphysical
  thresholds  \cite{PP1,PP2}.   Such  thresholds   may be particularly
  misleading if $\xi_Q$  is  chosen in the  vicinity of  unity, giving
  rise to distortions in the lineshape of the unstable particle.

\end{itemize}

We then proceed  to isolate the propagator-like  pinch parts  from the
BFG  boxes  and vertices,   for general   $\xi_Q$.  Clearly,   the box
contributions  are the same as   in the linear renormalizable  gauges;
they can be recovered from Eq.\ (\ref{BTOT}) by the simple replacement
$\xi_w\to \xi_Q$.   The  same is  true  for  the pinch   contributions
involving the $WW$ virtual states, {\em i.e.}, ${\cal V}_{WW}$ in Eq.\ 
(\ref{VWW}).  In this way,  the total  pinch box contribution,  ${\cal
  B}$, and ${\cal V}_{WW}$ may separately written down
\begin{eqnarray}
  \label{BTOTBFG}
{\cal B} (q^2) &=& \frac{\alpha_w}{16\pi M^2_W}\, \Big[\, B_0(q^2,M^2_W,M^2_W)
-B_0(q^2,\xi_Q M^2_W,\xi_Q M^2_W) \, \Big]\, (q^2-M_H^2)^2 ,\qquad\\
  \label{VWWBFG}
{\cal V}_{WW}(q^2) &=& - \frac{\alpha_w}{4\pi}
\Big[ \Big(1+\frac{q^2}{2M_W^2}\Big)
B_0(q^2,M^2_W,M^2_W) - \Big(1-\xi_Q+\frac{q^2}{M_W^2}\Big)
B_0(q^2,M^2_W,\xi_Q M^2_W)\nonumber\\
&&+\Big(\, \frac{q^2}{2M_W^2}-\xi_Q\Big) B_0(q^2,\xi_Q M^2_W,\xi_Q M^2_W)
\Big]\, (q^2-M_H^2)\, .
\end{eqnarray}
However,   the vertex graph  ${\cal V}_{GW}$  is  different, since the
coupling between $\widehat{H}G^\pm W^\mp $ in the BFG differs from the
respective $HG^\pm W^\mp$ coupling in the $R_{\xi}$ gauges ($2q_{\mu}$
as opposed to $(2k+q)_{\mu}$, respectively). Specifically, we have
\begin{eqnarray}
  \label{VGWBFG}
{\cal V}_{GW}(q^2) &=& -\, \frac{\alpha_w}{4\pi}
\Big[ \Big( 1-\xi_Q+\frac{q^2}{M_W^2}\Big) 
B_0(q^2,M^2_W,\xi_Q M^2_W)\, 
-\, \frac{q^2}{M_W^2} B_0(q^2,\xi_Q M^2_W,\xi_Q M^2_W)\Big]
\nonumber\\ 
&&\times\, (q^2-M_H^2)\, .
\end{eqnarray}
Adding both pinch terms in Eqs.\ (\ref{VWWBFG}) and (\ref{VGWBFG}), we
easily obtain
\begin{eqnarray}
  \label{VTOTBFG}
{\cal V}(q^2) &=& -\frac{\alpha_w}{4\pi}
\Big[\Big( 1+\frac{q^2}{2M_W^2}\Big) 
B_0(q^2,M^2_W,M^2_W)\, -\, \Big( \frac{q^2}{2M_W^2}+\xi_Q \Big)
B_0(q^2,\xi_Q M^2_W,\xi_Q M^2_W) \Big]\nonumber\\
&&\times\, (q^2-M_H^2)\, .
\end{eqnarray}
Finally, the total pinch  contribution to the Higgs boson self-energy,
which is obtained by  forming  the sum  of Eqs.\  (\ref{BTOTBFG})  and
(\ref{VTOTBFG}), is given by
\begin{eqnarray}
  \label{PTBFG}
\Pi^{\widehat{H}\widehat{H},P}(q^2) &=&  -\, \frac{\alpha_w}{4\pi}
 \Big[\Big(\, 1+\frac{q^2}{4M_W^2}+\frac{M_H^2}{4M_W^2}\Big)
B_0(q^2,M^2_W,M^2_W)\nonumber\\
&& -\, \Big(\, \frac{q^2}{4M_W^2} +
\frac{M_H^2}{4M_W^2}+\xi_Q \Big) B_0(q^2,\xi_Q M^2_W,\xi_Q M^2_W) 
\Big]\, (q^2-M_H^2)\, .
\end{eqnarray}
Adding Eq.\ (\ref{PTBFG}) to Eq.\ (\ref{DBFG}), we  arrive again at the
expression of Eq.\ (\ref{HPT})

In a similar  way,  we may compute   the  contributions of the   other
virtual channels  ($t\bar{t}$, $ZZ$, and  $HH$) to the effective Higgs
boson self-energy. They are given by
\begin{eqnarray}
  \label{ttpair}
{\widehat \Pi}^{HH}_{(tt)}(q^2) &=& \frac{3\alpha_w}{8\pi}
\frac{m_t^2}{M^2_W}(q^2-4m_t^2) B_0(q^2,\, m_t^2,\,m_t^2)\, ,\\
  \label{ZZpair}
{\widehat \Pi}^{HH}_{(ZZ)}(q^2) &=& \frac{\alpha_w}{32\pi}
\frac{M^4_H }{M^2_W}\Big[\, 1+4\frac{M^2_Z }{M^2_H}
-4\frac{M^2_Z }{M^4_H}(2q^2-3M^2_Z)\Big]\, B_0(q^2, M^2_Z,M^2_Z)\, ,\\
  \label{HHpair}
{\widehat \Pi}^{HH}_{(HH)}(q^2) &=& \frac{9\alpha_w}{32\pi}\frac{M^4_H }{M^2_W}
B_0(q^2, M^2_H, M^2_H)\, .
\end{eqnarray}
Note  that  ${\widehat  \Pi}_{(tt)} (q^2)$ and  ${\widehat \Pi}_{(HH)}
(q^2)$ are identical  to  their  conventionally  defined counterparts,
{\em i.e.}, they receive no pinch contributions.

\setcounter{equation}{0}
\section {The resonant Higgs boson and unitarity} 
\indent 

In this section we show how one can obtain the results of the previous
section by resorting to the fundamental properties of unitarity and
analyticity of $S$-matrix elements.

As  explained  in  detail in \cite{PP2},  a  close   connection exists
between  gauge invariance and  unitarity, which is best established by
looking at the two  sides of the equation for  the  OT.  The OT  for a
given process $\langle b|T|a \rangle$ is
\begin{equation}
  \label {OT}
\langle b|\,(T-T^\dagger)\,| a \rangle\ =\  i\sum_m 
(2\pi)^4 \delta^{4}(P_m-P_a)\langle m |T|b \rangle^*\,
                                            \langle m |T|a \rangle\, ,
\end{equation}
where  the sum  $\sum_m$ should be   understood to be  over the entire
phase space and spins  of all possible on-shell intermediate particles
$m$.   The  RHS   of Eq.\  (\ref{OT})   consists  of  the  product  of
GFP-independent    on    shell     amplitudes,   thus   enforcing  the
gauge-invariance of the imaginary part of the amplitude on the LHS. In
particular, even though the LHS contains unphysical particles, such as
ghosts and  would-be  Goldstone bosons,  which    could give rise   to
unphysical   thresholds,   Eq.(\ref{OT})  guarantees   that  all  such
contributions    will   vanish.    In    general,  the  aforementioned
cancellation  takes  place  after contributions from  the propagator-,
vertex-,     and box-diagrams have  been    combined.  There are field
theories however, such  as scalar theories, or  QED, which allow for a
stronger version of  the equality given  in Eq.(\ref{OT}): The optical
relationship holds  {\em  individually} for the  propagator-, vertex-,
and box-diagrams.  On  the other hand,  in  non-Abelian gauge theories
this  stronger version   of the OT  does {\em   not} hold  in  general
\cite{axun};  this  is  so because,  unlike  their  scalar  or Abelian
counterparts,  the conventional  self-energies,  vertex and boxes  are
{\em  gauge-dependent}. But if one   rearranges instead the amplitudes
according to the PT algorithm, the same stronger version of the OT can
also be realized  in the context of  non-Abelian gauge theories at one
loop,   as  has     been  demonstrated in    a   series   of    papers
\cite{PP1,PP2,PRW}.   Specifically, let us  apply the PT on both sides
of   Eq.(\ref{OT}):  The  PT rearrangement   of   the tree-level cross
sections appearing in the RHS   gives rise to new  process-independent
(self-energy-like) parts, which are equal to the imaginary part of the
effective self-energies  obtained by the application of  the PT on the
one-loop expression for  the  amplitude $\langle a|T|b\rangle$ on  the
LHS  .  The same  result is true for  the  vertex- and box-like parts,
defined by the  PT on either side of  Eq.(\ref{OT}).  In  other words,
effective  sub-amplitudes  obtained after the  application   of the PT
satisfy the OT {\em individually}, {\em e.g.},
\begin{equation}
  \label{OTPT}
\Im m \Big( \langle  a|T|a\rangle_{\rm PT}^{j}\Big)\  =\ \frac{1}{2}\,
\sum_{f}\int  \Big(   \langle  f|T|a\rangle   \langle f|T|a\rangle^{*}
\Big)_{\rm PT}^{j}\, ,
\end{equation}
where the subscript  ``PT''  indicates that the PT   rearrangement has
been carried  out, and   the  index $j=S,V,B$,  distinguishes  between
effective self-energy, vertex, and boxes, respectively.

Turning to a specific example involving the Higgs  boson, let us apply
the  previous arguments  to  the   case  of  the process  $t\bar{t}\to
t\bar{t}$.   At  the  tree-level  this process  can be  mediated  by a
photon, a $Z$ boson, and a possibly resonant  $H$ scalar.  We focus on
the sub-amplitude which contains two intermediate  $W$ bosons. In that
case the OT yields
\begin{equation}
  \label{OTWW}
\Im m \langle t\bar{t}|T|t\bar{t}\rangle\ =\ \frac{1}{2}
\int dX_{\rm LIPS}\ \langle t\bar{t}|T|W^{+}W^{-}\rangle 
\langle W^{+}W^{-}|T|t\bar{t}\rangle^{*}\, ,
\end{equation}
where the Lorentz-invariant phase-space (LIPS) measure is defined as
\begin{equation}
  \label{LIPS}
\int dX_{\rm LIPS} = \frac{1}{(2\pi)^2}\, 
\int d^{4}k_1 \int d^{4}k_2\, \delta_+ (k^2_1-M^2_W)\delta_+ (k^2_2-M^2_W)
\delta^{(4)}(q-k_1-k_2)\, ,
\end{equation}
and $\delta_+(k^2-m^2)  \equiv \theta (k^0) \delta(k^2-m^2)$.   We now
introduce the abbreviations ${\cal  M} = \langle t(p_1) \bar{t}(p_2) |
T  |  t(p_1)\bar{t}(p_2)\rangle$  and  ${\cal  T}  =   \langle  t(p_1)
\bar{t}(p_2)  | T | W^{+}(k_+)  W^{-}(k_-)\rangle $,  and focus on the
RHS of  Eq.(\ref{OTWW}).    Diagrammatically, the amplitude  $\cal  T$
consists  of two distinct parts:    an  s-channel amplitude, $   {\cal
  T}_{s\mu\nu}$,  which  is given  in Figs.\   3(a)  and 3(b),  and  a
$t$-channel amplitude, $ {\cal T}_{t\mu\nu}$, which depends on the $b$
quark  propagator, as shown in Fig.\  3(c).  The subscript ``$s$'' and
``$t$'' refers to the corresponding  Mandelstam variables, {\em i.e.},
$s =  q^2 = (p_1+p_2)^2   = (k_++k_-)^2$,  and    $t = (p_1-k_+)^2   =
(p_2-k_-)^2$.  ${\cal T}_{s\mu\nu}$ can be further decomposed into two
different  $s$-exchange amplitudes:  one mediated  by  a Higgs  boson,
denoted  by  $ {\cal   T}_{s\mu\nu}^H$,  and one   mediated by the two
neutral  gauge   bosons   $\gamma$ and      $Z$,  denoted by    ${\cal
  T}_{s\mu\nu}^V$, with $V=\gamma, Z$.  The explicit form of the above
amplitudes reads:
\begin{eqnarray}
  \label{TSV}
{\cal T}_{s\, \mu\nu}^V &=& - \frac{g^2_w}{2}
  \sum_{V=\gamma, Z}\bar{v}(p_2)\gamma_{\rho}(g_v^V+g_a^V\gamma_5)u(p_1)
  U_{V}^{\rho\lambda}(q)\Gamma_{\lambda\mu\nu}(q,k_+,k_-)\, ,\\
  \label{TSH}
{\cal   T}_{s\,   \mu\nu}^H   &=&    \frac{g^2_w}{2}\, m_t\, \bar{v}(p_2)u(p_1)
  \Delta_H(q)\, g_{\mu\nu}\, ,\\
  \label{Tt}
{\cal T}_{t}^{\mu\nu} &=& -\frac{g_w^2}{2}\, \bar{v}(p_2)\, \gamma^\nu P_L\, 
  \frac{1}{\not\! p_1-\not\! k_+ -m_b}\,  \gamma^\mu P_L\,  u(p_1)\, ,
\end{eqnarray}
with $\Delta_H (q)    = (q^2-M_H^2)^{-1}$, $g_v^{\gamma} =  4   \sin^2
\theta_w / 3$,  $g_a^{\gamma} = 0$,   $g_v^{Z} = 1/2  - g_v^{\gamma}$,
$g_a^{Z} = - 1/2$, and
\begin{equation}
  \label{3GV}
\Gamma_{\lambda\mu\nu}(q,k_+,k_-)= (k_--k_+)_{\lambda}g_{\mu\nu}
-(q+k_-)_{\mu}g_{\lambda\nu}+ (q+k_+)_{\nu}g_{\mu\lambda}\, .
\end{equation}
In Eq.(\ref{TSV}),  $U_Z^{\rho\lambda}(q)$  denotes the  propagator of
the $Z$ boson in the unitary gauge, and $U_\gamma^{\rho\lambda}(q)$ is
the photon propagator in an  arbitrary gauge.  The gauge dependence of
the    photon is trivially canceled,    as   soon as ${\cal   T}_{s\,
  \mu\nu}^V$ is  contracted with the  polarization  vectors of the $W$
bosons.    With    the definitions  given  above,   the   RHS of  Eq.\ 
(\ref{OTWW}) becomes
\begin{eqnarray}
  \label{Mew}
\Im m {\cal M} \ &=&\  
{\cal T}_{\mu\nu}Q^{\mu\rho}(k_+)Q^{\nu\sigma}(k_-){\cal T}^*_{\rho\sigma}
\nonumber\\
&=& [{\cal T}_{s\mu\nu}^V + {\cal T}_{s\mu\nu}^H +{\cal T}_{t \mu\nu}]
Q^{\mu\rho}(k_+)Q^{\nu\sigma}(k_-)
[{\cal T}_{s\rho\sigma}^V + {\cal T}_{s\rho\sigma}^H +{\cal
  T}_{t\rho\sigma}]^*\, ,
\end{eqnarray}
where 
\begin{equation}
  \label{WPol}
Q^{\mu\nu} (k)\ =\ -g^{\mu\nu}\, +\, \frac{k^\mu k^\nu}{M^2_W}\, 
\end{equation}
is the  $W$ polarization tensor.  Obviously, $k^{\mu}Q_{\mu\nu}(k)=0$,
when $k^2=  M^2_{W}$.  Furthermore, in  Eq.\ (\ref{Mew}), we   omit the
integration measure $1/2\int dX_{\rm LIPS}$.

Since  our main interest  lies in the Higgs-boson-mediated interaction
contained in the transition $t\bar{t}\to t\bar{t}$, we wish to isolate
the  part which depends  on the Higgs  boson. In  doing so particular
care   is  needed,   because, despite  appearances,   the  $t$-channel
amplitude ${\cal T}_{t}^{\mu\nu}$   contains contributions which   are
related to the Higgs-boson interaction.  These contributions emerge by
virtue of the following WI:
\begin{eqnarray}
  \label{TP}
\frac{k_{+}^{\mu}k_{-}^{\nu}}{M^2_W}
{\cal T}_{t\mu\nu} &=&
{\cal T}_{P}^{H}+ \dots \, , \nonumber\\ 
{\cal T}_{P}^{H} &=& - \frac{g_w^2}{4}\frac{m_t}{M^2_W}\bar{v}(p_2)u(p_1)\, ,
\end{eqnarray}
shown  schematically  in Fig.\ 2.  The  above WI   is triggered by the
longitudinal momenta $k_{+}^{\mu}$ and $k_{-}^{\nu}$, originating from
the  polarization tensors $Q^{\mu\rho}(k_+)$ and $Q^{\nu\sigma}(k_-)$,
respectively.   The ellipses   in  Eq.\  (\ref{TP}) denote  additional
contributions which are  not related to  the Higgs boson,  {\em i.e.},
their coupling  to the external fermions is  {\em not} proportional to
$m_t$.  Notice that the  combined action of  {\em both } $k_{+}^{\mu}$
and $k_{-}^{\nu}$ is necessary, in order for  the piece related to the
Higgs boson to appear.

We then proceed  to carry out  the multiplication  on the RHS  of Eq.\
(\ref{Mew}) (we   suppress Lorentz indices). To  begin  with, the term
${\cal T}_{s}^V Q(k_+)Q(k_-){\cal   T}_{s}^{V*}$ has no  dependence on
the Higgs boson, and we can discard it. In  addition, the terms ${\cal
  T}_{s}^H    Q(k_+)Q(k_-){\cal  T}_{s}^{V*}$ and    ${\cal   T}_{s}^V
Q(k_+)Q(k_-){\cal T}_{t}$  give Higgs-boson  related pieces, which are
however antisymmetric under the exchange $k_{+}\leftrightarrow k_{-}$,
and therefore vanish upon the  symmetric phase-space integration. This
may be readily verified, if one employs the following two identities:
\begin{eqnarray} 
  \label{GQQ}
\Gamma_{\lambda\mu\nu}(q,k_+,k_-)
Q^{\mu\rho}(k_+)Q^{\nu}_\rho (k_-) &=& 
\Big[\, \frac{(q^2)^2}{4M^4_W} -3\, \Big] (k_+ - k_-)_{\lambda} \, ,\\
  \label{kkG}
k_+^\mu k_-^\nu \Gamma_{\lambda\mu\nu}(q,k_+,k_-) &=&
\frac{q^2}{2}\, (k_+ - k_-)_{\lambda} \, .
\end{eqnarray} 
This  last  result   is    in  agreement  with  earlier   observations
\cite{APRL}, that any non-vanishing $Z H$ transition  would lead to CP
violation, and therefore, it     should be absent in   a  CP-invariant
theory,  such as the  bosonic part of the   bare Lagrangian of the SM.
Finally, the  part of Eq.\   (\ref{Mew})  related to  the Higgs  boson
reads:
\begin{equation}
{\Im m {\cal M}}_{\rm Higgs}= Q(k_+)Q(k_-){\cal T}_{s}^H{\cal T}_{s}^{H*}
+\big[ Q(k_+)Q(k_-) ({\cal T}_{s}^H {\cal T}_{t}^*
+{\cal T}_{t}{\cal T}_{s}^{H*}+{\cal T}_{t}{\cal T}_{t}^*)\big]_{\rm
  Higgs} \, .
\end{equation}
This last expression will be now separated into two distinct pieces as
follows:
\begin{equation}
  \label{HsHv}
{\Im m {\cal M}}_{\rm Higgs} = {\Im m \widehat {\cal M}}_{\rm self}^H+
{\Im m \widehat {\cal M}}_{\rm vert}^H \, ,
\end{equation}
{\em i.e.},  a universal,  self-energy-like  piece, ${\Im  m  \widehat
  {\cal M}}_{\rm self}^H$, which does not depend  on the propagator of
the $b$ quark, and a vertex-like piece ${\Im m \widehat {\cal M}}_{\rm
  vert}^H $, which explicitly contains the  $b$ quark propagator.  The
propagator-like contribution may be written as
\begin{eqnarray}
  \label{Hself}
{\Im m \widehat {\cal M}}_{\rm self}^H &=&
{\cal T}^H_{s\mu\nu}Q^{\mu\lambda}(k_+)Q^{\nu\rho}(k_-)
{\cal T}^{H*}_{s\lambda\rho}
+\Big(\,\frac{1}{2}g^{\mu\nu}+ \frac{k_{+}^{\mu}k_{-}^{\nu}}{M^2_W}\,\Big)
\Big({\cal T}^H_{s\mu\nu}{\cal T}_{P}^{H*} + 
{\cal T}_{P}^{H}{\cal T}^{H*}_{s\mu\nu}\Big)\nonumber\\
&&+\, {\cal T}_{P}^{H}{\cal T}_{P}^{H*} \, .
\end{eqnarray}
The closed expressions for the terms  on the RHS of Eq.\ (\ref{Hself})
are as follows:
\begin{eqnarray}
   \label{HS1}
\lefteqn{\hspace{-1.8cm} 
{\cal T}^H_{s\mu\nu}Q^{\mu\lambda}(k_+)Q^{\nu\rho}(k_-) 
{\cal T}^{H*}_{s\lambda\rho} \ =\
\Big(\frac{gm_t}{2M_W}\Big)^{2}\bar{v}(p_2)u(p_1)\Delta_H(q)} \nonumber\\
&&\times\, \Big(\frac{g^2}{4}\Big)\, [(q^2)^2-4q^2M_W^2+12M_W^4]\Delta_H(q)
\bar{u}(p_1)v(p_2)\, ,\\
\Big(\,\frac{1}{2}g^{\mu\nu}+ \frac{k_{+}^{\mu}k_{-}^{\nu}}{M^2_W}\,\Big)
   \label{HS2}
\lefteqn{ {\cal T}^H_{s\mu\nu}{\cal T}_{P}^{H*}+ \mbox{c.c.} \ =\
-\Big(\frac{gm_t}{2M_W}\Big)^{2}\bar{v}(p_2)u(p_1)\Delta_H(q) }\nonumber\\
&&\times \Big(\frac{g^2}{4}\Big)(q^2-M^2_H)(q^2+2M^2_W)
\Delta_H(q)\bar{u}(p_1)v(p_2)\, ,\\
   \label{HS3}
{\cal T}_{P}^{H}{\cal T}_{P}^{H*} & =& 
\Big(\frac{gm_t}{2M_W}\Big)^{2}\bar{v}(p_2)u(p_1)\Delta_H(q)\,
\Big(\frac{g^2}{4}\Big)\, (q^2-M^2_H)^2\bar{u}(p_1)v(p_2)\, ,
\end{eqnarray}
where  the abbreviation c.c.\   stands for complex conjugation.  After
adding the above  propagator-like  contributions and carrying out  the
two $W$-boson phase-space integration, we define the imaginary part of
the effective PT self-energy for  the Higgs boson in the  conventional
way,  {\em  i.e.},  as   the part of  the   above  amplitude which  is
sandwiched between the two bare Higgs boson propagators $\Delta_H(q)$.
In this way we obtain
\begin{equation}
  \label{Ufin}
\Im m \widehat{\Pi}^{HH}_{(WW)}(q^2)\ =\ 
\frac{\alpha_w}{16}\frac{M^4_H}{M^2_W}
\Big(1-\frac{4M^2_W}{q^2}\, \Big)^{1/2}\, \Big[\, 1+4\frac{M_W^2}{M_H^2}- 
4\frac{M_W^2}{M_H^4}(2q^2-3M_W^2)\, \Big] \, .
\end{equation}
Notice the crucial cancellation of the $(q^2)^2$ terms; had such terms
survived, they would  have given rise  to a running width which  would
grossly  contradict the ET  (see  also discussion in  Section 6). Eq.\
(\ref{Ufin}) is in agreement with the result reported in \cite{APRL}.

We can now easily  establish contact with  the results of the previous
section.   Starting  from Eq.\   (\ref{HPT}),   we can arrive at  Eq.\
(\ref{Ufin}) by using the following relation:
\begin{eqnarray}
  \label{imhz}
\frac{1}{16\pi^2}\, \Im m\ B_0(q^2,m_1^2,m^2_2) &=&
-\frac{1}{16\pi^2}\, \Im m\Big\{ \, \int^1_{0} dx
\ln[m_1^2x+m_2^2(1-x)-q^2x(1-x)]\Big\}
\nonumber\\
&=& \theta [ q^2-(m_1+m_2)^2 ]\, \frac{1}{16\pi q^2}\, 
\lambda^{1/2}(q^2,m^2_1,m^2_2)\nonumber\\
&=& \frac{1}{2}\int dX_{\rm LIPS}\, ,\qquad
\end{eqnarray}
with $\lambda (x,y,z) = (x-y-z)^2 - 4yz$.   Of course, for the case at
hand, we  have  $m_{1}=m_{2}=M_{W}$.  Conversely, we  can recover from
Eq.\ (\ref{Ufin}) the on-shell renormalized result of Eq.\ (\ref{HPT})
by means  of    a twice-subtracted   (on shell)   dispersion  relation
\cite{PP2,PRW}.

The contribution  to the PT   Higgs self-energy, which comes from  two
intermediate $Z$ bosons,  may be  obtained in  an  analogous way.  For
definiteness, in Fig.\ 4 we plot  separately the dependence of all the
kinematic  channels  involved in  $\Im m  \widehat{\Pi}^{HH}(s)$  as a
function of the  centre-of-mass  (c.m.)  energy $\sqrt{s}$.  The solid
line corresponds to  the total effect  of all intermediate states.  In
Fig.\ 4(a), we  have displayed the results  of a light  Higgs scenario
with a mass $M_H = 300$ GeV, whereas predictions  obtained for a heavy
Higgs  with $M_H = 700$  GeV are presented in  Fig.\ 4(b). Notice that
the   absorptive   part   of   the    bosonic    channels  $\Im      m
\widehat{\Pi}^{HH}_{(VV)}(s)  =  \Im  m \widehat{\Pi}^{HH}_{(WW)}(s) +
\Im m \widehat{\Pi}^{HH}_{(ZZ)}(s)$, represented by a dash-dotted line
in both plots, turns negative far  above the resonant point $s=M^2_H$,
as can be readily deduced  from the closed  expressions given in Eqs.\ 
(\ref{HPT})  and         (\ref{ZZpair}).    Specifically,   $\Im     m
\widehat{\Pi}^{HH}_{(VV)}(s)$ turns negative when $\sqrt{s} > 430$ GeV
for $M_H=300$ GeV,  and $\sqrt{s} > 2$ TeV  for $M_H =  700$ GeV.  The
dependence of $-\Im m  \widehat{\Pi}^{HH}_{(VV)}(s)$ on $\sqrt{s}$  is
indicated  by a long-dash-dotted  line.  However,  we must remark that
the total absorptive part of the Higgs  boson self-energy stays always
positive due to the large positive contribution of the heavy top quark
($m_t = 170$ GeV).  Thus, at c.m.\ energies $\sqrt{s} \gg M_H$, $\Im m
\widehat{\Pi}^{HH}(s)$ has the following asymptotic behaviour:
\begin{equation}
  \label{HHasym}
\Im m  \widehat{\Pi}^{HH}(s)\ \sim \ \frac{\alpha_w\, s}{8M^2_W}\,
(3m^2_t\, -\, 4M^2_W\, -\, 2M^2_Z)\, .
\end{equation}
The fact that the bosonic contributions to  the absorptive part of the
Higgs-boson self-energy is negative at large $s$ is reminiscent of the
PT gauge-boson self-energies  in   theories with  asymptotic  freedom,
whose     absorptive  parts are   also   negative.   For instance, the
absorptive part of the PT (or BFM) gluon self-energy in quark-less QCD
has the exact same  feature, and, as a result,  it does not admit  the
usual K\"allen-Lehmann spectral representation.  By analogy, far above
the resonant  point, the resummed    Higgs-boson propagator loses  its
meaning as a  description  of the BW  dynamics  of the unstable  Higgs
particle, but it rather   serves as the   ``effective charge'' of  the
universal   Higgs-mediated part of    the electroweak interaction.  In
Section 5.3, we will take a closer look at this issue.

In the following, we  study the  resonant behaviour of the
resummed Higgs-boson propagator
\begin{equation}
  \label{DelH}
\widehat{\Delta}^H (s) \ =\ [\, s\, -\, M^2_H\, +\, 
\widehat{\Pi}^{HH}(s)\, ]^{-1}\, ,
\end{equation}
within different approaches. For example, $\widehat{\Delta}^H (s)$ may
occur in the process $t\bar{t} \to H^*  \to t\bar{t}$.  In Fig.\ 5, we
display  the dependence of  the  modulus of  the  resummed Higgs-boson
propagator as a  function of the  c.m.\ energy  $\sqrt{s}$.  The solid
line   refers to the result obtained   in the PT resummation approach,
whereas  the  dashed,   dotted and   dash-dotted lines  correspond  to
resumming Higgs self-energies in the  BFG  with $\xi_Q = 100,\  1000$,
and  in the unitary   gauge, respectively.  Notice  the characteristic
presence of  unphysical  thresholds   in  the BFG,   which    manifest
themselves as artificial resonances.  As  can also be seen from Figs.\
5(a)  and 5(b)  (for $M_H =300$  and  $700$ GeV, respectively), in the
unitary gauge   the   width  increases as   $s^2$   and  distorts  the
Higgs-boson lineshape.  As a final remark, we point out that the usual
description of  unstable  particles by  means   of a  constant   width
approach,  where  $M^2_H - \widehat{\Pi}^{HH}(s)$   is replaced by the
complex pole $M^2_H  - iM_H\Gamma_H$  in $\widehat{\Delta}^H (s)$  for
any value of $s$, leads in the limit $s \to 0$ to a non-vanishing $\Im
m \widehat{\Delta}^H (s)$, and therefore violates the OT \cite{PP1}.

\setcounter{equation}{0}
\section{Renormalization group analysis}

The ultimate goal of this program is to provide a systematic framework
for  constructing  physically meaningful Born-improved  approximations
for  resonant  transition amplitudes.  In   doing  so, we  have mainly
focused   on gauge-invariance and  unitarity,   and shown how  one can
manifestly maintain such crucial properties even when resonant bosonic
contributions are  considered.  In the   next two sections  we turn to
another  important  property,   namely  the invariance  of    the Born
amplitudes under  the renormalization  group.  In particular,  we will
show explicitly that the amplitudes obtained by our resummation method
are built   out of  renormalization-group-invariant        structures.
Furthermore, we will demonstrate  how one can generalize the effective
charge, a familiar concept in the context of  gauge bosons such as the
$W$  and  $Z$ bosons, to  the  case of the scalar   Higgs boson.  This
scalar  ``effective charge'' constitutes a   common component in every
Higgs-boson mediated process, regardless of the nature of incoming and
outgoing  states,     and  can  thus  be    viewed   as   a universal,
process-independent entity, intrinsic to the Higgs boson.

This section is  organized as follows: We first  review the concept of
the effective charge  in the   context  of QED;  then  we discuss  its
generalization  to the case of a   non-Abelian gauge theories, such as
QCD.  The crux of this analysis is that  by virtue of the WI's present
in gauge   theories the effective charge  is  both invariant under the
renormalization group  and  process-independent.  At  the end  of this
section   we discuss  a counter-example,  {\em  i.e.}, the  case of an
asymptotically free   scalar model in   six space-time dimensions, and
analyze the reasons which make the construction of an effective charge
not possible. In particular, we explain why  in this theory one cannot
reconcile  invariance   under   renormalization   group   and  process
independence.  Interestingly enough,  the construction of an effective
charge in  a scalar context becomes  again  possible after adding, and
subsequently breaking  spontaneously, a global   symmetry, as we  will
show at the end of this section.

\subsection{Effective charge in QED}

We  start our discussion  with  the case   of QED.  The  Abelian gauge
symmetry  of the   theory   gives  rise   to  the    fundamental   WI:
$q_{\mu}\Gamma^{\mu , 0}(p,p+q)= S^{0\, -1}(p+q)-S^{0\, -1}(p)$, where
$\Gamma^{\mu,0}$ is the bare  photon-electron vertex and $S^0 (k)$ the
dressed    electron propagator.  The  above   identity  is valid  both
perturbatively to  all  orders, as well   as non-perturbatively.   The
requirement  that     the   renormalized vertex       $\Gamma^{\mu}  =
Z_{1}\Gamma^{\mu ,   0}$   and  the  renormalized  self-energy   $S  =
Z_{f}^{-1} S^0$ satisfy the same identity imposes the equality between
the  vertex  renormalization    constant $Z_{1}$   and   the  electron
wave-function renormalization constant  $Z_{f}$, namely $Z_{1}=Z_{f}$. 
As a result, the photon wave-function renormalization constant $Z_{A}$
and the   charge renormalization constant  $Z_{e}  =  Z_{1} Z_{2}^{-1}
Z_{3}^{-1/2}$ are related by the following fundamental equation:
\begin{equation}
Z_{e}\ =\ Z_{A}^{-1/2}.
\end{equation}

The unrenormalized photon self-energy   is $\Pi^0_{\mu\nu}(q) =  (-q^2
g_{\mu\nu}  +  q_{\mu}q_{\nu})\Pi^0(q^2)$,  where  $\Pi^0 (q^2)$ is  a
GFP-independent function to all  orders in perturbation theory.  After
performing the  standard  Dyson   summation,  we obtain   the  dressed
propagator between conserved external currents
\begin{equation}
  \label{Dqed}
\Delta^0_{\mu\nu}(q)\ =\ \frac{-g_{\mu\nu}}{q^2\, [1+\Pi^0(q^2)]} \, .
\end{equation}
The   above   quantity   is   universal,   in   the   sense  that   is
process independent. We can now form the following RGI combination:
\begin{equation}
  \label{RGICqed}
R^{e}_{\mu\nu}(q^2)\ \equiv\ \alpha_{\rm eff}(q^2)\, 
\frac{-g_{\mu\nu}}{q^2} \, ,
\end{equation}
where 
\begin{equation}
  \label{alphaqed}
\alpha_{\rm eff}(q^2)\ =\ \frac{(e^0)^2}{4\pi}\, \frac{1}{1+\Pi^0 (q^2)}
\ =\ \frac{e^2}{4\pi}\, \frac{1}{1+\Pi (q^2)} \, .
\end{equation}
The last equality in Eq.\ (\ref{alphaqed})  can be readily obtained if
one uses    the relations  between  renormalized    and unrenormalized
parameters:    $e^2 = (Z_f^2Z_A/Z_1^2)     (e^0)^2$, $1+\Pi (q^2)= Z_A
[1+\Pi^0  (q^2)]$,  and $Z_1=Z_f$.   For  $q^2\gg  m_e^2$, $\alpha_{\rm
  eff}(q^2)$ coincides  with  the  one-loop  running coupling  of  the
theory.  We  must remark that the effective  charge  has a non-trivial
dependence on the  masses of  the  particles appearing  in the quantum
loops,  which   allows its   reconstruction  from  physical amplitudes
\cite{PRW}.  In general, the transition  amplitude  of a QED  process,
such  as $e^+  e^-\to e^+ e^-$,  consists of  two RGI  combinations: a
process-independent one,   namely  the  effective  QED charge  defined
above, and a process-dependent  one, namely the sum  of vertex and box
diagrams.

\subsection{Effective charge in QCD}

In non-Abelian gauge theories  the crucial equality $Z_1=Z_f$ does not
hold in  general.  Furthermore, in contrast to   the photon  case, the
gluon vacuum  polarization depends on  the  GFP, already  at one  loop
order.  These  facts make the  non-Abelian  generalization  of the QED
concept of  the  effective  charge  non-trivial.  The  possibility  of
defining an effective  charge for QCD in  the framework of the PT  was
discussed first by Cornwall \cite{PT}, and was further investigated 
in a series of recent papers \cite{PP1,NJW1,NJW2}.

The PT   rearrangement  of   physical amplitudes  gives   rise   to  a
GFP-independent effective gluon self-energy,  and restores at the same
time the equalities
\begin{equation}
\widehat Z_1\ =\ \widehat Z_f\, ,\qquad \widehat Z_g\ =\ 
\widehat Z_{A}^{-1/2} \, ,
\end{equation}
where the carets denote the corresponding renormalization constants in
the PT, and $g$ is the QCD coupling. Having restored QED-like WI's and
GFP  independence, and    using the additional    fact  that  the   
one-loop
PT
self-energy is  process-independent \cite{NJW0}
and can be Dyson-resummed to all orders \cite{PP1,NJW2},
the  construction of
the  universal RGI combination   and the  corresponding QCD  effective
charge is immediate.  We have
\begin{equation}
  \label{RGICqcd}
\widehat R^{g}_{\mu\nu} (q^2)\ \equiv\
 \alpha_{s,\rm eff}(q^2)\frac{-g_{\mu\nu}}{q^2} \, ,
\end{equation}
where 
\begin{equation}
  \label{alphaqcd}
\alpha_{s, \rm eff}(q^2)\ =\ \frac{(g^0)^2}{4\pi}\, 
 \frac{1}{1+\widehat{\Pi}^0 (q^2)}\ =\ 
  \frac{g^2}{4\pi}\, \frac{1}{1 + \widehat{\Pi}(q^2)} \, .
\end{equation}
It is  interesting  to note  that in  the BFM formulation  of QCD the
Green's functions  satisfy by construction QED  WI's, to all orders in
perturbation theory.  On   the other hand,  the BFM  Green's functions
still  depend on  the GFP $\xi_Q$.   In  the case of the  gluon vacuum
polarization this dependence on $\xi_Q$ is trivial,  since it does not
affect the prefactor of the leading logarithm, and is just an additive
constant.  This constant may be considered as  an arbitrariness in the
renormalization  scheme, and will   hence  disappear when forming  the
scheme-independent RGI  quantity given in  Eq.\ (\ref{RGICqcd}).  This
is however not  true in the case of  massive gauge  bosons; there, the
dependence on  the GFP cannot  be removed  by means  of an appropriate
choice of renormalization scheme.

\subsection{The scalar case}

The ability to define a  {\em process-independent} RGI quantity is not
a common characteristic of all scalar field theories; for example this
is not what happens in the case of pure  scalar theories.  However, if
the scalar theory is spontaneously broken, then a RGI effective charge
may be defined for the Higgs boson, which is inversely proportional to
its vacuum expectation  value  (VEV).   In  the following,   we  shall
examine both situations.

Let us first study $(\phi^{3})_{6}$,  {\em i.e.}, scalar $\phi^{3}$ in
six  space-time dimensions.  The  theory is asymptotically free, gauge
invariance is of course not an issue, and just as in QED, the OT holds
for  individual Feynman diagrams, and  the self-energy can be formally
Dyson-resummed.  However, unlike QED, if  one was to use this formally
resummed  self-energy inside   a  tree-level amplitude, the  resulting
expression would  not be renormalization  group invariant.  The reason
is that there is no QED-like  WI enforcing the equality between vertex
and wave-function renormalization.  As a  result  of that, it is  only
after  the vertex  correction have  been included  that the  resulting
combination  becomes a RGI combination.   The drawback of this is that
the inclusion of the vertices introduces process-dependence.  In other
words, in the $(\phi^{3})_{6}$ case it  is not possible to construct a
RGI  quantity which is, at  the  same time, process independent,  {\em
  i.e.}, one cannot reconcile process independence and renormalization
group invariance \cite{Velt2}.

To see this in detail,  let us study the  Veltman model \cite{Velt} at
$d=6$ instead of $d=4$.  This  theory contains a light scalar, $\phi$,
and a  heavy scalar,  $\Phi$, with a  mass $M_\Phi  > 2 M_\phi$.   The
heavy scalar decays into two $\phi$'s, via the interaction term in the
Lagrangian
\begin{equation}
{\cal L}_{int}\ =\ \frac{\lambda}{2} \phi^2\Phi,
\end{equation}
where $\lambda$ is a non-zero   coupling constant.  The  wave-function
renormalization constants $Z_{\phi}$  and $Z_{\Phi}$, corresponding to
the  fields   $\phi$  and   $\Phi$,   respectively,  and    the vertex
renormalization  $Z_{\phi^2\Phi}$ have been calculated 
in the minimal subtraction scheme \cite{phi6}.  They are
\begin{equation} 
Z_{\phi}\ =\ Z_{\Phi}\ =\ 1\, +\, \Big(\frac{1}{6}\Big)\, 
\frac{g^2}{64\pi^3\epsilon}\, , \qquad
Z_{\lambda}\ =\ 1\, +\, \frac{g^2}{64\pi^3\epsilon} \, . 
\end{equation}
Clearly,  one has $Z_{\Phi}\neq Z_{\lambda}$.    Of course, since  the
pole terms of $Z_{\Phi}$   and $Z_{\lambda}$ are different, the  above
inequality  will  be  true   in  any other  renormalization   scheme.  
Consequently, the charge renormalization $Z_{\lambda}$, defined by the
equation   $\lambda^0=Z_{\lambda}\lambda$.   As  a  result,  for   the
combination $(\lambda^0)^{2} \Delta^0 (s)$, which is the direct analog
of the QED effective charge,  we have that $(\lambda^0)^2\Delta^0  (s)
\neq  \lambda^2 \Delta (s)$.   Therefore, in order  to arrive at a RGI
expression, the  vertex   correction must   be supplemented.  So,  the
combination
\begin{equation}
\phi^0\phi^0\, \Gamma^0\, \Delta^0\, \Gamma^0\, \phi^0\phi^0\ =\
 \phi\phi\,\Gamma\, \Delta\, \Gamma\, \phi\phi
\end{equation}
is a RGI quantity, but unlike the QED case, it {\em cannot} be written
as the product of a process-independent  and a process-dependent part,
which are individually RGI.

To explore further     how  the  process-dependence  enters,   it   is
instructive to add yet another set of  scalar fields $\chi$, such that
$M_\Phi > 2M_\chi > 2 M_\phi$, and an extra interaction term
\begin{equation}
{\cal L}'_{\rm int}\ =\ \frac{g}{2} \chi^2\Phi \, ,
\end{equation}
where $g$ is  another non-zero coupling constant,  with $g\neq\lambda$
in general. We will ignore for simplicity additional interaction terms
such as   $\phi^{3}$,   $\chi^{3}$,  $\phi^{2}\chi$,   $\chi^{2}\phi$.
Several  of  them  may be  eliminated  by imposing  an  extra  mirror
symmetry  of the type   $\phi \rightarrow -\phi$ and  $\chi\rightarrow
-\chi$; in any case   the presence of  such terms  does not alter  our
conclusions qualitatively \cite{coll}.

In order to  mimic gauge theories,  we  next set $g=\lambda$.  Let  us
consider  two different  processes,  $\phi\phi\rightarrow\phi\phi$ and
$\chi\chi\rightarrow\chi\chi$,    both  mediated    by  an $s$-channel
resonant $\Phi$.   The RGI    quantities for  the two  processes   are
$\phi\phi     \Gamma^{\phi^2\Phi}     (s,M_{\Phi},M_\phi)    \Delta(s)
\Gamma^{\phi^2\Phi}   (s,M_{\Phi},M_\phi)   \phi\phi$   and  $\chi\chi
\Gamma^{\chi^2\Phi} (s,M_{\Phi},M_\chi)  \Delta(s) \Gamma^{\chi^2\Phi}
(s,M_{\Phi},M_\chi) \chi\chi$,  where we have explicitly displayed the
dependence    of    the         vertices    $\Gamma^{\phi^2\Phi}$  and
$\Gamma^{\chi^2\Phi}$  on the masses.  It is  now easy to see that the
process   dependence   enters    through   the   simple   fact    that
$\Gamma^{\phi^2\Phi}$ depends on $M_\phi$ but not on $M_\chi$, whereas
the  reverse is   true  for  $\Gamma^{\chi^2\Phi}$.\footnote[1]{It  is
  elementary to  verify that the  functional  dependence of the vertex
  functions on the respective masses is non-trivial.} Evidently, there
is no RGI quantity common in these two amplitudes.

Let us now consider  a  four-dimensional $\Phi^4$ scalar theory  which
has a U(1)  global invariance and includes  a fermion $f$. The fermion
$f$ is introduced  in order to  prevent  the scalar theory from  being
super-renormalizable, so that   one  is able to  study   non-trivial
renormalization  effects. The  part of the   Lagrangian related to the
Higgs potential of the model has the form
\begin{equation}
  \label{SSBpot}
{\cal L}_V\ =\ \mu^2\Phi^*\Phi\, +\, \lambda (\Phi^*\Phi )^2\, .
\end{equation}
The interaction of the scalar $\Phi$ to the fermion $f$ is given by
\begin{equation}
  \label{ffPhi}
{\cal L}_{\rm int}\ =\ g\Phi\, \bar{f}_Lf_R\, +\quad \mbox{H.c.}
\end{equation}
The global U(1) symmetry of ${\cal L}_V$ in Eq.\ (\ref{SSBpot}) breaks
down spontaneously and  the  resulting theory resembles the  un-gauged
SM, where the  fermion $f$ may represent for   example the top  quark. 
Specifically, the field $\Phi$ must  be expanded around its VEV,  {\em
  i.e.},  $\Phi = (v + H   + iJ)/\sqrt{2}$, where the  field  $H$ is a
CP-even Higgs  boson  with mass  $M_H  = \sqrt{2}\mu$  and $J$  is the
massless CP-odd  Goldstone boson associated with   the breaking of the
continuous  U(1) symmetry.  After  the SSB of   the U(1) symmetry, the
fermion  $f$  acquires a mass $m_f  =  gv/\sqrt{2}$.  If $M_H > 2m_f$,
then the decay process   $H\to f\bar{f}$ is kinematically  allowed and
the Higgs boson becomes an unstable particle.

Beyond  the  Born approximation, the wave-function  renormalization of
the     Higgs field     $Z^{1/2}_{\Phi}$  renders  the    VEV $\langle
\Phi^0\rangle \equiv v^0$ UV finite, {\em viz.}, $v^0 = Z^{1/2}_{\Phi}
(v + \delta v)$, with vanishing divergent part $(\delta v)^{\rm div} =
0$. As a  consequence,   the  expression  $\Delta^0 (s)  /   (v^0)^2$,
involving the resummed Higgs-boson propagator is RGI, {\em i.e.},
\begin{equation}
  \label{RGIscalar}
\frac{1}{(v^0)^2}\, \Delta^0 (s)\ =\  \frac{1}{v^2}\, \Delta (s).
\end{equation}
It  is then obvious  that the VEV of the  field $\Phi$ in a SSB scalar
theory  plays an instrumental role in  defining a RGI effective charge
for the  Higgs interactions, very  much in analogy  to the QED and QCD
cases discussed  earlier.   The Higgs   field couples  universally  to
matter with  a ``charge''  inversely proportional  to  its VEV in  the
symmetric U(1)  limit.   If one now  wishes  to embed this  scalar SSB
model into a  gauge theory, the situation  becomes  more involved.  In
fact, within conventional gauge-fixing schemes such as $R_\xi$ gauges,
$\delta v$ is not UV finite and  hence, Eq.\ (\ref{RGIscalar}) does no
longer hold.  In the next section, we shall discuss the possibility of
identifying  RGI effective charges for the   gauge and Higgs bosons in
the electroweak sector of the SM.

\setcounter{equation}{0}
\section{Effective charges in the electroweak sector of \newline 
the SM}

In the previous section we established  in detail the conditions which
enable the  construction of {\em process-independent} RGI combinations
for the gauge bosons of both Abelian and non-Abelian theories (QED and
QCD, respectively).   In this section we extend  this  analysis to the
electroweak sector of the SM. First, we review how the construction of
effective charges associated  to the  gauge bosons  of the theory   is
possible by  virtue of  the  WI's relating the  PT effective $n$-point
functions \cite{PRW}.  Furthermore,   we  discuss for the first   time
various   subtleties related  to the   definition  of the $W$ and  $Z$
effective charges,    which  originate    from  the fact    that   the
corresponding gauge bosons are unstable.  We then  turn to the case of
the Higgs boson  and examine  the possibility  of constructing  a {\em
  process-independent}  RGI quantity for the case  of the Higgs boson.
The  answer to this  question is by no means  obvious, since the Higgs
boson results      from      the   SSB   of    the     gauge     group
SU(2)$_L\otimes$U(1)$_Y$, and  gauge-fixing and  ghost terms  spoil in
general the equality  (\ref{RGIscalar}).  However,  it turns out  that
because of  the PT  WI's, it  is possible to  construct a  Higgs-boson
``effective charge,'' in direct analogy to the gauge boson case.

\subsection{The PT Ward identities of the SM}

It is well-known that in the PT  effective $n$-point functions satisfy
(at least at one loop) naive, tree-level like WI's, as happens in QED.
This is to  be  contrasted to  the conventional  $n$-point  functions,
which in general satisfy Slavnov-Taylor  identities, which involve the
Green's functions of the unphysical ghosts of the theory.  The PT WI's
are a  direct consequence of the requirement  that the S-matrix be GFP
independent.   The PT WI's for the  electroweak sector  of the SM have
been derived in \cite{JP90,PP1}.  In fact, based  on these WI's, it is
possible to prove  a stronger version of  this GFP independence of the
S-matrix: The S-matrix satisfies  the ``dual gauge-fixing  property,''
which states that one is free to choose  different GFP's for the gauge
bosons  inside  and outside the  quantum  loops \cite{DGF}.  The above
property  is  intrinsic to the  S-matrix,   and is  not linked  to any
special  gauge-fixing procedure.     Its derivation is   based on  the
observation that the PT rearrangement gives rise to one-loop $n$-point
functions for which all dependence on  the GFP stemming from the gauge
bosons inside  the quantum loops has  disappeared, {\em regardless} of
the choice of the GFP for the gauge bosons outside the loops.  For the
final cancellation of this latter  gauge dependence to go through, the
$n$-point  functions constructed  via the  PT must satisfy  tree-level
WI's.  Explicit calculations have demonstrated that this is indeed the
case.  The   above    ``dual gauge-fixing property''   holds  for  the
unrenormalized S-matrix.  If   one  imposes the requirement   that the
renormalized PT $n$-point functions satisfy exactly the same set of WI
as  their unrenormalized counterparts,   one  then concludes that  the
``dual gauge-fixing property''   holds  also after  renormalization.   
After enforcing this last requirement one finally  arrives at a set of
conditions   relating   the   various wave-function-   and   coupling-
renormalization constants of the theory.

To see  this in detail, we start  by listing the relations between the
bare and renormalized parameters for the electroweak sector of the SM.
We indicate all  (bare) unrenormalized quantities with the superscript
`0'. For the masses we have
\begin{eqnarray}
  \label{massren}
(M^0_W)^2 &=& M_W^2\ +\ \delta M_W^2\, , \qquad
(M^0_Z)^2\ =\ M_Z^2\ +\ \delta M_Z^2\, , \nonumber\\
(M^0_H)^2 &=& M_H^2\ +\ \delta M_H^2\, ,\qquad
m^0_f\ =\ m_f\ +\ \delta m_{f}\, .
\end{eqnarray}
In addition, the wave-function renormalizations are given by
\begin{eqnarray}
  \label{WFR}
W^{\pm,0}_\mu &=& \widehat{Z}_W^{1/2} W^{\pm}_\mu\, , \qquad
Z^0_\mu\ =\ \widehat{Z}_{Z}^{1/2} Z_\mu\, , \nonumber\\
G^{\pm,0}&=& \widehat{Z}_{G^{\pm}}^{1/2} G^{\pm}\, , \qquad
G^{0,0}\ =\ \widehat{Z}_{G^0}^{1/2}G^0\, , \nonumber\\
H^0 &=& \widehat{Z}_H^{1/2} H\, , \qquad
f^0_{L(R)}\ =\ \widehat{Z}^{L(R)1/2}_f f_{L(R)}\, , \nonumber\\
g^0_w &=&\widehat{Z}_{g_w}\, g_w\, , \qquad
c^0_w\ =\ \widehat{Z}_{c_{w}}c_{w}\, ,
\end{eqnarray}
with
\begin{equation}
\widehat{Z}_{c_w}\ =\ \Big(\, 1\, +\, \frac{\delta M_W^2}{M_W^2}\, \Big)^{1/2}
\Big(\, 1\, +\, \frac{\delta M_Z^2}{M_Z^2}\, \Big)^{-1/2}\, .
\end{equation}
If we expand $\widehat{Z}_{c_{w}}$ perturbatively, we have
\begin{equation}
\widehat{Z}_{c_{w}}\ =\ 1+\frac{1}{2}\frac{\delta c^2_w}{c^2_w} + \dots\,,
\end{equation}
with
\begin{equation}
\frac{\delta c^2_w}{c^2_w}\ =\ \frac{\delta M_W^2}{M_W^2}-
\frac{\delta M_Z^2}{M_Z^2}\, ,
\end{equation}
which is the usual one-loop result.  The  carets in the above formulas
indicate  as   usual that  both  the   calculations as  well   as  the
renormalization procedure are carried out in the PT framework.

Imposing the requirement that the  PT Green's functions should respect
the  same  WI's  before and   after renormalization  we  arrive at the
following relations:
\begin{eqnarray}
  \label{W1} 
\widehat{Z}_W &=& {\widehat{Z}}_{g_w}^{-2}\, ,\\
  \label{W2}
\widehat{Z}_Z &=&  \widehat{Z}_W \widehat{Z}_{c_{w}}^2 \, ,\\
  \label{W3}
\widehat{Z}_{G^0} &=& \widehat{Z}_{G^{\pm}}\ = \ 
\widehat{Z}_W\ \Big(\, 1\, +\, \frac{\delta M^2_W}{M^2_W }\, \Big)  \, , \\
  \label{W4}
\widehat{Z}^L_u&=& \widehat{Z}^L_d \, .
\end{eqnarray}
In deriving the above expressions, the following {\em exact} algebraic
identity may be used
\begin{equation}
\Big(\, 1\,+\,\frac{\delta M^2_W}{M^2_W }\,\Big)\, \Big(\, 1\,-\,
\frac{\delta M^2_W}{(M^0_W)^2}\, \Big)\ =\ 1\, .
\end{equation}

It is important to notice that the relations listed above are exact to
{\em  all  orders} in   perturbation theory.   Instead, in  the  usual
perturbative treatment, one sets $\widehat{Z}_i^{1/2}= 1 + \frac{1}{2}
\delta  \widehat{Z}_i$,  for $i=W,Z,H,f$,   and neglects higher  order
terms.  For  example, at  one loop order,  {\em  i.e.}, if we  neglect
terms of  order  $g^{4}$  and higher,  the  relation  Eq.\  (\ref{W3})
becomes
\begin{equation}
  \label{W3pert}
\delta\widehat{Z}_{G^{0}}\ =\ \delta\widehat{Z}_{G^{\pm}}\ =\ 
\delta\widehat{Z}_W\, +\,  \frac{\delta M^2_W}{M^2_W }\, .
\end{equation}

It is instructive  to show with an  explicit example how the relations
in Eq.\ (\ref{W3}) may be  derived. To this  end, we first define  the
proper unrenormalized one-loop vertices:
\begin{eqnarray}
  \label{overWdu}
\overline{\Gamma}_\mu^{W^+ \bar{u}d,0}(q,p_{\bar{u}},p_d) &\equiv& 
\Gamma_{0\mu}^{W^+ \bar{u}d,0}\, +\, \widehat{\Gamma}_\mu^{W^+
  \bar{u}d,0}(q,p_{\bar{u}},p_d)\, ,\nonumber\\
\overline{\Gamma}^{G^+ \bar{u}d,0} (q,p_{\bar{u}},p_{d}) &\equiv& 
\Gamma_0^{G^+ \bar{u}d,0}\, +\, \widehat{\Gamma}^{G^+ \bar{u}d,0}
(q,p_{\bar{u}},p_d)\, ,
\end{eqnarray}
where $\Gamma_{0\mu}^{W^+ \bar{u}d,0}$  and $\widehat{\Gamma}_\mu^{W^+
  \bar{u}d,0}$ are    the  tree-level and  one-loop   PT $W^+\bar{u}d$
vertices, respectively. Correspondingly, $\Gamma_0^{G^+   \bar{u}d,0}$
and   $\widehat{\Gamma}^{G^+  \bar{u}d,0}$  are   the   Born-level and
one-loop PT $G^+\bar{u}d$ vertices. Furthermore, if one neglects quark
mixing, the  bare dressed  $u$- and $d$-  type quark   propagators are
given by
\begin{eqnarray}
  \label{SuSd}
\widehat{S}^0_u (p_u) & =& [\, \not\! p_u\, -\, m^0_u\, +\, 
\widehat{\Sigma}^{\bar{u}u,0}(p_u)\, ]^{-1}\, ,\nonumber\\
\widehat{S}^0_d (p_d) & =& [\, \not\! p_d\, -\, m^0_d\, +\, 
\widehat{\Sigma}^{\bar{d}d,0}(p_d)\, ]^{-1}\, .
\end{eqnarray}
Because of the  fact that the bare effective PT vertices and self-energies
satisfy tree-level WI's, one then has 
\begin{equation}
  \label{WIbare}
q^\mu\,   \overline{\Gamma}_\mu^{W^+\bar{u}d,0}(q,p_{\bar{u}},p_d)\,
+\,   M_W^0\overline{\Gamma}^{G^+\bar{u}d,0}(q,p_{\bar{u}},p_d)\  =\
-\, \frac{ig^0_w}{\sqrt{2}}\, \Big[\, \widehat{S}^{0-1}_u (p_{\bar{u}})P_L\,
-\, P_R\widehat{S}^{0-1}_d(p_d)\, \Big]\, .
\end{equation}
The renormalized quantities are defined as follows:
\begin{eqnarray}
  \label{ZWG}
\frac{1}{g_w}\, \overline{\Gamma}_\mu^{W^+\bar{u}d} &=&
\widehat{Z}_{W^+\bar{u}d}\, \frac{1}{g^0_w}\, 
\overline{\Gamma}_\mu^{W^+\bar{u}d,0}\, ,\qquad
\frac{1}{g_w}\,\overline{\Gamma}^{G^+\bar{u}d}\ =\ 
\widehat{Z}_{G^+\bar{u}d} \frac{1}{g^0_w}\,
\overline{\Gamma}^{G^+\bar{u}d,0}\, , \\
  \label{ZZZ1}
\widehat{S}^0_d (p_d) &=& \widehat{Z}^{1/2}_d
\widehat{S}_d (p_d)\widehat{Z}^{1/2\dagger}_d\, ,\qquad
\widehat{S}^0_u (p_{\bar{u}})\ =\ \widehat{Z}^{1/2}_u
\widehat{S}_u (p_{\bar{u}}) \widehat{Z}^{1/2\dagger}_u\, ,
\end{eqnarray}
where $\widehat{Z}^{1/2}_f = \widehat{Z}^{L1/2}_f P_L + \widehat{Z}^{R
  1/2}_f   P_R$ and  $\widehat{Z}_{G^+\bar{u}d}\ =\ \widehat{Z}^L_{G^+
  \bar{u} d} P_L +\widehat{Z}^R_{G^+\bar{u}d}  P_R$.  Then, the vertex
renormalization   constants   $\widehat{Z}_{W^+     \bar{u}d}$     and
$\widehat{Z}^{L,R}_{G^+ \bar{u}d}$ are  related to the renormalization
constants introduced in Eq.\ (\ref{WFR}) as follows:
\begin{eqnarray}
  \label{ZWud}
\widehat{Z}_{W^+\bar{u}d} &=&
\widehat{Z}_{g_w}\widehat{Z}_W^{1/2}\widehat{Z}_d^{L1/2}
\widehat{Z}_u^{L1/2}\, ,\\
  \label{ZGud}
\widehat{Z}^L_{G^+\bar{u}d} &=&  
\widehat{Z}_{g_w}\widehat{Z}_{G^+}^{1/2}\widehat{Z}_d^{L1/2}
\widehat{Z}_u^{R1/2}\, ,\qquad
\widehat{Z}^R_{G^+\bar{u}d} \ =\  
\widehat{Z}_{g_w}\widehat{Z}_{G^+}^{1/2}\widehat{Z}_u^{L1/2}
\widehat{Z}_d^{R1/2}\, .
\end{eqnarray}
After  replacing  the  bare  by  the  renormalized quantities in  Eq.\
(\ref{WIbare})   by    means of    Eqs.\ (\ref{ZZZ1}),   (\ref{ZWud}),
(\ref{ZGud}) and   (\ref{massren}) for  the  mass renormalization,  we
require that  the  renormalized  WI retains  its  original form,  {\em
  i.e.},
\begin{equation}
  \label{WIren}
q^{\mu}\overline{\Gamma}_{\mu}^{W^+\bar{u}d}(q,p_{\bar{u}},p_d)\, 
+\, M_W\overline{\Gamma}^{G^+\bar{u}d}(q,p_{\bar{u}},p_d)\ =\
-\, \frac{ig_w}{\sqrt{2}}\, \Big[\, \widehat{S}^{-1}_u (p_{\bar{u}})P_L\,
-\, P_R\widehat{S}^{-1}_d(p_d)\, \Big]\, .
\end{equation}
The  above requirement  leads to  relations among the  renormalization
constants within the framework  of perturbation theory.  {}From the WI
involving  the  chirality  structure $P_R\overline{\Gamma}_\mu^{   W^+
  \bar{u} d, 0}  P_L$ in   Eq.\ (\ref{WIbare}),  we thus obtain   that
$\widehat{Z}_{g_w}   =   \widehat{Z}_W^{-1/2}$ and $\widehat{Z}^L_u  =
\widehat{Z}^L_d$,   which  are   Eqs.\    (\ref{W1})  and  (\ref{W4}),
respectively.     Furthermore, imposing   that  the  form   of  WI for
$P_L\overline{\Gamma}_\mu^{   W^+   \bar{u}      d,   0}    P_L$    or
$P_R\overline{\Gamma}_\mu^{ W^+  \bar{u} d, 0}  P_R$ remains  the same
after     renormalization    yields     $\widehat{Z}_{G^+}^{1/2}     =
\widehat{Z}_W^{1/2} (1+  \delta  M^2_W /  M^2_W)^{1/2}$,  which is the
last equality of Eq.\ (\ref{W3}).

Following an exactly  similar procedure we can  derive the rest of the
relations listed in Eqs.\ (\ref{W2}) and (\ref{W3}).

\subsection{Effective charges for the gauge bosons}

In    this  sub-section we     show  how   the  relations  among   the
renormalization constants derived above enable one to construct a {\em
  process-independent} RGI quantity     for the gauge  bosons   of the
theory.   For  definiteness, we  concentrate  on  the  $W$ boson,  but
similar arguments apply for the photon and the $Z$ boson.

First, we shall show that the  bare and renormalized PT {\em resummed}
$W$-boson       propagators,      ${\Delta}^{W,0}_{\mu\nu}(q)$     and
$\widehat{\Delta}^{W}_{\mu\nu}(q)$,  respectively,     satisfy     the
following relation
\begin{equation}
  \label{resummult}
{\Delta}^{W,0}_{\mu\nu}(q)\ =\ \widehat{Z}_W\, 
\widehat{\Delta}^{W}_{\mu\nu}(q)\, .
\end{equation}

We start with  the  most general form of  $\widehat{\Delta}^{W,0}_{\mu
  \nu} (q)$ given by
\begin{equation}
  \label{DTDL}
\widehat{\Delta}^{W,0}_{\mu\nu}(q)\ =\
\widehat{\Delta}^{W,0}_{T}(q^2)t_{\mu\nu}\, +\, 
\widehat{\Delta}^{W,0}_{L}(q^2)\ell_{\mu\nu} \, ,
\end{equation}
where 
\begin{displaymath} 
t_{\mu\nu}(q)\ =\ {}- g_{\mu\nu} + \frac{q_\mu q_\nu}{q^2}\, ,\quad
\ell_{\mu\nu}(q)\ =\ \frac{q_\mu q_\nu}{q^2}\ .
\end{displaymath}
As  was shown  in detail   in \cite{PP1}, if   we  decompose the  bare
one-loop PT $W$-boson self-energy $\widehat{\Pi}_{\mu\nu}^{W,0}(q)$ in
the form
\begin{equation}
\widehat{\Pi}_{\mu\nu}^{W,0}(q)\ =\ \widehat{\Pi}_{T}^{W,0}(q^2)t_{\mu\nu}\, 
+\, \widehat{\Pi}_{L}^{W,0}(q^2)\ell_{\mu\nu} \, ,
\end{equation}
then in the PT resummation formalism the quantities 
$\widehat{\Delta}^{W,0}_{T}(q^2)$ and 
$\widehat{\Delta}^{W,0}_{L}(q^2)$ are given by
\begin{eqnarray}
  \label{DTexp}
\widehat{\Delta}^{W,0}_{T}(q^2) &=& [\, q^2\, -\, (M^0_W)^2\, +\,
\widehat{\Pi}_{T}^{W,0}(q^2)\,]^{-1}\, ,\\
  \label{DLexp}
\widehat{\Delta}^{W,0}_{L}(q^2) &=&
[\, (M^0_W)^2\, -\, \widehat{\Pi}_{L}^{W,0}(q^2)\, ]^{-1}\, . 
\end{eqnarray}
The standard renormalization procedure  is to define the wave function
renormalization, $\widehat{Z}_W$, by means of  the transverse part  of
the resummed $W$-boson propagator:
\begin{equation}
  \label{TransRen}
\widehat{Z}_W\, [\, q^2\, -\, (M^0_W)^2\, +\,
\widehat{\Pi}_{T}^{W,0}(q^2)\, ]\ =\ q^2\, -\, M_W^2\, +\, 
\widehat{\Pi}_{T}^{W}(q^2)\, ,
\end{equation} 
where  the   explicit   form   of  $\widehat{Z}_W$   depends   on  the
renormalization  scheme. Similarly, the   propagator of the associated
would-be Goldstone boson $G^+$ is renormalized as usual, {\em i.e.}, 
\begin{equation}
  \label{GoldPropRen}
\widehat{D}^{G^+,0}(q^2)\ =\ [\, q^2\, -\, \widehat{\Omega}^0(q^2)\, ]^{-1}
\ =\ \widehat{Z}_{G^+}\, [\,q^2\,-\,\widehat{\Omega}(q^2)\, ]^{-1}\, ,
\end{equation}
with 
\begin{equation}
  \label{GoldRen}
G^{\pm,0}\ =\ \widehat{Z}_{G^{+}}G^{\pm}\, .
\end{equation} 
Note  that we only need  to  carry out a wave-function renormalization
for the Goldstone  boson self-energy ${\widehat{\Omega}}^0(q^2)$ (with
tadpole and  seagull graphs included), since ${\widehat{\Omega}}^0 (0)
= 0$,  in  agreement with  the  Goldstone theorem,  which states  that
Goldstone bosons are massless  to all orders  in perturbation theory.  
The latter is a result of the gauge  invariance of the diagrammatic PT
method.

{}From the PT WI's involving the self-energies, we have
\begin{equation}
  \label{PLOmega}
\widehat{\Pi}_{L}^{W,0}(q^2)\ =\ \frac{(M^0_W)^2}{q^2}\,
\widehat{\Omega}^0(q^2)\, ,
\end{equation}
which implies that
\begin{equation}
  \label{DLDGRel}
\widehat{\Delta}^{W,0}_{L}(q^2)\ =\ \frac{q^2}{(M^0_W)^2}\,
{\widehat{D}}^{G^{+},0}(q^2)\, .
\end{equation}
This last identity allows us to write the resummed $W$-boson
propagator in the form
\begin{eqnarray}
  \label{DMRInt}
\widehat{\Delta}^{W,0}_{\mu\nu}(q) &=&
\widehat{Z}_W [\,
\widehat{\Delta}^{W}_{T}(q^2)t_{\mu\nu}\, +\,
\widehat{Z}_W^{-1}\widehat{\Delta}^{W,0}_{L}(q^2)\ell_{\mu\nu}\,]
\nonumber\\
&=& \widehat{Z}_W [\, \widehat{\Delta}^{W}_{T}(q^2)t_{\mu\nu}\, +\,
\widehat{Z}_W^{-1}\widehat{Z}_{G^+}
\big(1+\frac{\delta M_W^2}{M_W^2}\big)^{-1}\frac{q^2}{M_W^2}
{\widehat{D}}^{G^+}(q^2)\ell_{\mu\nu}\,]\, .
\end{eqnarray}
Imposing  that      Eq.\   (\ref{PLOmega}),   or    equivalently  Eq.\
(\ref{DLDGRel}), holds  for the  renormalized quantities as   well, we
find
\begin{equation}
\widehat{\Delta}^{W,0}_{\mu\nu}(q)\ =\
\widehat{Z}_W [\,\widehat{\Delta}^{W}_{T}(q^2)t_{\mu\nu}\, +\,
\widehat{Z}_W^{-1}\widehat{Z}_{G^+}
\big(1+\frac{\delta M_W^2}{M_W^2}\big)^{-1}
\widehat{\Delta}^W_{L}(q^2)
\ell_{\mu\nu}\,]\, .
\end{equation}
Finally, using the last equality of  Eq.\ (\ref{W3}) we arrive at Eq.\
(\ref{resummult}).

It is  now straightforward to  see  that the {\em process-independent}
RGI quantity for the $W$ boson is given by
\begin{eqnarray}
  \label{RW}
\widehat{R}^{W,0}_{\mu\nu} (q) &=& \frac{(g^0_w)^2}{4\pi}
\widehat{\Delta}^{W,0}_{\mu\nu}(q)\ =\ \frac{(g_w)^2}{4\pi}
\widehat{\Delta}^{W}_{\mu\nu}(q)\nonumber\\
&=& \widehat{R}^W_{\mu\nu} (q)\, .
\end{eqnarray}

At this point one might be tempted to separate the above quantity into
the product  of a dimension-full kinematic  factor and a dimensionless
quantity,  which could  be identified  with   an effective  charge, in
direct analogy to the  QED and QCD  cases.  This kind of factorization
may however introduce artifacts into both components, which are absent
from the original RGI expression.  For  example, take the simple case
where  $\widehat{R}^W_{\mu\nu}$  is sandwiched      between  conserved
external  currents (massless external fermions),  and let us decompose
$\widehat{R}^W_{\mu\nu}$ in the form
\begin{equation}
\widehat{R}^W_{\mu\nu}(q)\ =\ \alpha_{w,\rm eff}(q^2)\
\frac{- g_{\mu\nu}}{q^2 - \bar{s}(q^2)}\, ,
\end{equation}
where $\bar{s}(q^2)$ denotes the position of the physical complex pole
of  the  $W$ boson which  appears  on the second  Riemann  sheet.  Two
possible parameterizations of   the pole  are   $\bar{s} = M_{W}^2   -
iM_{W}\Gamma_W$ (constant imaginary part) or $\bar{s}(q^2) = M_{W}^2 -
iq^2  \Gamma_W/M_{W} $    ($q^2$-dependent  imaginary   part),   where
$\Gamma_W$ is the constant width of the $W$ boson on the pole.  In the
first case, we see that at $q^2 = 0$ for example, the kinematic factor
is complex,  whereas the  RGI quantity is   real,  {\em i.e.}, $\Im  m
\widehat{R}^W_{\mu\nu}   (0)  = 0$, since,    by construction, it only
develops an imaginary  part  at the  lowest physical threshold  $q^2 >
m_e^2$.   Consequently, $\alpha_{w,\rm  eff}  (0)$ is  also imaginary,
{\em  i.e.},  it  contains thresholds    which are  artifacts  of  the
decomposition.  The second parameterization of $\bar{s}(q^2)$ does not
have the above  problems, but still,  the effective charge so  defined
contains erroneous information about the position of the various decay
channels.  Similar considerations apply to the case  of the $Z$ boson,
or other possible gauge bosons appearing in extensions of the SM.

Apart from  identifying the  process-independent RGI quantity  for the
$W$  boson $\widehat{R}^W_{\mu\nu} (q)$ in   Eq.\ (\ref{RW}), one  can
also  introduce process-dependent  RGI  quantities.  For instance,  to
one-loop, the $\gamma W^-W^+$ vertex may be written in the form
\begin{equation}
  \label{RgamWW}
\widehat{R}^{\gamma W^-W^+}_{\mu\nu\lambda} (q,k_-,k_+)\ =\ 
\Big(\frac{4\pi}{g^2_w}\Big)\, \frac{1}{e}\, 
\overline{\Gamma}_{\mu\nu\lambda}^{\gamma W^-W^+}(q,k_-,k_+)\, ,
\end{equation}
with 
\begin{equation}
\overline{\Gamma}_{\mu\nu\lambda}^{\gamma W^-W^+}(q,k_-,k_+)\ =\ 
\Gamma_{0\mu\nu\lambda}^{\gamma W^-W^+}\, +\, 
\widehat{\Gamma}_{\mu\nu\lambda}^{\gamma W^-W^+}(q,k_-,k_+)\, .
\end{equation}
Here,   $\Gamma_{0\mu\nu\lambda}^{\gamma W^-W^+}$ and $\widehat{\Gamma
  }_{\mu \nu \lambda}^{\gamma W^- W^+} (q,k_-,k_+)$ are the tree-level
and one-loop PT $\gamma W^-W^+$ couplings, respectively.  The quantity
$\widehat{R}^{\gamma W^-W^+}_{\mu\nu\lambda} (q,k_-,k_+)$ is UV finite
and invariant  under  the renormalization group.   This  can easily be
shown by means of the one-loop  PT WI, which can also  be written in a
RGI form, {\em viz.},
\begin{equation}
  \label{RgamWI}
q^\mu\, \widehat{R}_{\mu\nu\lambda}^{\gamma W^-W^+}(q,k_-,k_+)
\ =\ \widehat{R}^{W-1}_{\nu\lambda}(k_-)\, -\, 
\widehat{R}^{W-1}_{\nu\lambda}(k_+)\, .
\end{equation}
This    last    equation shows  how     the   action  of    $q^\mu$ on
$\widehat{R}_{\mu\nu\lambda}^{\gamma  W^-W^+}$    projects   out   the
process-independent   part of the   $\gamma  W^-W^+$ vertex, which  is
related to  the $W$-boson effective charge. Similarly, one
can construct RGI  combinations  for all  the PT vertices  related  to
couplings,  {\em e.g.},  $\gamma  W^-G^+$,  $ZW^-W^+$, $ZG^-G^+$, {\em
  etc}.  In particular, at  LEP2, it is very  advantageous to use  the
RGI   expressions $\widehat{R}_{\mu\nu\lambda}^{\gamma  W^-W^+}$   and
$\widehat{R}_{\mu\nu\lambda}^{Z W^-W^+}$,   which lead to   UV  finite
form-factors for the $\gamma W^-W^+$ and $ZW^-W^+$ vertices.

\subsection{The effective charge of the Higgs boson}

We  now proceed to extend  the notion of the  effective  charge to the
case  of the  Higgs boson.   For this   purpose, we  first express the
unrenormalized Higgs-boson propagator in terms of the renormalized one
as follows:
\begin{eqnarray}
  \label{DeltH}
\hat{\Delta}^{H,0}(q^2) &=& [\,q^2\, -\, (M^0_H)^2\, +\,
\widehat{\Pi}^{HH,0}(q^2)\, ]^{-1}\ =\ \widehat{Z}_H [\,q^2\, -\, M_H^2\, +\,
\widehat{\Pi}^{HH}(q^2)\,]^{-1}\nonumber\\ 
&=& \widehat{Z}_H\, \hat{\Delta}^H(q^2)\, ,
\end{eqnarray}
where $M_H$ may  be defined to be  the real part  of the complex  pole
position  of   $\hat{\Delta}^H(q^2)$.   Following  a procedure  rather
similar   to that given  in Section  5.1,  we should exploit the gauge
symmetry of   the  SM,   in    order to   deduce   relations   between
$\widehat{Z}_H$ and the other renormalization constants.

We start again with  the PT WI relating  the one-loop vertices $HZG^0$
and $HG^0G^0$, {\em i.e.},
\begin{equation}
  \label{HZG0}
k^\mu_1 \overline{\Gamma}_\mu^{HZG^0,0}
(q,k_1,k_2)\, +\, iM^0_Z \overline{\Gamma}^{HG^0G^0,0}(q,k_1,k_2)
\ =\ -\, \frac{g^0_w}{2c^0_w}\, 
\Big[\, \Big(\widehat{\Delta}^{H,0}(q^2)\Big)^{-1}\, -\, 
\Big(\widehat{D}^{G^0,0}(k^2_2)\Big)^{-1}\, \Big]\, ,  
\end{equation}
where $\widehat{D}^{G^0,0}(k^2) = [k^2- \widehat{\Pi}^{G^0G^0,0}(k^2)
]^{-1} = \widehat{Z}_{G^0} [k^2- \widehat{\Pi}^{G^0G^0}(k^2)]^{-1}$ and 
\begin{equation}
  \label{overHZG}
\overline{\Gamma}_\mu^{HZG^0,0}\ \equiv\ 
\Gamma_{0\mu}^{HZG^0,0}\, +\, \widehat{\Gamma}_\mu^{HZG^0,0}\, ,\qquad
\overline{\Gamma}^{HG^0G^0,0}\ \equiv\ \Gamma_0^{HG^0G^0,0}\, +\, 
\widehat{\Gamma}^{HG^0G^0,0}\, .
\end{equation}
As before,   in Eq.\ (\ref{overHZG}), the   subscript `0' denotes tree
level couplings, while  the caret indicates one-loop vertices obtained
in the PT.

As  usual, we write the  unrenormalized $HZG^0$ and $HG^0G^0$ vertices
as a  product of  the  renormalized ones and a  vertex renormalization
constant,
\begin{equation}
  \label{ZZHG}
\frac{2c_w}{g_w}\, \overline{\Gamma}_\mu^{HZG^0}\ =\ \widehat{Z}_{HZG^0}\,
\frac{2c^0_w}{g^0_w}\,  \overline{\Gamma}_\mu^{HZG^0,0}\, ,\qquad
\frac{2c_w}{g_w}\,\overline{\Gamma}^{HG^0G^0}\ =\ \widehat{Z}_{HG^0G^0}\,
\frac{2c^0_w}{g^0_w}\,  \overline{\Gamma}^{HG^0G^0,0}\, .
\end{equation}
Making   use  of the fact   that the  effective $HZG^0$  and $HG^0G^0$
vertices are completely renormalized  by a redefinition of  the fields
and the couplings given in Eq.\ (\ref{WFR}), we find
\begin{eqnarray}
  \label{ZZG0r}
\widehat{Z}_{HZG^0} &=&
\widehat{Z}_{g_w}\widehat{Z}^{-1}_{c_w} \widehat{Z}^{1/2}_Z
\widehat{Z}^{1/2}_H \widehat{Z}^{1/2}_{G^0}\ =\ 
\widehat{Z}^{1/2}_H \widehat{Z}^{1/2}_{G^0}\, ,\\
  \label{ZG0G0r}
\widehat{Z}_{HG^0G^0} &=& \widehat{Z}_{g_w}\widehat{Z}^{-1}_{c_w}
\widehat{Z}^{1/2}_H \widehat{Z}_{G^0}\ =\ 
\widehat{Z}^{1/2}_H \widehat{Z}^{1/2}_{G^0}\, \Big(\, 1\, +\,
\frac{\delta M^2_Z}{M^2_Z}\, \Big)^{1/2}\, .
\end{eqnarray}
In the  last step of  Eqs.\ (\ref{ZZG0r}) and (\ref{ZG0G0r}),  we have
used the relations given in Eqs.\ (\ref{W1})--(\ref{W3}).

In order that the PT WI in Eq.\ (\ref{HZG0}) maintains the same form
after renormalization, it is necessary to have
\begin{equation}
  \label{ZHZG}
\widehat{Z}_H\ = \ \widehat{Z}_{G^0}\, .
\end{equation}
Employing Eqs.\ (\ref{W1}), (\ref{W3}) and (\ref{ZHZG}), it is easy to
show that
\begin{eqnarray}
  \label{RGIC}
\widehat{R}^{H,0}(q^2) &=& \frac{(g^0_w)^2}{(M^0_W)^2}
\hat{\Delta}^{H,0}(q^2)
\ =\ \Big[\, \frac{g^2_w}{M_W^2}\, \hat{\Delta}^{H}(q^2)\,\Big]\, 
\widehat{Z}_g^2 \widehat{Z}_H\, 
        \Big( 1+\frac{\delta M_W^2}{M_W^2}\, \Big)^{-1} \nonumber\\
&=& \widehat{R}^H (q^2)
\end{eqnarray}
is a {\em process-independent}  RGI quantity, in  close analogy to the
RGI quantity   of the $W$   boson  $\widehat{R}^W_{\mu\nu}(q)$.  As  a
byproduct of   this, we  also   find  that  $\widehat{R}^{G^+}(k^2)  =
(g^2_w/M_W^2) \widehat{D}^{G^+}  (k^2)$  and $\widehat{R}^{G^0}(k^2) =
(g^2_w/M_W^2)   \widehat{D}^{G^0}   (k^2)$  are   invariant  under the
renormalization   group.   Hence,   we   conclude  that  the  quantity
$\widehat{R}^H(q^2)$ provides a  natural generalization of the concept
of  the  effective charge in   the case  of the  Higgs  boson.  It  is
interesting to notice the exact analogy between the  form of the Higgs
boson effective charge of Eq.\  (\ref{RGIC}) obtained within a  theory
with a non-Abelian gauge  symmetry such as the   SM, and that of  Eq.\ 
(\ref{RGIscalar}) derived in the context of a  much simpler model with
Abelian global (un-gauged) symmetry.

Finally, a direct   derivation  of the  above general   result may  be
obtained  if one adopts  the   symmetric formulation of the  classical
action in the BFM \cite{SBFM}.  Within  this formulation one is led to
the minimal on-shell    scheme,  with  the   relevant  renormalization
constants satisfying
\begin{equation}
\hat{\Phi}^0\ =\ Z_{\hat{\Phi}}^{1/2}\, \hat{\Phi}\, ,\qquad
\hat{v}^0\ =\ Z_{\hat{\Phi}}^{1/2}\,(\hat{v} + \delta\hat{v})\, .
\end{equation}
Using the above  relations, and the additional fact  that, due  to the
background symmetry, in this formulation $ \delta \hat{v} =0$, one can
immediately show that
\begin{equation}
(\hat{v}^0)^{-2}\, \langle 0|\, T:\hat{\Phi}^0(x)\, \hat{\Phi}^0(y):
                     \, |0\rangle\
 =\  \hat{v}^{-2}\, \langle 0 |\,  T:\hat{\Phi}(x)\, \hat{\Phi}(y):
                     \, |0\rangle\, ,
\end{equation}
which  is Eq.\ (\ref{RGIC}).      Even though the  analysis   of  this
subsection which  led  to Eq.\  (\ref{RGIC}) is  more general since it
does  not  rely on   any      particular gauge-fixing  procedure    or
renormalization scheme,  this  latter derivation within  the symmetric
BFM    framework has  the  advantage   of   directly generalizing  the
construction of the scalar effective charge given for the toy model of
Section 4.3 to the realistic case of the SM.
 
\setcounter{equation}{0}
\section{Diagrammatic analysis of the equivalence \newline 
 theorem} 
\indent

Cornwall,  Levin  and  Tiktopoulos, and shortly  afterwards Vayonakis,
showed  that  at very  high  energies the   amplitude for emission  or
absorption of a longitudinally polarized  gauge boson becomes equal to
the    amplitude in   which the    gauge  boson   is   replaced by the
corresponding  would-be Goldstone    boson  \cite{EqTh}.  The    above
statement is a consequence of the underlying local gauge invariance of
the SM, and  is  known as the  equivalence  theorem (ET);  it has been
proven to  hold  to  all  orders in perturbation   theory for multiple
absorptions and    emissions   of massive vector    bosons  \cite{CG}.
Compliance  with  this  theorem is  a  necessary  requirement for  any
resummation algorithm, since any  Born-improved amplitude  which fails
to  satisfy it is bound  to be missing important physical information.
The reason  why most resummation  methods are at odds  with  the ET is
that in the usual diagrammatic analysis the underlying symmetry of the
amplitudes is not manifest; just as happens in the case of the OT, the
conventional sub-amplitudes defined  in  terms of Feynman diagrams  do
{\em  not}  satisfy the ET individually.   The  resummation of  such a
sub-amplitude will in turn  distort several subtle cancellations, thus
giving rise to artifacts and  unphysical effects.  Instead, as we will
show in detail in this  section, the PT  sub-amplitudes satisfy the ET
{\em individually}.   As  is common  in  the  PT  framework,  the only
non-trivial step for accomplishing  this is the proper exploitation of
elementary WI's.  In addition, the part of the Born-improved amplitude
containing the  resummed Higgs boson self-energy  (or the RGI quantity
defined in  Section 5.3) satisfies the  ET {\em independently}  of the
rest of the (non-resummed) amplitude.  This is explicitly demonstrated
by resorting almost exclusively to the fact that, in contrast to their
conventional counterparts, the  one-loop  Higgs-boson vertices defined
within the PT satisfy naive, tree-level WI's.

The formal derivation of the ET  is based on the observation \cite{CG}
that, by  virtue of the  Becchi--Rouet--Stora  (BRS) invariance of the
theory, the connected transition amplitude  between physical states of
any   number  $n$  of insertions  of   the  gauge-fixing  term $F^{a}$
vanishes, {\em i.e.},
\begin{equation}
  \label{BRSET}
\langle f |\,  T: F^{a_1} (x_1) F^{a_2}(x_2) \dots F^{a_n} (x_n):\, |
           i\rangle_{\rm con.} \ =\ 0 \, .
\end{equation}
In the renormalizable $R_\xi$ gauges, $F^{a}$ assumes the form
\begin{equation}
  \label{Fa}
F^a (x)\ =\ \partial^\mu V^a_\mu (x)\, +\, \xi_a M_{V^a} G^a (x)\, ,
\end{equation}
where $V^a_\mu$ denotes the  massive gauge boson, {\em e.g.},  $W^\pm$
or $Z$, $G^a$ its corresponding  would-be Goldstone boson, {\em e.g.},
$G^\pm$ or $G^0$, $\xi_a$ its GFP,  and $M_{V^a}$ its mass.  Since for
energies   $E_V  \gg    M_V$   the  longitudinal  polarization  vector
$\varepsilon^\mu_L (k)$ of the gauge boson $V$ behaves as
\begin{equation}
  \label{emuL}
\varepsilon^\mu_L (k) \ =\ \frac{k^\mu}{M_V}\, +\, v^\mu (k)\, ,
\end{equation}
with  $v_\mu (k) =  {\cal  O} (M_V/E_V)$,  in the configuration  space
$\varepsilon^\mu_L (k)$ may  be represented naively by  the derivative
$\partial_\mu/M_V$,  which in turn enables   one to use the identities
derived from Eq.\  (\ref{BRSET}). Beyond  the tree  level, one  has in
general to  include  correction  factors \cite{YY},  denoted  here  as
$K^{a_i}$, which take into account  renormalization effects.  Finally,
given that, due  to the unitarity of the  SM, amplitudes  
involving only
would-be Goldstone bosons cannot  grow faster than  a constant at high
energies, one finally arrives at
\begin{equation}
  \label{ETnaive}
{\cal T}(V^{a_1}_L\dots V^{a_n}_L;X )\ =\ \prod_{i=1}^n\, K^{a_i}\,  
{\cal T}(G^{a_1}\dots G^{a_n};X )\ +\ {\cal O}(M/E)\, ,
\end{equation}
where $V_L^{\alpha}\equiv \varepsilon^\mu_L V_{\mu}^{\alpha}$, and $X$
denotes all other fields.  The above equality represents the ET in its
most basic form.   Note   that the ET   cannot give   any  interesting
information for amplitudes which decrease in magnitude as $1/\sqrt{s}$
or  faster with increasing   energy $E\approx \sqrt{s}$.   In order to
obtain non-trivial  information for the energetically suppressed terms
of order $M/\sqrt{s}$  and their higher  powers, one has to invoke the
so-called  generalized  ET  (GET) \cite{CG},  whose  derivation  again
relies on  the  identities  of Eq.\  (\ref{BRSET}).   In  the case  of
amplitudes with two  longitudinal  $W^+$ bosons for  example,  the GET
establishes the following relation:
\begin{eqnarray}
  \label{GETWW}
\lefteqn{\hspace{-1.5cm}
{\cal T}[W^{+,\mu}_L(k_+) W^{-,\nu}_L(k_-);X ]}\nonumber\\ 
& =&  K^+K^-\,
{\cal T}[G^+(k_+)G^-(k_-);X] \, +\, K^+\,{\cal T}[G^+(k_+) w^{-,\nu}
(k_-);X] \nonumber\\
&& +\, K^-\, {\cal T}[w^{+,\mu}(k_+) G^-(k_-);X]\, +\, 
{\cal T}[w^{+,\mu}(k_+) w^{-,\nu}(k_-);X ]\, ,
\end{eqnarray}
and for two longitudinal $Z$ bosons: 
\begin{eqnarray}
  \label{GETZZ}
{\cal T}[Z^\mu_L(k_1) Z^\nu_L(k_2);X ] & =&  (K^0)^2\,
{\cal T}[G^0(k_1)G^0(k_2);X] \, +\, K^0\,{\cal T}[G^0(k_1) z^\nu (k_2);X]
\nonumber\\ 
&&+\, K^0\,{\cal T}[z^\mu (k_1) G^0(k_2);X]\, +\, 
{\cal T}[z^\mu (k_1) z^\nu (k_2);X ]\, ,
\end{eqnarray}
where $w^{\pm,\mu} (k_\pm) = \varepsilon^\mu_L(k_\pm) - k^\mu_\pm/M_Z$
and  $z^\mu (k_{1,2}) =  \varepsilon^\mu_L(k_{1,2}) - k^\mu_{1,2}/M_W$
are the energetically suppressed parts of the longitudinally polarized
$W^\pm$ and $Z$ bosons,  respectively.  In addition, $K^\pm$ and $K^0$
are renormalization correction  factors mentioned above.   In the Born
approximation, they take the  values $K^+=-1$, $K^-=1$, and $K^0= -i$,
if  the four-momenta of  the gauge bosons  are incoming \cite{BL}, and
reverse  their sign in the    opposite  case.  Formulas analogous   to
(\ref{GETWW}) and (\ref{GETZZ}) can be derived for an arbitrary number
of longitudinally polarized $W$ and  $Z$ bosons.  In the following, we
will   restrict ourselves to  amplitudes   involving two vector bosons
only.

Let us consider the process $\nu (p_1)\bar{\nu}(p_2) \to Z^\mu_L (k_1)
Z^\nu_L (k_2)$ at the tree level, where $\nu$ is a Dirac neutrino with
mass  $m$ and the  four-momenta of the $Z$ boson  are defined to enter
the interaction   vertices as shown   in  Fig.\ 6.   The total  matrix
element   ${\cal   T}(\nu\bar{\nu}\to Z_L   Z_L)$ is the   sum  of two
amplitudes:
\begin{equation}
  \label{TnuZZ}
{\cal T}(\nu\bar{\nu}\to Z_L Z_L)\ =\
{\cal T}^H_s (Z_LZ_L)\, +\, {\cal T}_t (Z_LZ_L)\, ,
\end{equation}
where
\begin{eqnarray}
 \label{THsZZLL}
{\cal T}^H_s (Z_LZ_L) &=&
\varepsilon^\mu_L(k_1)\varepsilon^\nu_L(k_2)\, {\cal T}^H_{s\,
  \mu\nu} (ZZ)\, , \\
\label{TtZZLL}
{\cal T}_t (Z_LZ_L)&=&
\varepsilon^\mu_L(k_1)\varepsilon^\nu_L(k_2)\, {\cal T}_{t\,
  \mu\nu} (ZZ)\, ,
\end{eqnarray}
with
\begin{eqnarray}
  \label{THsZZ}
{\cal T}^H_{s\, \mu\nu} (ZZ) &=& -i\Gamma^{HZZ}_{0\mu\nu}\, 
\Delta_H (k_1+k_2)\, \Big(\frac{g_w m}{2M_W}\Big)\, \bar{v}(p_2)u(p_1)\, ,\\
  \label{TtZZ}
{\cal T}_{t\, \mu\nu} (ZZ) &=& - \Big(\frac{g_w}{2c_w}\Big)^2\,
\bar{v}(p_2)\Big( \gamma_\nu P_L\, \frac{1}{\not\! p_1 + \not\! k_1 - m}\, 
\gamma_\mu P_L\nonumber\\ 
&&+\, \gamma_\mu P_L\, \frac{1}{\not\! p_1 + \not\! k_2 - m}\,
                                           \gamma_\nu P_L \Big)u(p_1)\, .
\end{eqnarray}
In  Eq.\ (\ref{THsZZ}), the  tree-level  $HZZ$ coupling is defined  as
$\Gamma^{HZZ}_{0\mu\nu} =  ig_w\,  M^2_Z/M_W g_{\mu\nu}$.  The  entire
amplitude ${\cal T}(\nu\bar{\nu}\to Z_L Z_L)$  satisfies of course the
GET (and hence  the ET).  What we  wish to investigate here is whether
the   GET holds for the   Higgs-mediated  part of  the amplitude  {\em
  independently}.  The reason we  turn directly to  the GET instead of
the ET,  is  simply that both   the  ${\cal T}^H_s (Z_LZ_L)$   of Eq.\
(\ref{THsZZ}) and the amplitude ${\cal T}^H_s (G^0G^0)$ given by
\begin{equation}
  \label{THsGG}
{\cal T}^H_s (G^0G^0)\ =\  -i\Gamma_0^{HG^0G^0}\,  
\Delta_H (k_1+k_2)\, \Big(\frac{g_w m}{2M_W}\Big)\, \bar{v}(p_2)u(p_1)\, ,
\end{equation}
where $\Gamma_0^{HG^0G^0} = -ig_w M^2_H/(2M_W)$, decrease at very high
energies  as  $1/s$,   because of the    presence  of the  Higgs-boson
propagator  in the $s$ channel.  So,   the ET in  this  case will only
furnish trivial information.  Instead, according to the GET \cite{CG},
additional amplitudes should be taken  into  account if one wishes  to
keep track of energetically suppressed terms to order $M^2_Z/s$.

In order to address  the question raised  above, we will calculate the
LHS of Eq.\  (\ref{GETZZ}) explicitly, using  full expressions for the
longitudinal  polarization vectors   involved, and check  whether  the
result  so      obtained coincides or   not  with     the sum   of the
Higgs-boson-dependent parts of the amplitudes  appearing on the RHS of
Eq.\ (\ref{GETZZ})  (with $X=  \nu\bar{\nu}$).   For that  purpose, we
first write the  longitudinal  polarization vector of the  gauge boson
$V$ in the covariant form
\begin{equation}
  \label{epsL}
\varepsilon^\mu_L (k)\ =\ \frac{1}{2\beta M_V}\, [ (1+\beta^2)k^\mu\, 
-\, (1-\beta^2)\tilde{k}^\mu ]\, ,
\end{equation}
where  $k^\mu  = (E_V, \vec{k}_V)$  is  the four-momentum   of the $V$
boson, $\tilde{k}^\mu =  k_\mu$ and $\beta  = |\vec{k}_V|/E_V$ is  the
$V$-boson velocity.  It is  convenient to work  in the c.m.\ system of
the process $\nu\bar{\nu}\to Z^\mu_L(k_1) Z^\nu_L (k_2)$; in that case
the polarization vector $\varepsilon^\mu_L (k_1)$ of the $Z^\mu$ boson
can   be  expressed   in  terms of    the  four-momenta $k^\mu_1$  and
$\tilde{k}^\mu_1 =  k^\mu_2$, and $\beta  = (1  - 4M^2_Z  / s)^{1/2}$. 
Likewise, $\varepsilon^\nu_L (k_2)$  is written in  terms of $k^\nu_2$
and $\tilde{k}_2^\nu  =     k^\nu_1$.  To   order $M^4_Z/s^2$,     the
energetically subleading  part   $z^\mu  (k_1)$  of $\varepsilon^\mu_L
(k_1)$ is obtained by
\begin{equation}
  \label{zmuL}
z^\mu (k_1)\ =\ \varepsilon^\mu_L (k_1)\, -\, \frac{k^\mu_1}{M_Z}\
 =\ -\, \frac{2M_Z}{s}\, k^\mu_2\ +\ {\cal O}\Big(\,
 \frac{M^4_Z}{s^2}\,\Big)\, .
\end{equation}
Furthermore, the residual vector $z^\mu (k_1)$ has the following
properties:
\begin{equation}
  \label{zmuprop}
z_\mu (k_1)\, k^\mu_1\ =\ -M_Z\, ,\qquad z_\mu (k_1)\, z^\mu (k_1)\ =\
0\, . 
\end{equation}
Exactly analogous formulas  and relations hold  for $\varepsilon^\nu_L
(k_2)$.  Using the  decomposition (\ref{emuL}) for $\varepsilon^{\mu ,
  \nu}_L  (k_{1,2})$ and the  properties (\ref{zmuprop}) for $z^{\mu ,
  \nu}(k_{1,2})$, we  find for the part  of the amplitude depending on
the Higgs boson
\begin{equation}
  \label{TsH1}
{\cal T}^H_s (Z_LZ_L)\ = \ -\, {\cal T}^H_s(G^0G^0)\, 
+\, \Delta {\cal T}^H_s\, +\, {\cal T}^H_s (zz)\, -\, {\cal T}^H_P (Z_LZ_L)\, ,
\end{equation}
where ${\cal T}^H_s (zz) = z^\mu z^\nu{\cal T}^H_{s\,\mu\nu}(ZZ)$ and
\begin{equation}
  \label{DTsH}
\Delta {\cal T}^H_s\ =\ - \Big( \frac{g_w m}{2M_W}\Big)\, 
\Big(\frac{g_wM^2_Z}{M_W}\Big)\, \Delta_H (k_1+k_2)\,
\bar{v}(p_2)u(p_1)\, ,  
\end{equation}
and  ${\cal  T}^H_P  (Z_LZ_L)$   is   the expression   given  in  Eq.\
(\ref{TP}), with $m_t$ replaced by $m$.   It is now straightforward to
verify that
\begin{equation}
  \label{GETsub}
\Delta {\cal T}^H_s\ =\ -\, i{\cal T}^H_s (zG^0)\, -\, i{\cal T}^H_s(G^0z)\, ,
\end{equation}
with
\begin{eqnarray}
  \label{THsZG}
{\cal T}^H_s (zG^0)\, +\, {\cal T}^H_s(G^0z) &=& z^\mu (k_1)\, 
{\cal T}^H_{s\,\mu} (ZG^0)\ +\ z^\nu (k_2)\,{\cal T}^H_{s\,\nu}
(G^0Z) \nonumber\\
&=& -i[z^\mu (k_1)\, \Gamma^{HZG^0}_{0\mu}\, +\, 
z^\nu (k_2)\, \Gamma^{HG^0Z}_{0\nu}]\nonumber\\ 
&&\times\, \Big( \frac{g_w m}{2M_W}\Big)\, \Delta_H (k_1+k_2)\,
\bar{v}(p_2)u(p_1)\, .
\end{eqnarray}
The  tree-level    $H(q)   Z_\mu(k_1)  G^0(k_2)$    coupling   in Eq.\
(\ref{THsZG})  is given by $\Gamma^{HZG^0}_{0\mu}  = g_w (q-k_2)_\mu /
(2c_w)$, with all momenta  defined  as incoming.  Clearly, the   first
three terms on the RHS of Eq.\ (\ref{TsH1}) are nothing but the sum of
the Higgs-boson-dependent  parts of the amplitudes ${\cal  T}[G^0(k_1)
G^0(k_2) ; \nu\bar{\nu}]$, ${\cal T}[z(k_1) G^0(k_2) ; \nu\bar{\nu}] +
{\cal T}[G^0(k_1) z(k_2); \nu\bar{\nu}]$ and ${\cal T} [z(k_1) z(k_2);
\nu\bar{\nu}]$.  Evidently,  the only reason preventing  ${\cal T}^H_s
(Z_LZ_L)$ from satisfying individually the GET is  the presence of the
term ${\cal T}^H_P (Z_LZ_L) $ on the RHS of Eq.\ (\ref{TsH1}).

However, according to  the PT, the genuine Higgs-boson-dependent  part
of the amplitude, $\widehat{{\cal T}}^H_s (Z_LZ_L)$, is obtained after
recognizing that  the momenta  $k_{1}^{\mu}$ and  $k_{2}^{\nu}$ coming
from  the polarization vectors    of the longitudinal  $Z$ bosons  can
extract   a  $s$-channel-like,   Higgs-boson-dependent  part  from the
non-resonant amplitude ${\cal T}_{t\,   \mu\nu}$, in exactly the  same
way as happens in Eq.\ (\ref{TP}), namely
\begin{equation}
  \label{TtP}
\frac{k^\mu_1}{M_Z}\,\frac{k^\nu_2}{M_Z}\, 
               {\cal T}_{t\,\mu\nu} (Z Z)\ =\ {\cal T}^H_P (Z_LZ_L)\ 
+ \dots\, ,
\end{equation}
where,   as in    Eq.\  (\ref{TP}), the   ellipses  denote  additional
contributions not  related  to the Higgs boson.  Thus, $\widehat{{\cal
    T}}^H_s (Z_LZ_L)= {\cal T}^H_s (Z_LZ_L)+{\cal T}^H_P (Z_LZ_L)$.

We  now want to check  if $\widehat{{\cal  T}}^H_s (Z_LZ_L)$ satisfies
the GET; for that to happen one must show that
\begin{eqnarray}
  \label{GETalmost}
\widehat{{\cal T}}^H_s (Z_LZ_L) &=&
\widehat{{\cal T}}^H_s (G^0G^0)+
\widehat{{\cal T}}^H_s (G^0z)+
\widehat{{\cal T}}^H_s (zG^0)+
\widehat{{\cal T}}^H_s (zz)\nonumber\\
&=& 
\ -\, {\cal T}^H_s(G^0G^0)\, 
  -\, i{\cal T}^H_s (zG^0)\, -\, i{\cal T}^H_s(G^0z)\, +\, 
                                           {\cal T}^H_s (zz)\nonumber\\
&& -\, {\cal T}^H_P(G^0G^0)\, -\, i{\cal T}^H_P(G^0z)
-\, i{\cal T}^H_P(zG^0)\, +\, {\cal T}^H_P(zz) \, ,
\end{eqnarray}
where the  amplitudes  ${\cal  T}^H_P(G^0G^0)$,  ${\cal T}^H_P(G^0z)$,
${\cal  T}^H_P(zG^0)$,    and  ${\cal   T}^H_P(zz)$    denote possible
Higgs-boson-  dependent   $s$-  channel pinch parts    coming from the
$t$-channel  amplitudes  ${\cal T}_t  (G^0G^0)$, ${\cal  T}_t (G^0z)$,
${\cal T}_t  (zG^0)$, and ${\cal  T}_t (zz)$, respectively. It is easy
to   convince    oneself   however  that  ${\cal  T}^H_P(G^0G^0)={\cal
  T}^H_P(G^0z)={\cal T}^H_P(zG^0)=  {\cal   T}^H_P(zz)=0$; this  is so
because, according  to Eq.\ (\ref{zmuL}), the energetically subleading
parts  $z^{\mu}(k_1)$   and $z^{\nu}(k_2)$  are   proportional to  the
``wrong''     momenta,   {\em  i.e.}\     $k^{\mu}_2$  and $k^{\nu}_1$,
respectively, instead  of  $k^{\mu}_1$   and $k^{\nu}_2$,   which  are
necessary for pinching.  Therefore, Eq.\ (\ref{GETalmost}) reduces to
\begin{equation}
  \label{GETtree}
\widehat{{\cal T}}^H_s (Z_LZ_L)=
\ -\, {\cal T}^H_s(G^0G^0)\, 
  -\, i{\cal T}^H_s (zG^0)\, -\, i{\cal T}^H_s(G^0z)\, +\, 
                                                {\cal T}^H_s(zz)\,  .
\end{equation}
But    this last  equation  is  immediately  true, by   virtue of Eq.\
(\ref{TsH1}) and the definition of $\widehat{{\cal T}}^H_s (Z_LZ_L)$.

It  is important   to  stress that  Eq.\ (\ref{GETtree})  demonstrates
explicitly how  the  tree-level Higgs-mediated  part of  the amplitude
${\cal T}^H_s$  satisfies  the GET {\em  independently},  provided the
pinching   contribution ${\cal  T}^H_P$ residing  in  the non-resonant
amplitude is taken properly into  consideration.  This fact reveals an
underlying  relation between the  PT and  the ET  at  the diagrammatic
level, and constitutes a major result of this paper.

We will    now  show  that  within  the   PT framework   the  equality
(\ref{GETtree})  remains  valid even after  the Higgs-boson propagator
has been resummed.  As    explained in \cite{PP1,PP2},  the  effective
one-loop PT $HZZ$ vertex $\widehat{\Gamma}_{\mu\nu}^{HZZ} (q,k_1,k_2)$
must be included in the amplitude  containing the resummed Higgs boson
propagator $\widehat{\Delta}^H (q)$;  this is so, because the one-loop
PT $H(q) Z^\mu    (k_1) Z^\nu (k_2)$  vertex  satisfies   a  number of
tree-level WI's which are crucial for ensuring the gauge invariance of
the resummed Higgs-mediated part of the amplitude
\begin{equation}
  \label{THsres}
\overline{{\cal T}}^H_{s\, \mu\nu} (Z_LZ_L) \ =\ -i
[\Gamma^{HZZ}_{0\mu\nu} + \widehat{\Gamma}_{\mu\nu}^{HZZ}
(q,k_1,k_2)]\, \Big(\frac{g_w m}{2M_W}\Big)\, 
\widehat{\Delta}^H (q)\, \bar{v}(p_2)u(p_1)\, .
\end{equation}
The PT WI's identities are
\begin{eqnarray}
  \label{PTHZZ1}
k^\nu_2 \widehat{\Gamma}_{\mu\nu}^{HZZ}
(q,k_1,k_2)\, +\, iM_Z \widehat{\Gamma}^{HZG^0}_\mu (q,k_1,k_2)
&=& -\, \frac{g_w}{2c_w}\, \widehat{\Pi}^{ZG^0}_\mu (k_1)\, ,\\
  \label{PTHZZ2}
k^\mu_1 \widehat{\Gamma}_\mu^{HZG^0}
(q,k_1,k_2)\, +\, iM_Z \widehat{\Gamma}^{HG^0G^0}(q,k_1,k_2)
&=& -\, \frac{g_w}{2c_w}\, \Big[\, \widehat{\Pi}^{HH}(q^2)\, +\, 
\widehat{\Pi}^{G^0G^0}(k^2_2)\, \Big]\, ,\qquad\\
  \label{PTHZZ3}
k^\mu_1 k^\nu_2 \widehat{\Gamma}_{\mu\nu}^{HZZ}
(q,k_1,k_2)\, +\, M^2_Z \widehat{\Gamma}^{HG^0G^0}(q,k_1,k_2)
&=& \frac{ig_wM_Z}{2c_w}\, \Big[\, \widehat{\Pi}^{HH}(q^2)\, +\, 
\widehat{\Pi}^{G^0G^0}(k^2_1)     \nonumber\\
&&+\, \widehat{\Pi}^{G^0G^0}(k^2_2)\, \Big]\, .
\end{eqnarray}
As before, we define all momenta to flow into the $HZZ$ vertex with $q
+ k_1 + k_2 = 0$.  The closed form  of the effective one-loop PT $HZZ$
coupling is given in Appendix B.  Note that exactly the same WI's hold
true for the tree-level $HZZ$ coupling before quantizing the classical
action by introducing   gauge-fixing terms and  ghost  fields.   To be
specific, the  tree-level WI's derived from  the classical  action are
recovered    from    Eqs.\     (\ref{PTHZZ1})--(\ref{PTHZZ3})       if
$\widehat{\Pi}^{HH}(q^2)$   and    $-\widehat{\Pi}^{G^0G^0}(k^2)$  are
replaced with   the   inverse free  propagators  of   the Higgs  boson
$\Delta_H^{-1}  (q)  = q^2  - M^2_H$  and   the $G^0$ Goldstone  boson
$\Delta_{G^0}^{-1}(k)        =    k^2$,       respectively,      while
$\widehat{\Pi}^{ZG^0}_\mu (k)$  is substituted  by $iM_Z k_\mu$, which
represents the $G^0Z$ mixing. Of  course, in the $R_\xi$ gauges  there
is no $G^0Z$  mixing at  tree level  because  it cancels against   the
corresponding gauge-fixing term.

Within our Born-improved approximation the neutrino-exchange amplitude
${\cal T}_t (Z_LZ_L)$ retains its  tree-level form; its only  function
is to  provide the PT term  ${\cal T}^H_P (Z_LZ_L)$.  This latter term
is responsible for the bad high-energy  behaviour of both the resonant
and  non-resonant  amplitudes, which  violate the  ET separately.  The
validity of the ET for the individual  amplitudes can be restored only
after the PT term ${\cal T}^H_P  (Z_LZ_L)$ is added to the $s$-channel
amplitude  $\overline{{\cal T}}^H_s  (Z_LZ_L)$,  exactly as happens in
the case of  the  tree-level  (non-resummed)  $\widehat{{\cal  T}}^H_s
(Z_LZ_L)$.  Indeed, it is not   difficult to show that the  amplitudes
$\overline{{\cal T}}^H_s (Z_LZ_L)  +   {\cal T}^H_P$ and  ${\cal  T}_t
(Z_LZ_L) -  {\cal  T}^H_P$ satisfy  the  GET  and hence  the  ET  {\em
  individually}.    For example,  by   employing the  PT  WI's in Eq.\
(\ref{PTHZZ1})--(\ref{PTHZZ3}), we have that
\begin{equation}
  \label{THsGET}
\overline{\cal T}^H_s (Z_LZ_L)\, +\, {\cal T}^H_P\ =\ 
-\, \overline{\cal T}^H_s (G^0G^0)\, -i\overline{\cal T}^H_s (zG^0)\,
-\, i\overline{\cal T}^H_s (G^0z)\, +\, \overline{\cal T}^H_s (zz)\, ,
\end{equation}
where 
\begin{equation}
  \label{THsGres}
\overline{\cal T}^H_s (G^0G^0)\ =\ -i[ \Gamma_0^{HG^0G^0} + 
\widehat{\Gamma}^{HG^0G^0}(q,k_1,k_2)]\,  
\widehat{\Delta}^H (q)\, \Big(\frac{g_w m}{2M_W}\Big)\,\bar{v}(p_2)u(p_1)\, ,
\end{equation}
and the sum of the resummed amplitudes $\overline{\cal T}^H_s (zG^0) +
\overline{\cal   T}^H_s  (G^0z)$   is  defined   analogously  to  Eq.\
(\ref{THsZG}), {\em i.e.},
\begin{eqnarray}
  \label{THsZGres}
\overline{\cal T}^H_s (zG^0)\, +\, \overline{\cal T}^H_s(G^0z) &=& 
z^\mu (k_1)\, \overline{\cal T}^H_{s\,\mu} (ZG^0)\ +\ z^\nu (k_2)\,
\overline{\cal T}^H_{s\,\nu} (G^0Z) \nonumber\\
&=& -i\Big\{\, z^\mu (k_1)\, [ \Gamma^{HZG^0}_{0\mu} +
\widehat{\Gamma}^{HZG^0}_\mu (q,k_1,k_2)]\, +\, 
z^\nu (k_2)\, [ \Gamma^{HG^0Z}_{0\nu}  \nonumber\\
&&+ \widehat{\Gamma}^{HG^0Z}_\nu 
(q,k_1,k_2)]\, \Big\}\, \Big( \frac{g_w m}{2M_W}\Big)\, 
                                              \widehat{\Delta}^H (q)\,
\bar{v}(p_2)u(p_1)\, .
\end{eqnarray}
Finally, we have  defined  $\overline{\cal T}^H_s (zz) =   z^\mu (k_1)
z^\nu  (k_2) \overline{{\cal   T}}^H_{s\, \mu\nu} (Z_LZ_L)$.   In  the
derivation  of  Eq.\  (\ref{THsGET}), we  have  also  used the PT  WI:
$\widehat{\Pi}^{ZG^0}_\mu (k)=-iM_Z k_\mu \widehat{\Pi}^{G^0G^0}(k^2)/
k^2$.

The above  considerations  can  be  straightforwardly  extended   to
processes  involving  the  $HWW$   vertex,  {\em e.g.},  the  reaction
$t\bar{t} \to H^*\to W^+_L W^-_L$. As has been discussed in Section 3,
one has to  extract from  the   $b$-quark exchange graph the  PT  term
related to the  Higgs-mediated part of the  amplitude ({\em cf.}\ Eq.\ 
(\ref{TP})).  Similarly, after adding the   PT term ${\cal T}^H_P$  to
the  resummed Higgs-exchange  amplitude $\overline{\cal  T}^H_s (W^+_L
W^-_L)$, we can show that
\begin{equation}
  \label{THsGETWW}
\overline{\cal T}^H_s (W^+_L W^-_L)\, +\, {\cal T}^H_P\ =\ 
-\, \overline{\cal T}^H_s (G^+G^-)\, +\overline{\cal T}^H_s (w^+G^-)\,
-\, \overline{\cal T}^H_s (G^+w^-)\, +\, \overline{\cal T}^H_s (w^+ w^-)\, ,
\end{equation}
which is  in  agreement with GET in   Eq.\ (\ref{GETWW}).   Again, the
derivation relies on effective one-loop PT WI's, which are the same as
those  naively   deduced  from  the  classical   action  in   the Born
approximation.    In this case, the   PT WI's pertaining  to the $H(q)
W^+(k_+) W^-(k_-)$ vertex are given by
\begin{eqnarray}
  \label{PTHWW1}
k^\mu_+ \widehat{\Gamma}_{\mu\nu}^{HW^+ W^-}
(q,k_+,k_- )\, +\, M_W \widehat{\Gamma}_\nu^{HG^+ W^-}(q,k_+,k_-)
&=& -\, \frac{ig_w}{2}\, \widehat{\Pi}^{W^-G^+}_\nu (k_-)\, ,\qquad\\
  \label{PTHWW2}
k^\nu_- \widehat{\Gamma}_{\mu\nu}^{HW^+ W^-}
(q,k_+,k_- )\, -\, M_W \widehat{\Gamma}_\mu^{HW^+ G^-}(q,k_+,k_-)
&=& \frac{ig_w}{2}\, \widehat{\Pi}^{W^+G^-}_\mu (k_+)\, ,\qquad\\
  \label{PTHWW3}
k^\mu_\pm \widehat{\Gamma}_\mu^{HW^\pm G^\mp}
(q,k_\pm,k_\mp )\, \pm\, M_W \widehat{\Gamma}^{HG^+ G^-}(q,k_+,k_-)
&=& \pm \, \frac{ig_w}{2}\, \Big[ \, \widehat{\Pi}^{HH}(q^2)\, +\, 
\widehat{\Omega} (k^2_\mp)\, \Big]\, ,\qquad\\
  \label{PTHWW4}
k^\mu_+ k^\nu_- \widehat{\Gamma}_{\mu\nu}^{HW^+W^-}
(q,k_+,k_-)\, +\, M^2_W \widehat{\Gamma}^{HG^+G^-}(q,k_+,k_-)
&=& \frac{ig_wM_W}{2}\, \Big[\, \widehat{\Pi}^{HH}(q^2)\, +\, 
\widehat{\Omega}(k^2_+)     \nonumber\\
&&+\, \widehat{\Omega}(k^2_-)\, \Big]\, .
\end{eqnarray}
The analytic form of the  PT vertex $\widehat{\Gamma}_{\mu\nu}^{H  W^+
  W^-} (q,k_+,k_- )$ is given in Appendix A.  The one-loop PT vertices
$\widehat{\Gamma}_\mu^{HW^\pm   G^\mp}  (q,   k_\pm,    k_\mp  )$  and
$\widehat{\Gamma}^{HG^+G^-}(q,  k_+, k_-)$ may be  gained by using the
WI's in Eqs.\ (\ref{PTHWW1})--(\ref{PTHWW4}) and known expressions for
the PT Higgs- and $G^+$- boson self-energies \cite{JP90}.

At this  point we should   note that the GET  is  still valid for  the
Higgs-mediated part of the amplitude even if we use the RGI expression
for the resummed Higgs  boson propagator $\widehat{R}^H(s)$ defined in
Section 5.3.   Similarly,  one    can    define   the
process-dependent  RGI   combinations   involving,  {\em   e.g.},  the
$HW^+W^-$ vertex:
\begin{eqnarray}
  \label{RHWW}
\widehat{R}^{HW^+W^-}_{\mu\nu}(q,k_+,k_-) &=& \Big(\frac{M^2_W}{g^2_w}\Big)\, 
\frac{1}{g_wM_W}\, \overline{\Gamma}^{HW^+W^-}_{\mu\nu}(q,k_+,k_-),\nonumber\\
\widehat{R}^{HW^+G^-}_\mu (q,k_+,k_-) &=& \Big(\frac{M^2_W}{g^2_w}\Big)\, 
\frac{1}{g_w}\, \overline{\Gamma}^{HW^+G^-}_\mu (q,k_+,k_-),\nonumber\\
\widehat{R}^{HG^+G^-} (q,k_+,k_-) &=& \Big(\frac{M^2_W}{g^2_w}\Big)\, 
\frac{M_W}{g_w}\, \overline{\Gamma}^{HG^+G^-} (q,k_+,k_-)\, .
\end{eqnarray}
As  before, ``barred'' quantities  denote the  sum over the tree-level
and   one-loop    PT vertices.   The     UV   finite, RGI   quantities
$\widehat{R}^H$,                                  $\widehat{R}^{G^+}$,
$\widehat{R}^{HW^+W^-}_{\mu\nu}$,  $\widehat{R}^{H   W^+G^-}_\mu$  and
$\widehat{R}^{HG^+G^-}$  satisfy  tree-level-type PT   WI's in  direct
analogy to  those given  in  Eqs.\ (\ref{PTHWW1})--(\ref{PTHWW4}).  In
this formulation, any resummed transition amplitude  can be written in
terms of a product of RGI  quantities, where the vertices are replaced
by the respective $\widehat{R}$ expressions.  As a consequence of this
formulation, the factors $K^\pm$   and $K^0$ retain their   tree-level
values      after    renormalization    provided   the   wave-function
renormalizations for the external  Goldstone bosons are properly taken
into account.

In summary, we have shown how the  diagrammatic method based on the PT
enables  the   decomposition   of  the amplitude   into    a  resummed
propagator-like amplitude  and a non-resonant background which satisfy
the GET as  well as the ET {\em  individually}.  This feature provides
an additional non-trivial   consistency check for the PT   resummation
approach, and,  at the same  time,  renders  the ET conceptually  more
intuitive.

\setcounter{equation}{0}
\section{Conclusions}

The formulation of the PT resummation  approach is extended to analyze
resonant transition amplitudes which involve the SM Higgs boson as an
intermediate state.  The main  results of our  study may be summarized
as follows:

\begin{itemize}

\item[(i)] The PT   rearrangement of the amplitude   gives rise  to  a
  self-energy for the  Higgs boson which  is {\em independent}  of the
  GFP   in  every gauge-fixing  scheme.     This  self-energy is  {\em
    universal}, in the sense that it is {\em process independent}, and
  may be  {\em  resummed}  following the   method presented in   Ref.\
  \cite{PP1}. In addition, it only  displays {\em physical}  fermionic
  and  bosonic   thresholds,   in contrast    to  the  gauge-dependent
  self-energies obtained   by the   conventional methods,  where  {\em
    unphysical} bosonic thresholds  appear.  Furthermore, it satisfies
  {\em individually} the  OT, {\em  both}   for fermionic  as well  as
  bosonic contributions.
  
\item[(ii)] When the resummed Higgs  boson propagator is multiplied by
  the universal   quantity $g^2_w/M^2_W$,  or, equivalently,    by the
  inverse square of the  vacuum expectation value  of the Higgs field,
  it gives  rise to  a {\em renormalization-group-invariant} quantity,
  in direct analogy  to the  {\em effective  charge} of  the photon in
  QED.  The above  construction becomes  possible   by virtue  of  the
  naive,   tree-level  WI's   satisfied  by   the   GFP-independent PT
  sub-amplitudes.
   
\item[(iii)] At high  energies any amplitude involving  longitudinally
  polarized  gauge bosons satisfies    the   ET, but  the   individual
  $s$-channel and $t$-channel contributions of  the amplitude do  not.
  Instead, the PT decomposition of such an amplitude gives rise to two
  kinematically  distinct pieces, a  genuine $s$-channel and a genuine
  $t$-channel,  which  satisfy   the ET    {\em individually}.    Most
  importantly, the above property {\em persists}  even {\em after} the
  $s$-channel Higgs boson  self-energy has been resummed, thus solving
  a long-standing problem.

\end{itemize}

The significance of the  above analysis when computing the theoretical
predictions for the Higgs-boson lineshape is clear.  The Born-improved
amplitudes constructed with the above formalism are in accordance with
all physical requirements imposed, and reliably capture the underlying
dynamics.   Most noticeably,   the  ability to  construct a  universal
Higgs-mediated   component, in direct    analogy to the  QED effective
charge, is rather  intriguing. This universal part  is common to every
Higgs-boson-mediated process, and,  even though  the process-dependent
background must be eventually  taken  into account, it determines  the
Higgs-boson lineshape comfortably  away from the resonance.  It  would
be of  great phenomenological importance  to confront  the predictions
for   Higgs- production and  decay   processes computed within the  PT
resummation  approach    against  future data   obtained  from planned
high-energy colliders  such  as   the LHC, the   next-linear  $e^+e^-$
collider with c.m.\ energy 500 GeV, and the first muon collider.

\vspace{0.7cm}\noindent   {\bf   Acknowledgments.}    
JP acknowledges financial support  from the Department of  Physics and
Astronomy of the    University of Manchester  (PPARC  grant),  and the
Centre de Physique Th\'eorique (CPT)  at Marseille (CNRS grant), while
parts of  this  work were  been  completed, and thanks the  Max Planck
Institute for   its warm hospitality.   AP thanks  the theory group of
Manchester University for the kind hospitality extended to him.

\newpage

\def\theequation{\Alph{section}.\arabic{equation}}
\begin{appendix}
\setcounter{equation}{0}
\section{One-loop absorptive $HWW$ coupling in the PT}

Based on the established equivalence between the  PT and the covariant
background field gauge for $\xi_Q=1$, we calculate the absorptive part
of the effective PT $HWW$ vertex at one  loop, using the Feynman rules
listed in \cite{DDW}.  The analytic results  are expressed in terms of
standard loop integrals  introduced by 't-Hooft and Veltman \cite{HV}.
For definiteness, we use the conventions of Ref.\ \cite{BAK}.

If  one  assumes that   the  external $W$ bosons  are  contracted with
physical polarization vectors or  conserved currents, the one-loop  PT
$HWW$ coupling $\widehat{\Gamma}^{HW^+W^-}_{\mu\nu} (q,k_+,k_-)$   may
then be decomposed in general as follows:
\begin{eqnarray}
  \label{HWWdec}
\widehat{\Gamma}^{HW^+W^-}_{\mu\nu}(Q,p,k) &=& g_w M_W\, \Big[\, 
\Big(1\, +\, A(Q^2)\Big)\, g_{\mu\nu}\, +\, B(Q^2)\, \frac{k_\mu\, 
p_\nu}{M^2_W} \nonumber\\  
&& +\, iC(Q^2)\, \frac{1}{M^2_W}\,\varepsilon_{\mu\nu\lambda\rho} 
k^\lambda p^\rho\, \Big] ,
\end{eqnarray}
where $Q + p + k = 0$ and  $A(Q^2)$, $B(Q^2)$ and $C(Q^2)$ are general
form-factors.  Only $A(Q^2)$  must be  renormalized, whereas  $B(Q^2)$
and  $C(Q^2)$ are UV   finite.  The  form-factor $C(Q^2)$ occurs    in
CP-violating scenarios only, {\em i.e.}, $C(Q^2) = 0$.

In the improved Born-level approximation, only the absorptive parts of
the form-factors $A(Q^2)$  and    $B(Q^2)$ are of  relevance,  as  the
dispersive parts  participate  in the  one-loop  renormalization.  The
diagrams contributing  to  the  absorptive form-factors $\bar{A}$  and
$\bar{B}$ are shown in Fig.\ 7.  To a good approximation, the external
$W$ bosons  are considered to  be  stable and  the $b$ quark massless.
The analytic results for the absorptive form-factor $\bar{A}(Q^2)$ are
then given by
\begin{eqnarray}
  \label{AWa}
i \bar{A}_{(a)} &=& \frac{\alpha_w}{16\pi}\, \frac{m^2_t}{M^2_W}\,
\Big[ 8\bar{C}_{24}\, +\, (Q^2+k^2-p^2+4m^2_t)\bar{C}_0\,
+\, (3Q^2-3k^2-p^2)\bar{C}_{11}\nonumber\\
&& +(Q^2+5k^2-p^2)\bar{C}_{12}\Big](m^2_t,0,m^2_t)\, ,\\
  \label{AWb1}
i \bar{A}_{(b1)} &=& -\, \frac{\alpha_w}{\pi}\, \Big[\bar{B}_0(Q^2,M^2_W,M^2_W)
+2\frac{M^2_W}{M^2_Z}\bar{B}_0(Q^2,M^2_Z,M^2_Z)\Big]\, , \\
  \label{AWb2c5c6}
i\bar{A}_{(b2)} &=& i\bar{A}_{(c5)}\ =\ i\bar{A}_{(c6)}\ =\ 0\, ,\\
  \label{AWb3}
i\bar{A}_{(b3)} &=& -\, \frac{\alpha_w}{16\pi} \Big[\Big(
\frac{M^2_H}{M^2_W}+2\Big)\bar{B}_0(Q^2,M^2_W,M^2_W)\, +\,
\frac{1}{2}\Big(\frac{M^2_H}{M^2_W}\nonumber\\
&&+2\frac{M^2_Z}{M^2_W}\Big)\bar{B}_0(Q^2,M^2_Z,M^2_Z)\, +\, 
\frac{3}{2}\frac{M^2_H}{M^2_W}\bar{B}_0(Q^2,M^2_H,M^2_H)\Big]\, ,\\
  \label{AWb4}
i\bar{A}_{(b4)} &=& \frac{\alpha_w}{2\pi}\Big[\bar{B}_0(Q^2,M^2_W,M^2_W)\,
+\, \bar{B}_0(Q^2,M^2_Z,M^2_Z)\Big]\, ,\\
  \label{AWc1}
i\bar{A}_{(c1)} &=& \frac{\alpha_w}{\pi}\Big[ \Big( 1-\frac{M^2_W}{M^2_Z}\Big)
\Big( 4\bar{C}_{24}+(Q^2-k^2-p^2)\bar{C}_0\Big)(M^2_W,0,M^2_W)
\nonumber\\
&&+\, \frac{M^2_W}{M^2_Z}\Big( 4\bar{C}_{24}+(Q^2-k^2-p^2)
\bar{C}_0\Big)(M^2_W,M^2_Z,M^2_W)\nonumber\\
&& +\, \frac{M^2_Z}{M^2_W}\Big( 4\bar{C}_{24}
 +(Q^2-k^2-p^2) \bar{C}_0\Big)(M^2_Z,M^2_W,M^2_Z)\, \Big]\, ,\\
  \label{AWc2c3}
i\bar{A}_{(c2)}+i\bar{A}_{(c3)} &=& \frac{\alpha_w}{2\pi} Q^2\, \Big[
\Big( 2\frac{M^2_W}{M^2_Z}-1\Big)\bar{C}_0(M^2_W,M^2_Z,M^2_W)\, +\,
\Big( 2\frac{M^2_W}{M^2_Z}-1\Big)\bar{C}_0(M^2_W,0,M^2_W)
\nonumber\\
&&+\, \bar{C}_0(M^2_Z,M^2_W,M^2_Z)\, \Big]\, ,\\
  \label{AWc4}
i\bar{A}_{(c4)} &=& -\, \frac{\alpha_w}{4\pi}\, \Big[\frac{M^4_Z}{M^2_W}
\Big( 2\frac{M^2_W}{M^2_Z}-1\Big)^2\bar{C}_0(M^2_Z,M^2_W,M^2_Z)\, +\,
M^2_W\bar{C}_0(M^2_W,M^2_Z,M^2_W)\nonumber\\
&&+\, M^2_W\bar{C}_0(M^2_W,M^2_H,M^2_W)\Big]\, ,\\
  \label{AWc7}
i\bar{A}_{(c7)} &=&  -\, \frac{\alpha_w}{8\pi}\, \Big[ M^2_Z
\Big(\frac{M^2_H}{M^2_W}+2\Big)\Big( 2\frac{M^2_W}{M^2_Z}-1\Big)^2
\bar{C}_0(M^2_W,M^2_Z,M^2_W) \nonumber\\
&&+\, 4M^2_W \Big(\frac{M^2_H}{M^2_W}+2 \Big)\Big( 1-\frac{M^2_W}{M^2_Z}\Big)
\bar{C}_0(M^2_W,0,M^2_W)\, +\, M^2_W\Big(\frac{M^2_H}{M^2_W}\nonumber\\
&&+2\frac{M^2_Z}{M^2_W}\Big)\bar{C}_0(M^2_Z,M^2_W,M^2_Z)\,
+\, 3M^2_H \bar{C}_0(M^2_H,M^2_W,M^2_H)\Big]\, ,\\
  \label{AWc8}
i\bar{A}_{(c8)} &=&  -\, \frac{\alpha_w}{8\pi}\, \Big[ 
\Big(\frac{M^2_H}{M^2_W}+2\Big)\Big(\bar{C}_{24}(M^2_W,M^2_Z,M^2_W)
+\bar{C}_{24}(M^2_W,M^2_H,M^2_W)\Big)\nonumber\\
&&+\Big(\frac{M^2_H}{M^2_W}+2\frac{M^2_Z}{M^2_W}\Big)
\bar{C}_{24}(M^2_Z,M^2_W,M^2_Z)\, +\, 3\frac{M^2_H}{M^2_W}
\bar{C}_{24}(M^2_H,M^2_W,M^2_H)\Big]\, ,\nonumber\\
\\
  \label{AWc9c10} 
i\bar{A}_{(c9)}+i\bar{A}_{(c10)}
&=& \frac{2\alpha_w}{\pi}\, \Big[ \frac{M^2_W}{M^2_Z}
\bar{C}_{24}(M^2_W,M^2_Z,M^2_W)\, +\, \Big( 1-\frac{M^2_W}{M^2_Z}\Big)
\bar{C}_{24}(M^2_W,0,M^2_W)\nonumber\\
&&+\, \bar{C}_{24}(M^2_Z,M^2_W,M^2_Z)\Big]\, .
\end{eqnarray}
Here and in the following, we do not display the first three arguments
of the $C$ functions $(p^2,k^2,Q^2)$, which are common. The bar on the
loop  functions symbolizes that only   the  absorptive part should  be
considered.

Similarly, the individual  contributions to the absorptive form-factor
$\bar{B}(Q^2)$ are found to give
\begin{eqnarray}
  \label{BWa}
i\bar{B}_{(a)} &=& -\, \frac{\alpha_w}{8\pi}\, m^2_t\, 
(4\bar{C}_{23} + \bar{C}_0 + 3\bar{C}_{11} + \bar{C}_{12})(m^2_t,0,m^2_t)\,
,\\
  \label{BWb1234c47}
i\bar{B}_{(b1)} &=& i\bar{B}_{(b2)}\ =\ i\bar{B}_{(b3)}\ =\ i\bar{B}_{(b4)}\
=\ i\bar{B}_{(c4)}\ =\ i\bar{B}_{(c7)}\ =\ 0\, ,\\
  \label{BWc1}
i\bar{B}_{(c1)} &=& \frac{2\alpha_w}{\pi }\Big[ M^2_W \Big(1-
\frac{M^2_W}{M^2_Z}\Big) (2\bar{C}_{11}+2\bar{C}_{23}+\bar{C}_0)(M^2_W,0,M^2_W)
\nonumber\\
&&+\, \frac{M^4_W}{M^2_Z}\, (2\bar{C}_{11}+2\bar{C}_{23}+\bar{C}_0)
(M^2_W,M^2_Z,M^2_W)\, +\, M^2_W (2\bar{C}_{11}+2\bar{C}_{23}\nonumber\\
&&+\bar{C}_0)(M^2_Z,M^2_W,M^2_Z)\, \Big]\, ,\\
  \label{BWc2c3}
i\bar{B}_{(c2)}+i\bar{B}_{(c3)} &=& \frac{\alpha_w}{2\pi}\, \Big[
M^2_W\Big( 2\frac{M^2_W}{M^2_Z}-1\Big) (\bar{C}_0+\bar{C}_{11}+
\bar{C}_{12})(M^2_W,M^2_Z,M^2_W)\nonumber\\
&& +\, 2M^2_W \Big( 1-\frac{M^2_W}{M^2_Z}\Big)(\bar{C}_0+\bar{C}_{11}+
\bar{C}_{12})(M^2_W,0,M^2_W)\nonumber\\
&& +\, M^2_W (\bar{C}_0+\bar{C}_{11}+\bar{C}_{12})(M^2_Z,M^2_W,M^2_Z)\, 
\Big]\, ,\\
  \label{BWc5c6}
i\bar{B}_{(c5)}+i\bar{B}_{(c6)} &=& \frac{\alpha_w}{4\pi}\, \Big[
(M^2_Z-2M^2_W)(\bar{C}_0+\bar{C}_{12}-\bar{C}_{11})(M^2_Z,M^2_W,M^2_Z)\, -\,
M^2_W (\bar{C}_0+\bar{C}_{12}\nonumber\\
&&-\bar{C}_{11})(M^2_W,M^2_Z,M^2_W)\, +\, M^2_W 
(\bar{C}_0+\bar{C}_{12}-
\bar{C}_{11})(M^2_W,M^2_H,M^2_W) \Big],\qquad\ \\
  \label{BWc8}
i\bar{B}_{(c8)} &=& \frac{\alpha_w}{8\pi}\, \Big\{
(M^2_H+2M^2_W)\Big[(\bar{C}_{11}+\bar{C}_{23})(M^2_W,M^2_Z,M^2_W)\,
+\, (\bar{C}_{11}\nonumber\\
&&+\bar{C}_{23})(M^2_W,M^2_H,M^2_W)\Big]\, +\, 
(M^2_H+2M^2_Z)(\bar{C}_{11}+\bar{C}_{23})(M^2_Z,M^2_W,M^2_Z)\nonumber\\
&&+\, 3M^2_H(\bar{C}_{11}+\bar{C}_{23})(M^2_H,M^2_W,M^2_H)\Big\}\, ,\\
  \label{BWc9c10}
i\bar{B}_{(c9)} + i\bar{B}_{(c10)} &=& -\, \frac{2\alpha_w}{\pi}\,
\Big[ \frac{M^4_W}{M^2_Z}\, (\bar{C}_{11}+\bar{C}_{23})(M^2_W,M^2_Z,M^2_W)\, 
+\, M^2_W\Big(1-\frac{M^2_W}{M^2_Z}\Big)(\bar{C}_{11}\nonumber\\
&&+\bar{C}_{23})(M^2_W,0,M^2_W)\, +\, M^2_W (\bar{C}_{11}+\bar{C}_{23})
(M^2_Z,M^2_W,M^2_Z)\Big]\, .
\end{eqnarray}

\setcounter{equation}{0}
\section{One-loop absorptive $HZZ$ coupling in the PT}

Here we  present the one-loop results  for the absorptive form-factors
$\bar{A}$  and $\bar{B}$  of the $HZZ$   coupling in terms of standard
loop integrals. We consider the  general decomposition of the one-loop
$HZZ$ vertex
\begin{equation}
  \label{HZZdec}
\widehat{\Gamma}^{HZZ}_{\mu\nu}(Q,p,k)\ =\ \frac{g_w}{c_w} M_Z\, \Big[\, 
\Big(1\, +\, A(Q^2)\Big)\, g_{\mu\nu}\, +\, B(Q^2)\, \frac{k_\mu\, 
p_\nu}{M^2_Z}\, \Big]\, ,
\end{equation}
where  the  CP-violating  form-factor   $C(Q^2)$  analogous  to   Eq.\
(\ref{HWWdec}) is absent at one-loop in  the SM. In particular, we are
interested  in the absorptive  part  of  the form-factors  $A(Q^2)$ and
$B(Q^2)$. Calculating the graphs shown in Fig.\ 8, we obtain
\begin{eqnarray}
  \label{AZa}
i \bar{A}_{(a)} &=& \frac{\alpha_w}{16\pi}\, \frac{m^2_t}{M^2_W}\, \Big\{
(g^2_L+g^2_R)\, 
\Big[ 8\bar{C}_{24}\, +\, (Q^2+k^2-p^2+4m^2_t)\bar{C}_0\nonumber\\
&& +\, (3Q^2-3k^2-p^2)\bar{C}_{11}\, +\, 
(Q^2+5k^2-p^2)\bar{C}_{12}\Big](m^2_t,m^2_t,m^2_t)\nonumber\\
&& +\, 2g_Lg_R\, \Big[ 4m^2_t\bar{C}_0\, +\, (Q^2-k^2+p^2)\bar{C}_{11}
\nonumber\\
&&+\, (Q^2+k^2-p^2)\bar{C}_{12}\Big](m^2_t,m^2_t,m^2_t)\Big\},\\
  \label{AZb1}
i\bar{A}_{(b1)} &=& -\, \frac{2\alpha_w}{\pi}\, \frac{M^4_W}{M^4_Z}\, 
\bar{B}_0(Q^2,M^2_W,M^2_W)\, ,\\
  \label{AZb2c5c6}
i\bar{A}_{(b2)} &=& i\bar{A}_{(c5)}\ =\ i\bar{A}_{(c6)}\ =\ 0\, ,\\
  \label{AZb3}
i\bar{A}_{(b3)} &=& -\, \frac{\alpha_w}{16\pi} \Big[\Big(
\frac{M^2_H}{M^2_W}+2\Big)\Big( 2\frac{M^2_W}{M^2_Z} -1\Big)^2
\bar{B}_0(Q^2,M^2_W,M^2_W)\, +\, \frac{1}{2}\Big(\frac{M^2_H}{M^2_W}\nonumber\\
&&+\, 2\frac{M^2_Z}{M^2_W}\Big)
\bar{B}_0(Q^2,M^2_Z,M^2_Z)\, +\, 
\frac{3}{2}\frac{M^2_H}{M^2_W}\bar{B}_0(Q^2,M^2_H,M^2_H)\Big]\, ,\\
  \label{AZb4}
i\bar{A}_{(b4)} &=& \frac{\alpha_w}{\pi}\, \frac{M^4_W}{M^4_Z}\,
\bar{B}_0(Q^2,M^2_W,M^2_W)\, ,\\
  \label{AZc1}
i\bar{A}_{(c1)} &=& \frac{2\alpha_w}{\pi}\, \frac{M^4_W}{M^4_Z}\,
\Big[ 4\bar{C}_{24}+ (Q^2-k^2-p^2)\bar{C}_0\Big](M^2_W,M^2_W,M^2_W)\, ,\\
  \label{AZc2c3}
i\bar{A}_{(c2)}+i\bar{A}_{(c3)} &=& -\, \frac{\alpha_w}{2\pi}\,
\frac{M^2_W}{M^2_Z}\, Q^2\, \bar{C}_0(M^2_W,M^2_W,M^2_W)\, ,\\
  \label{AZc4}
i\bar{A}_{(c4)} &=& -\, \frac{\alpha_w}{4\pi}\, \Big[2M^2_W
\bar{C}_0(M^2_W,M^2_W,M^2_W)\, +\, \frac{M^4_Z}{M^2_W}
\bar{C}_0(M^2_Z,M^2_H,M^2_Z)\Big]\, ,\\
  \label{AZc7}
i\bar{A}_{(c7)} &=&  -\, \frac{\alpha_w}{8\pi}\, \Big[ 2(M^2_H+2M^2_W)
\bar{C}_0(M^2_W,M^2_W,M^2_W)\nonumber\\
&& +\, 3M^2_Z\frac{M^2_H}{M^2_W}\bar{C}_0(M^2_H,M^2_Z,M^2_H)\Big]\, ,\\
  \label{AZc8}
i\bar{A}_{(c8)} &=&  -\, \frac{\alpha_w}{4\pi}\, \Big[ 
\Big(2\frac{M^2_W}{M^2_Z}-1\Big)^2\Big(\frac{M^2_H}{M^2_W}+2\Big)
\bar{C}_{24}(M^2_W,M^2_W,M^2_W)\, +\, \frac{1}{2}\,
\Big(\frac{M^2_H}{M^2_W}\nonumber\\
&&+\, 2\frac{M^2_Z}{M^2_W}\Big)
\bar{C}_{24}(M^2_Z,M^2_H,M^2_Z)\, +\, \frac{3}{2}\, \frac{M^2_H}{M^2_W}
\bar{C}_{24}(M^2_H,M^2_Z,M^2_H)\Big]\, ,\\
  \label{AZc9c10} 
i\bar{A}_{(c9)}+i\bar{A}_{(c10)}
&=& \frac{4\alpha_w}{\pi}\, \frac{M^4_W}{M^4_Z}\, 
\bar{C}_{24}(M^2_W,M^2_W,M^2_W)\, .
\end{eqnarray}
In  Eq.\ (\ref{AZa}), we have  defined  as $g_L=(2M^2_W/M^2_Z) -1$ and
$g_R =-2(1-M^2_W/M^2_Z)$.

Furthermore, the individual contributions  to the $B$  form-factor are
given by
\begin{eqnarray}
  \label{BZa}
i\bar{B}_{(a)} &=& -\, \frac{\alpha_w}{8\pi}\, M^2_Z\frac{m^2_t}{M^2_W}\, 
\Big\{ (g^2_L+g^2_R)\, (4\bar{C}_{23} + \bar{C}_0 + 3\bar{C}_{11} + 
\bar{C}_{12})(m^2_t,m^2_t,m^2_t)\nonumber\\
&&+\, 2g_Lg_R\, (\bar{C}_{11}-\bar{C}_{12})(m^2_t,m^2_t,m^2_t)\Big\} ,\\
  \label{BZb1234c47}
i\bar{B}_{(b1)} &=& i\bar{B}_{(b2)}\ =\ i\bar{B}_{(b3)}\ =\ i\bar{B}_{(b4)}\
=\ i\bar{B}_{(c4)}\ =\ i\bar{B}_{(c7)}\ =\ 0\, ,\\
  \label{BZc1}
i\bar{B}_{(c1)} &=& \frac{4\alpha_w}{\pi }\frac{M^4_W}{M^2_Z}\, 
(2\bar{C}_{11}+2\bar{C}_{23}+\bar{C}_0)(M^2_W,M^2_W,M^2_W)\, ,\\
  \label{BZc2c3}
i\bar{B}_{(c2)}+i\bar{B}_{(c3)} &=& -\, \frac{\alpha_w}{2\pi}\, 
M^2_W\, (\bar{C}_0+\bar{C}_{11}+\bar{C}_{12})(M^2_W,M^2_W,M^2_W)\, ,\\
  \label{BZc5c6}
i\bar{B}_{(c5)}+i\bar{B}_{(c6)} &=& \frac{\alpha_w}{4\pi}\, \Big[
2(2M^2_W-M^2_Z)(\bar{C}_0+\bar{C}_{12}-\bar{C}_{11})(M^2_W,M^2_W,M^2_W)
\nonumber\\
&& -\, \frac{M^4_Z}{M^2_W}\, 
(\bar{C}_0+\bar{C}_{12}-\bar{C}_{11})(M^2_Z,M^2_H,M^2_Z)\Big], \\
  \label{BZc8}
i\bar{B}_{(c8)} &=& \frac{\alpha_w}{4\pi}\, M^2_Z\, \Big[\, 
\Big(2\frac{M^2_W}{M^2_Z}-1\Big)^2 \Big(\frac{M^2_H}{M^2_W}+2\Big)
(\bar{C}_{11}+\bar{C}_{23})(M^2_W,M^2_W,M^2_W)\nonumber\\
&&+\, \frac{1}{2}\Big(\frac{M^2_H}{M^2_W}+2\frac{M^2_Z}{M^2_W}\Big)
(\bar{C}_{11}+\bar{C}_{23})(M^2_Z,M^2_H,M^2_Z)\Big]\nonumber\\
&& +\, \frac{3}{2}\, \frac{M^2_H}{M^2_W}\, (\bar{C}_{11}+
\bar{C}_{23})(M^2_H,M^2_Z,M^2_H)\, \Big]\, ,\\
  \label{BZc9c10}
i\bar{B}_{(c9)} + i\bar{B}_{(c10)} &=& -\, \frac{4\alpha_w}{\pi}\,
\frac{M^4_W}{M^2_Z}\, (\bar{C}_{11}+\bar{C}_{23})(M^2_W,M^2_W,M^2_W)\, . 
\end{eqnarray}

\end{appendix}

\newpage

\begin{center}
\begin{picture}(400,400)(0,0)
\SetWidth{0.8}

\ArrowLine(0,370)(10,350)\ArrowLine(10,350)(0,330)
\ArrowLine(70,350)(80,370)\ArrowLine(80,330)(70,350)
\DashLine(10,350)(25,350){3}\DashLine(55,350)(70,350){3}
\PhotonArc(40,350)(15,0,360){2}{12}
\Text(0,380)[]{$t$}\Text(0,320)[]{$\bar{t}$}
\Text(80,380)[r]{$t$}\Text(80,320)[r]{$\bar{t}$}
\Text(16,360)[]{$H$}\Text(63,360)[]{$H$}
\Text(40,377)[]{$W^+$}\Text(40,325)[]{$W^-$}

\Text(40,300)[]{\bf (a)}

\ArrowLine(100,370)(110,350)\ArrowLine(110,350)(100,330)
\ArrowLine(170,350)(180,370)\ArrowLine(180,330)(170,350)
\DashLine(110,350)(125,350){3}\DashLine(155,350)(170,350){3}
\PhotonArc(140,350)(15,0,180){2}{6}\DashArrowArc(140,350)(15,180,360){3}
\Text(116,360)[]{$H$}\Text(163,360)[]{$H$}
\Text(140,377)[]{$W^\pm$}\Text(140,325)[]{$G^\mp$}

\Text(140,300)[]{\bf (b)}

\ArrowLine(200,370)(210,350)\ArrowLine(210,350)(200,330)
\ArrowLine(270,350)(280,370)\ArrowLine(280,330)(270,350)
\DashLine(210,350)(225,350){3}\DashLine(255,350)(270,350){3}
\DashArrowArc(240,350)(15,180,360){3}\DashArrowArcn(240,350)(15,180,360){3}
\Text(216,360)[]{$H$}\Text(263,360)[]{$H$}
\Text(240,377)[]{$G^+$}\Text(240,325)[]{$G^-$}

\Text(240,300)[]{\bf (c)}

\ArrowLine(300,370)(310,350)\ArrowLine(310,350)(300,330)
\ArrowLine(370,350)(380,370)\ArrowLine(380,330)(370,350)
\DashLine(310,350)(325,350){3}\DashLine(355,350)(370,350){3}
\DashArrowArc(340,350)(15,180,360){1}\DashArrowArc(340,350)(15,0,180){1}
\Text(316,360)[]{$H$}\Text(363,360)[]{$H$}
\Text(340,377)[]{$c^\pm$}\Text(340,325)[]{$c^\pm$}

\Text(340,300)[]{\bf (d)}

\ArrowLine(0,270)(10,250)\ArrowLine(10,250)(0,230)
\ArrowLine(60,270)(80,270)\ArrowLine(80,230)(60,230)
\ArrowLine(60,230)(60,270)
\DashLine(10,250)(30,250){3}\Photon(30,250)(60,270){2}{4}
\Photon(30,250)(60,230){2}{4}
\Text(20,260)[]{$H$}\Text(65,250)[l]{$b$}
\Text(55,272)[r]{$W^+$}\Text(55,228)[r]{$W^-$}

\Text(40,200)[]{\bf (e)}

\ArrowLine(100,270)(110,250)\ArrowLine(110,250)(100,230)
\ArrowLine(160,270)(180,270)\ArrowLine(180,230)(160,230)
\ArrowLine(160,230)(160,270)
\DashLine(110,250)(130,250){3}\Photon(130,250)(160,270){2}{4}
\DashArrowLine(130,250)(160,230){3}
\Text(120,260)[]{$H$}\Text(165,250)[l]{$b$}
\Text(155,272)[r]{$W^+$}\Text(155,228)[r]{$G^-$}

\Text(140,200)[]{\bf (f)}

\ArrowLine(200,270)(210,250)\ArrowLine(210,250)(200,230)
\ArrowLine(260,270)(280,270)\ArrowLine(280,230)(260,230)
\ArrowLine(260,230)(260,270)
\DashLine(210,250)(230,250){3}\DashArrowLine(230,250)(260,270){3}
\Photon(230,250)(260,230){2}{4}
\Text(220,260)[]{$H$}\Text(265,250)[l]{$b$}
\Text(255,272)[r]{$G^+$}\Text(255,228)[r]{$W^-$}

\Text(240,200)[]{\bf (g)}

\ArrowLine(300,270)(310,250)\ArrowLine(310,250)(300,230)
\ArrowLine(360,270)(380,270)\ArrowLine(380,230)(360,230)
\ArrowLine(360,230)(360,270)
\DashLine(310,250)(330,250){3}\DashArrowLine(330,250)(360,270){3}
\DashArrowLine(330,250)(360,230){3}
\Text(320,260)[]{$H$}\Text(365,250)[l]{$b$}
\Text(355,272)[r]{$G^+$}\Text(355,228)[r]{$G^-$}

\Text(340,200)[]{\bf (h)}

\ArrowLine(0,170)(20,170)\ArrowLine(20,130)(0,130)
\ArrowLine(20,170)(20,130)
\ArrowLine(70,150)(80,170)\ArrowLine(80,130)(70,150)
\DashLine(50,150)(70,150){3}\Photon(20,170)(50,150){2}{4}
\Photon(20,130)(50,150){2}{4}
\Text(60,160)[]{$H$}\Text(15,150)[r]{$b$}
\Text(25,175)[l]{$W^+$}\Text(25,125)[l]{$W^-$}

\Text(40,100)[]{\bf (i)}

\ArrowLine(100,170)(120,170)\ArrowLine(120,130)(100,130)
\ArrowLine(120,170)(120,130)
\ArrowLine(170,150)(180,170)\ArrowLine(180,130)(170,150)
\DashLine(150,150)(170,150){3}\Photon(120,170)(150,150){2}{4}
\DashArrowLine(120,130)(150,150){3}
\Text(160,160)[]{$H$}\Text(115,150)[r]{$b$}
\Text(125,175)[l]{$W^+$}\Text(125,125)[l]{$G^-$}

\Text(140,100)[]{\bf (j)}

\ArrowLine(200,170)(220,170)\ArrowLine(220,130)(200,130)
\ArrowLine(220,170)(220,130)
\ArrowLine(270,150)(280,170)\ArrowLine(280,130)(270,150)
\DashLine(250,150)(270,150){3}\DashArrowLine(220,170)(250,150){3}
\Photon(220,130)(250,150){2}{4}
\Text(260,160)[]{$H$}\Text(215,150)[r]{$b$}
\Text(225,175)[l]{$G^+$}\Text(225,125)[l]{$W^-$}

\Text(240,100)[]{\bf (k)}

\ArrowLine(300,170)(320,170)\ArrowLine(320,130)(300,130)
\ArrowLine(320,170)(320,130)
\ArrowLine(370,150)(380,170)\ArrowLine(380,130)(370,150)
\DashLine(350,150)(370,150){3}\DashArrowLine(320,170)(350,150){3}
\DashArrowLine(320,130)(350,150){3}
\Text(360,160)[]{$H$}\Text(315,150)[r]{$b$}
\Text(325,175)[l]{$G^+$}\Text(325,125)[l]{$G^-$}

\Text(340,100)[]{\bf (l)}

\ArrowLine(0,70)(20,70)\ArrowLine(20,30)(0,30)
\ArrowLine(20,70)(20,30)
\ArrowLine(60,70)(80,70)\ArrowLine(80,30)(60,30)
\ArrowLine(60,30)(60,70)
\Photon(20,70)(60,70){2}{5}\Photon(20,30)(60,30){2}{5}
\Text(15,50)[r]{$b$}\Text(65,50)[l]{$b$} 
\Text(45,80)[]{$W^+$}\Text(45,20)[]{$W^-$}

\Text(40,0)[]{\bf (m)}

\ArrowLine(100,70)(120,70)\ArrowLine(120,30)(100,30)
\ArrowLine(120,70)(120,30)
\ArrowLine(160,70)(180,70)\ArrowLine(180,30)(160,30)
\ArrowLine(160,30)(160,70)
\Photon(120,70)(160,70){2}{5}\DashArrowLine(120,30)(160,30){3}
\Text(115,50)[r]{$b$}\Text(165,50)[l]{$b$} 
\Text(145,80)[]{$W^+$}\Text(145,20)[]{$G^-$}

\Text(140,0)[]{\bf (n)}

\ArrowLine(200,70)(220,70)\ArrowLine(220,30)(200,30)
\ArrowLine(220,70)(220,30)
\ArrowLine(260,70)(280,70)\ArrowLine(280,30)(260,30)
\ArrowLine(260,30)(260,70)
\DashArrowLine(220,70)(260,70){3} \Photon(220,30)(260,30){2}{5}
\Text(215,50)[r]{$b$}\Text(265,50)[l]{$b$} 
\Text(245,80)[]{$G^+$}\Text(245,20)[]{$W^-$}

\Text(240,0)[]{\bf (o)}

\ArrowLine(300,70)(320,70)\ArrowLine(320,30)(300,30)
\ArrowLine(320,70)(320,30)
\ArrowLine(360,70)(380,70)\ArrowLine(380,30)(360,30)
\ArrowLine(360,30)(360,70)
\DashArrowLine(320,70)(360,70){3} \DashArrowLine(320,30)(360,30){3}
\Text(315,50)[r]{$b$}\Text(365,50)[l]{$b$} 
\Text(345,80)[]{$G^+$}\Text(345,20)[]{$G^-$}

\Text(340,0)[]{\bf (p)}

\end{picture}\\[0.7cm]
{\small {\bf Fig.\ 1:} ~The Higgs-mediated part of the one-loop
  amplitude $t\bar{t} \to t\bar{t}$.}

\end{center}

\newpage

\begin{center}
\begin{picture}(300,150)(0,0)
\SetWidth{0.8}

\ArrowLine(0,100)(30,100)\ArrowLine(30,100)(30,60)
\ArrowLine(30,60)(0,60)\Photon(30,100)(60,100){2}{4}
\Photon(30,60)(60,60){2}{4}
\Text(0,110)[l]{$t$}\Text(0,50)[l]{$\bar{t}$}\Text(35,80)[l]{$b$}
\Text(60,110)[r]{$W^+_\mu$}\Text(60,50)[r]{$W^-_\nu$}
\Text(70,100)[l]{$\frac{\displaystyle k^\mu_+}{\displaystyle M_W}$}
\Text(70,60)[l]{$\frac{\displaystyle k^\nu_-}{\displaystyle M_W}$}

\Text(40,20)[]{\bf (a)}

\LongArrow(105,80)(125,80)\Text(112.5,87)[]{PT}

\ArrowLine(140,100)(160,80)\ArrowLine(160,80)(140,60)
\DashLine(160,80)(190,80){3}\Vertex(190,80){3}
\Photon(190,80)(210,100){2}{3}\Photon(190,80)(210,60){2}{3}
\Text(140,110)[l]{$t$}\Text(140,50)[l]{$\bar{t}$}\Text(175,90)[]{$H(q)$}
\Text(210,105)[r]{$W^+$}\Text(210,55)[r]{$W^-$}
\Text(220,80)[l]{$\Big(-\, \frac{\displaystyle g_w}{\displaystyle
    2M_W}\,\Big)\, (q^2 - M^2_H)$}

\Text(215,20)[]{\bf (b)}

\end{picture}\\
{\small {\bf Fig.\ 2:} ~Higgs-like contribution from the $t$-channel graph.}
\end{center}

\vspace{4.cm}

\begin{center}
\begin{picture}(400,150)(0,0)
\SetWidth{0.8}

\ArrowLine(0,100)(20,80)\ArrowLine(20,80)(0,60)
\Photon(20,80)(50,80){2}{4}
\Photon(50,80)(70,100){2}{3}\Photon(50,80)(70,60){2}{3}
\Text(10,110)[r]{$t(p_1)$}\Text(10,50)[r]{$\bar{t}(p_2)$}
\Text(35,90)[]{$\gamma, Z$}\Text(35,70)[]{$(q)$}
\Text(90,110)[r]{$W^+_\mu (k_+)$}\Text(90,50)[r]{$W^-_\nu (k_-)$}

\Text(35,20)[]{\bf (a)}

\ArrowLine(150,100)(170,80)\ArrowLine(170,80)(150,60)
\DashArrowLine(170,80)(200,80){3}
\Photon(200,80)(220,100){2}{3}\Photon(200,80)(220,60){2}{3}
\Text(185,90)[]{$H$}
\Text(150,110)[r]{$t$}\Text(150,50)[r]{$\bar{t}$}
\Text(240,110)[r]{$W^+_\mu$}\Text(240,50)[r]{$W^-_\nu$}

\Text(185,20)[]{\bf (b)}

\ArrowLine(300,100)(330,100)\ArrowLine(330,100)(330,60)
\ArrowLine(330,60)(300,60)\Photon(330,100)(360,100){2}{4}
\Photon(330,60)(360,60){2}{4}
\Text(300,110)[l]{$t$}\Text(300,50)[l]{$\bar{t}$}\Text(335,80)[l]{$b$}
\Text(380,110)[r]{$W^+_\mu$}\Text(380,50)[r]{$W^-_\nu$}

\Text(335,20)[]{\bf (c)}

\end{picture}\\
{\small {\bf Fig.\ 3:} ~Feynman diagrams pertaining to the process
 $t\bar{t} \to W^+W^-$.}
\end{center}

\begin{figure}[p]
   \leavevmode
 \begin{center}
   \epsfxsize=15.0cm
   \epsffile[0 0 539 652]{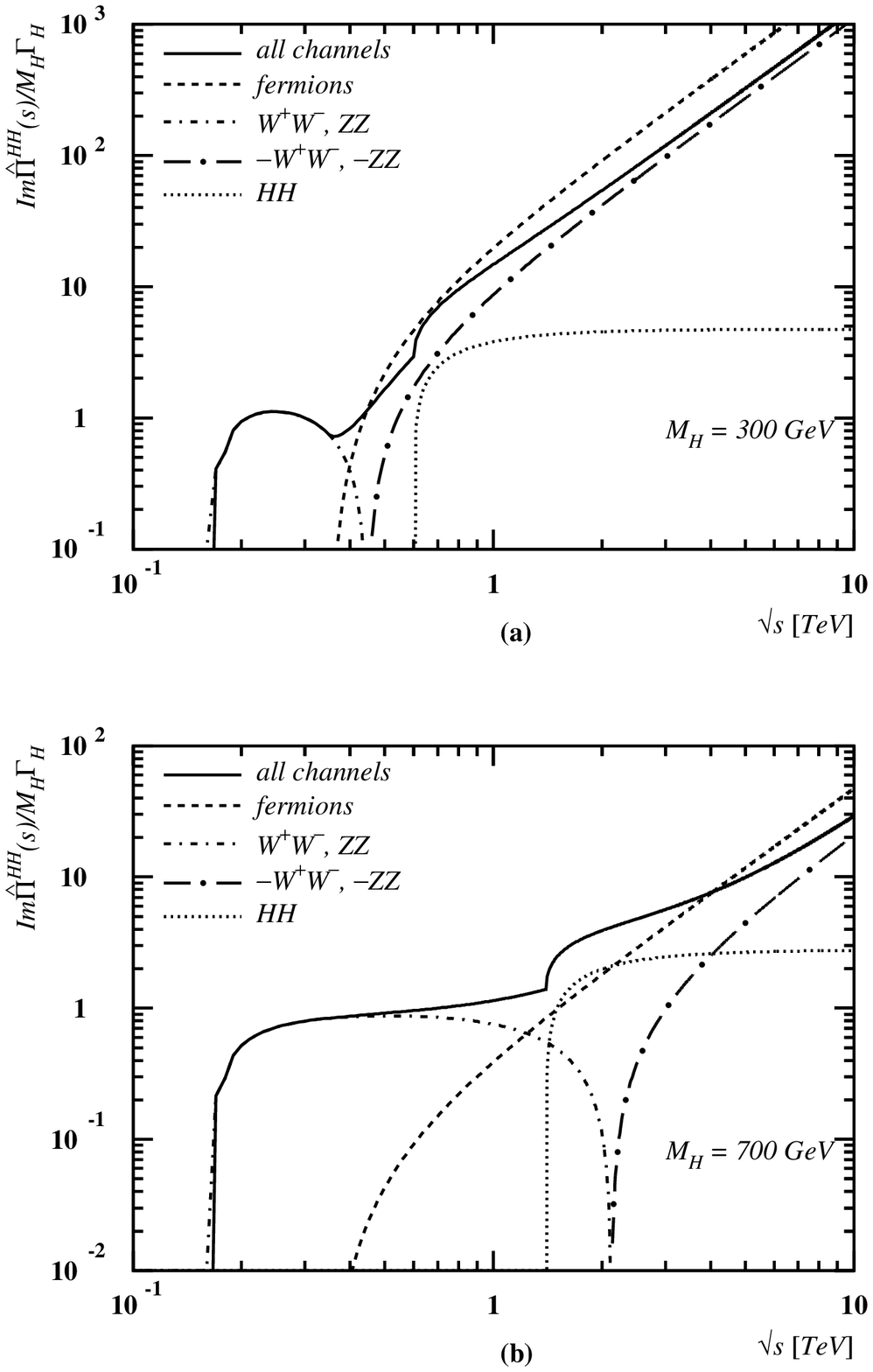}
{\small {\bf Fig.\ 4:} ~Dependence of $Im\widehat{\Pi}^{HH}(s)/M_H\Gamma_H$ 
on $\sqrt{s}$ for individual intermediate states.}
 \end{center}
\end{figure}

\begin{figure}[p]
   \leavevmode
 \begin{center}
   \epsfxsize=15.0cm
   \epsffile[0 0 539 652]{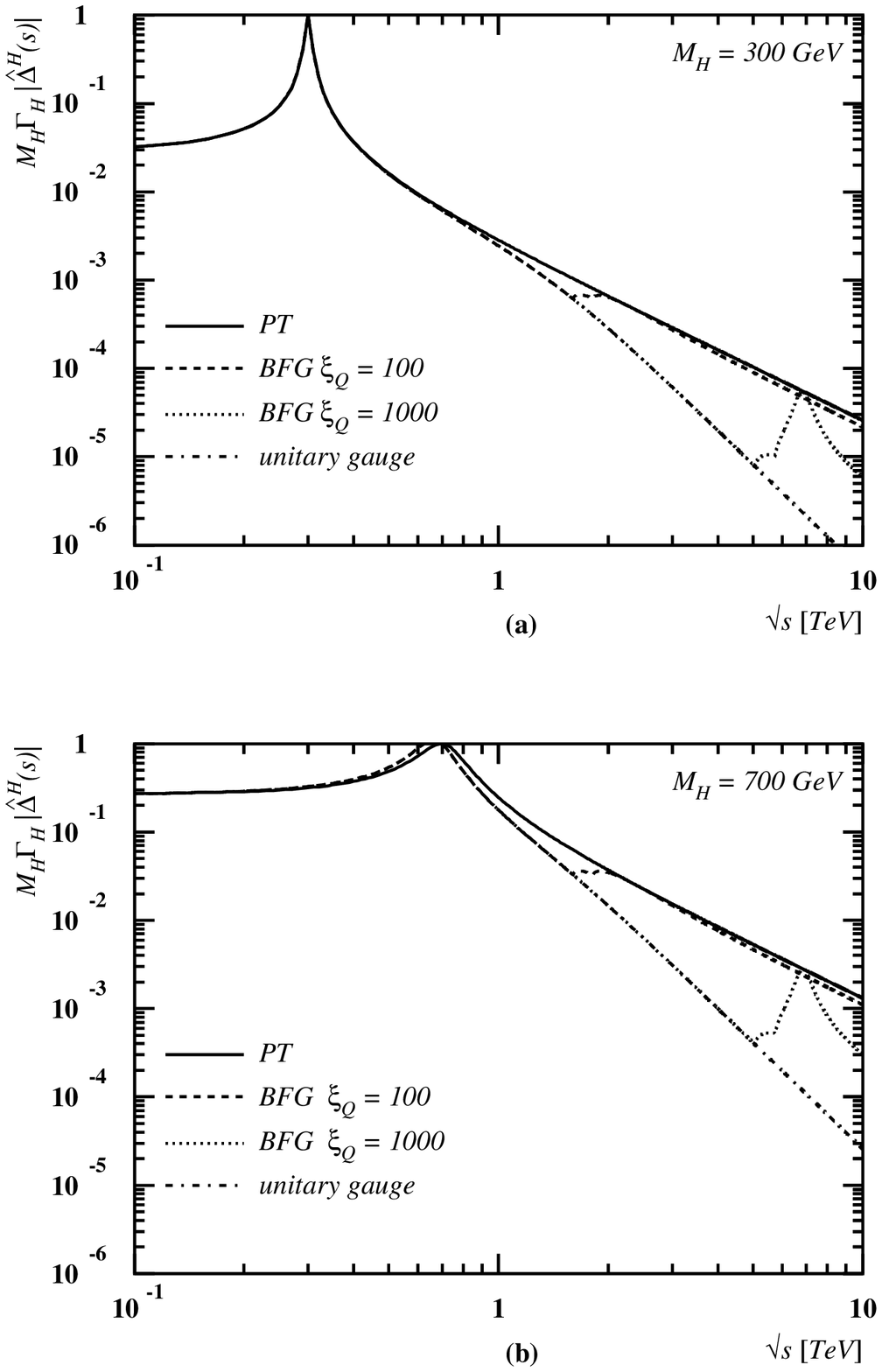}
{\small {\bf Fig.\ 5:} ~$M_H\Gamma_H\, |\hat{\Delta}^H(s)|$ 
versus $\sqrt{s}$ in different resummation schemes.}
 \end{center}
\end{figure}

\newpage

${}$
\vspace{3.cm}

\begin{center}
\begin{picture}(400,200)(0,0)
\SetWidth{0.8}

\ArrowLine(0,130)(30,100)\ArrowLine(30,100)(0,70)
\DashArrowLine(30,100)(50,100){3}\GCirc(65,100){15}{0.8}
\DashArrowLine(80,100)(100,100){3}\GCirc(115,100){15}{0.8}
\Photon(120,115)(140,130){3}{2}\Photon(120,85)(140,70){3}{2}
\Text(10,140)[r]{$\nu (p_1)$}\Text(10,60)[r]{$\bar{\nu}(p_2)$}
\Text(160,140)[r]{$Z_\mu (k_1)$}\Text(160,60)[r]{$Z_\nu (k_2)$}
\Text(40,110)[]{$H$}\Text(90,110)[]{$H$}
\Text(65,125)[]{$\hat{\Delta}^H(q)$}
\Text(135,100)[l]{$\Gamma^{HZZ}_{0\mu\nu} + \widehat{\Gamma}^{HZZ}_{\mu\nu}$}
\LongArrow(130,132)(122,124)\LongArrow(130,68)(122,76)

\Text(65,20)[]{\bf (a)}

\ArrowLine(240,120)(270,120)\ArrowLine(270,120)(270,80)
\ArrowLine(270,80)(240,80)\Photon(270,120)(300,120){2}{4}
\Photon(270,80)(300,80){2}{4}
\Text(240,130)[l]{$\nu$}\Text(240,70)[l]{$\bar{\nu}$}
\Text(275,100)[l]{$\nu (p_1+k_1)$}
\Text(300,130)[r]{$Z_\mu$}\Text(300,70)[r]{$Z_\nu$}

\Text(270,20)[]{\bf (b)}

\ArrowLine(340,120)(370,120)\ArrowLine(370,120)(370,80)
\ArrowLine(370,80)(340,80)\Photon(370,120)(400,120){2}{4}
\Photon(370,80)(400,80){2}{4}
\Text(340,130)[l]{$\nu$}\Text(340,70)[l]{$\bar{\nu}$}
\Text(375,100)[l]{$\nu (p_1+k_2)$}
\Text(400,130)[r]{$Z_\nu$}\Text(400,70)[r]{$Z_\mu$}

\Text(370,20)[]{\bf (c)}

\end{picture}\\
{\small {\bf Fig.\ 6:} ~Resummation of the Higgs-mediated amplitude 
pertinent to $\nu\bar{\nu} \to ZZ$.}
\end{center}

\begin{center}
\begin{picture}(400,500)(0,0)
\SetWidth{0.8}

\DashArrowLine(0,470)(20,470){4}\ArrowLine(20,470)(50,490)
\ArrowLine(50,490)(50,450)\ArrowLine(50,450)(20,470)
\Photon(50,490)(70,490){3}{2.5}\Photon(50,450)(70,450){3}{2.5}
\Text(5,480)[r]{{\small $\widehat{H}(Q)$}}
\Text(100,500)[r]{\small $\widehat{W}^+_\mu(p)$}
\Text(100,440)[r]{\small $\widehat{W}^-_\nu(k)$}
\Text(30,490)[r]{\small $t$}
\Text(30,450)[r]{\small $t$}
\Text(55,470)[l]{\small $b$}
\Text(35,420)[]{\bf (a)}

\DashArrowLine(150,470)(170,470){4}
\PhotonArc(185,470)(15,0,180){3}{5}\PhotonArc(185,470)(15,180,360){3}{5}
\Photon(200,470)(220,490){-3}{2.5}\Photon(200,470)(220,450){3}{2.5}
\Text(185,495)[]{\small $W^+,Z$}
\Text(185,445)[]{\small $W^-,Z$}
\Text(185,420)[]{\bf (b1)}

\DashArrowLine(300,470)(320,470){4}
\PhotonArc(335,470)(15,0,180){3}{5}\DashArrowArc(335,470)(15,180,360){3}
\Photon(350,470)(370,490){-3}{2.5}\Photon(350,470)(370,450){3}{2.5}
\Text(335,495)[]{\small $W^\pm,Z$}
\Text(335,445)[]{\small $G^\mp,G^0$}
\Text(335,420)[]{\bf (b2)}

\DashArrowLine(0,370)(20,370){4}
\DashArrowArcn(35,370)(15,180,360){3}\DashArrowArc(35,370)(15,180,360){3}
\Photon(50,370)(70,390){-3}{2.5}\Photon(50,370)(70,350){3}{2.5}
\Text(35,395)[]{\small $G^+,G^0,H$}
\Text(35,345)[]{\small $G^-,G^0,H$}
\Text(35,320)[]{\bf (b3)}

\DashArrowLine(150,370)(170,370){4}
\DashArrowArc(185,370)(15,0,180){1.5}\DashArrowArc(185,370)(15,180,360){1.5}
\Photon(200,370)(220,390){-3}{2.5}\Photon(200,370)(220,350){3}{2.5}
\Text(185,395)[]{\small $c^\pm,c_Z$}
\Text(185,345)[]{\small $c^\pm,c_Z$}
\Text(185,320)[]{\bf (b4)}

\DashArrowLine(300,370)(320,370){4}\Photon(320,370)(350,390){2}{4}
\Photon(350,390)(350,350){2}{4}\Photon(350,350)(320,370){2}{4}
\Photon(350,390)(370,390){3}{2.5}\Photon(350,350)(370,350){3}{2.5}
\Text(345,395)[r]{\small $W^+,W^+,Z$}
\Text(345,345)[r]{\small $W^-,W^-,Z$}
\Text(355,370)[l]{\small $Z,\gamma,W^-$}
\Text(335,320)[]{\bf (c1)}

\DashArrowLine(0,270)(20,270){4}\Photon(20,270)(50,290){2}{4}
\Photon(50,290)(50,250){2}{4}\DashArrowLine(20,270)(50,250){4}
\Photon(50,290)(70,290){3}{2.5}\Photon(50,250)(70,250){3}{2.5}
\Text(45,295)[r]{\small $W^+,W^+,Z$}
\Text(45,245)[r]{\small $G^-,G^-,G^0$}
\Text(55,270)[l]{\small $Z,\gamma,W^-$}
\Text(35,220)[]{\bf (c2)}

\DashArrowLine(150,270)(170,270){4}\DashArrowLine(170,270)(200,290){4}
\Photon(200,290)(200,250){2}{4}\Photon(170,270)(200,250){2}{4}
\Photon(200,290)(220,290){3}{2.5}\Photon(200,250)(220,250){3}{2.5}
\Text(195,295)[r]{\small $G^+,G^+,G^0$}
\Text(195,245)[r]{\small $W^-,W^-,Z$}
\Text(205,270)[l]{\small $Z,\gamma,W^-$}
\Text(185,220)[]{\bf (c3)}

\DashArrowLine(300,270)(320,270){4}\Photon(320,270)(350,290){2}{4}
\DashArrowLine(350,290)(350,250){4}\Photon(350,250)(320,270){2}{4}
\Photon(350,290)(370,290){3}{2.5}\Photon(350,250)(370,250){3}{2.5}
\Text(345,295)[r]{\small $W^+,W^+,Z$}
\Text(345,245)[r]{\small $W^-,W^-,Z$}
\Text(355,270)[l]{\small $G^0,H,G^-$}
\Text(335,220)[]{\bf (c4)}

\DashArrowLine(0,170)(20,170){4}\DashArrowLine(20,170)(50,190){4}
\DashArrowLine(50,190)(50,150){4}\Photon(50,150)(20,170){2}{4}
\Photon(50,190)(70,190){3}{2.5}\Photon(50,150)(70,150){3}{2.5}
\Text(45,195)[r]{\small $G^+,G^+,G^0$}
\Text(45,145)[r]{\small $W^-,W^-,Z$}
\Text(55,170)[l]{\small $G^0,H,G^-$}
\Text(35,120)[]{\bf (c5)}

\DashArrowLine(150,170)(170,170){4}\Photon(170,170)(200,190){2}{4}
\DashArrowLine(200,190)(200,150){4}\DashArrowLine(170,170)(200,150){4}
\Photon(200,190)(220,190){3}{2.5}\Photon(200,150)(220,150){3}{2.5}
\Text(195,195)[r]{\small $W^+,W^+,Z$}
\Text(195,145)[r]{\small $G^-,G^-,G^0$}
\Text(205,170)[l]{\small $G^0,H,G^-$}
\Text(185,120)[]{\bf (c6)}

\DashArrowLine(300,170)(320,170){4}\DashArrowLine(320,170)(350,190){4}
\Photon(350,190)(350,150){2}{4}\DashArrowLine(320,170)(350,150){4}
\Photon(350,190)(370,190){3}{2.5}\Photon(350,150)(370,150){3}{2.5}
\Text(350,195)[r]{\small $G^+,G^+,G^0,H$}
\Text(350,145)[r]{\small $G^-,G^-,G^0,H$}
\Text(355,170)[l]{\small $Z,\gamma,W^-,W^-$}
\Text(335,120)[]{\bf (c7)}

\DashArrowLine(0,70)(20,70){4}\DashArrowLine(20,70)(50,90){4}
\DashArrowLine(50,90)(50,50){4}\DashArrowLine(20,70)(50,50){4}
\Photon(50,90)(70,90){3}{2.5}\Photon(50,50)(70,50){3}{2.5}
\Text(50,95)[r]{\small $G^+,G^+,G^0,H$}
\Text(50,45)[r]{\small $G^-,G^-,G^0,H$}
\Text(55,70)[l]{\small $G^0,H,G^-,G^-$}
\Text(35,20)[]{\bf (c8)}

\DashArrowLine(150,70)(170,70){4}\DashArrowLine(170,70)(200,90){1.5}
\DashArrowLine(200,90)(200,50){1.5}\DashArrowLine(200,50)(170,70){1.5}
\Photon(200,90)(220,90){3}{2.5}\Photon(200,50)(220,50){3}{2.5}
\Text(195,95)[r]{\small $c^+,c^+,c_Z$}
\Text(195,45)[r]{\small $c^+,c^+,c_Z$}
\Text(205,70)[l]{\small $c_Z,c_\gamma,c^-$}
\Text(185,20)[]{\bf (c9)}

\DashArrowLine(300,70)(320,70){4}\DashArrowLine(350,90)(320,70){1.5}
\DashArrowLine(350,50)(350,90){1.5}\DashArrowLine(320,70)(350,50){1.5}
\Photon(350,90)(370,90){3}{2.5}\Photon(350,50)(370,50){3}{2.5}
\Text(345,95)[r]{\small $c^-,c^-,c_Z$}
\Text(345,45)[r]{\small $c^-,c^-,c_Z$}
\Text(355,70)[l]{\small $c_Z,c_\gamma,c^+$}
\Text(335,20)[]{\bf (c10)}
\end{picture}\\
{\small {\bf Fig.\ 7:} ~Graphs contributing to the absorptive part of
the $HWW$ coupling in the BFG.}

\end{center}

\newpage

\begin{center}
\begin{picture}(400,500)(0,0)
\SetWidth{0.8}

\DashArrowLine(0,470)(20,470){4}\ArrowLine(20,470)(50,490)
\ArrowLine(50,490)(50,450)\ArrowLine(50,450)(20,470)
\Photon(50,490)(70,490){3}{2.5}\Photon(50,450)(70,450){3}{2.5}
\Text(5,480)[r]{{\small $\widehat{H}(Q)$}}
\Text(100,500)[r]{\small $\widehat{Z}_\mu(p)$}
\Text(100,440)[r]{\small $\widehat{Z}_\nu(k)$}
\Text(30,490)[r]{\small $f,\bar{f}$}
\Text(30,450)[r]{\small $f,\bar{f}$}
\Text(55,470)[l]{\small $f,\bar{f}$}
\Text(35,420)[]{\bf (a)}

\DashArrowLine(150,470)(170,470){4}
\PhotonArc(185,470)(15,0,180){3}{5}\PhotonArc(185,470)(15,180,360){3}{5}
\Photon(200,470)(220,490){-3}{2.5}\Photon(200,470)(220,450){3}{2.5}
\Text(185,495)[]{\small $W^+$}
\Text(185,445)[]{\small $W^-$}
\Text(185,420)[]{\bf (b1)}

\DashArrowLine(300,470)(320,470){4}
\PhotonArc(335,470)(15,0,180){3}{5}\DashArrowArc(335,470)(15,180,360){3}
\Photon(350,470)(370,490){-3}{2.5}\Photon(350,470)(370,450){3}{2.5}
\Text(335,495)[]{\small $W^\pm$}
\Text(335,445)[]{\small $G^\mp$}
\Text(335,420)[]{\bf (b2)}

\DashArrowLine(0,370)(20,370){4}
\DashArrowArcn(35,370)(15,180,360){3}\DashArrowArc(35,370)(15,180,360){3}
\Photon(50,370)(70,390){-3}{2.5}\Photon(50,370)(70,350){3}{2.5}
\Text(35,395)[]{\small $G^+,G^0,H$}
\Text(35,345)[]{\small $G^-,G^0,H$}
\Text(35,320)[]{\bf (b3)}

\DashArrowLine(150,370)(170,370){4}
\DashArrowArc(185,370)(15,0,180){1.5}\DashArrowArc(185,370)(15,180,360){1.5}
\Photon(200,370)(220,390){-3}{2.5}\Photon(200,370)(220,350){3}{2.5}
\Text(185,395)[]{\small $c^\pm$}
\Text(185,345)[]{\small $c^\pm$}
\Text(185,320)[]{\bf (b4)}

\DashArrowLine(300,370)(320,370){4}\Photon(320,370)(350,390){2}{4}
\Photon(350,390)(350,350){2}{4}\Photon(350,350)(320,370){2}{4}
\Photon(350,390)(370,390){3}{2.5}\Photon(350,350)(370,350){3}{2.5}
\Text(345,395)[r]{\small $W^\pm$}
\Text(345,345)[r]{\small $W^\mp$}
\Text(355,370)[l]{\small $W^\pm$}
\Text(335,320)[]{\bf (c1)}

\DashArrowLine(0,270)(20,270){4}\Photon(20,270)(50,290){2}{4}
\Photon(50,290)(50,250){2}{4}\DashArrowLine(20,270)(50,250){4}
\Photon(50,290)(70,290){3}{2.5}\Photon(50,250)(70,250){3}{2.5}
\Text(45,295)[r]{\small $W^\pm$}
\Text(45,245)[r]{\small $G^\mp$}
\Text(55,270)[l]{\small $W^\pm$}
\Text(35,220)[]{\bf (c2)}

\DashArrowLine(150,270)(170,270){4}\DashArrowLine(170,270)(200,290){4}
\Photon(200,290)(200,250){2}{4}\Photon(170,270)(200,250){2}{4}
\Photon(200,290)(220,290){3}{2.5}\Photon(200,250)(220,250){3}{2.5}
\Text(195,295)[r]{\small $G^\pm$}
\Text(195,245)[r]{\small $W^\mp$}
\Text(205,270)[l]{\small $W^\pm$}
\Text(185,220)[]{\bf (c3)}

\DashArrowLine(300,270)(320,270){4}\Photon(320,270)(350,290){2}{4}
\DashArrowLine(350,290)(350,250){4}\Photon(350,250)(320,270){2}{4}
\Photon(350,290)(370,290){3}{2.5}\Photon(350,250)(370,250){3}{2.5}
\Text(345,295)[r]{\small $W^\pm,Z$}
\Text(345,245)[r]{\small $W^\mp,Z$}
\Text(355,270)[l]{\small $G^\pm,H$}
\Text(335,220)[]{\bf (c4)}

\DashArrowLine(0,170)(20,170){4}\DashArrowLine(20,170)(50,190){4}
\DashArrowLine(50,190)(50,150){4}\Photon(50,150)(20,170){2}{4}
\Photon(50,190)(70,190){3}{2.5}\Photon(50,150)(70,150){3}{2.5}
\Text(45,195)[r]{\small $G^\pm,G^0$}
\Text(45,145)[r]{\small $W^\mp,Z$}
\Text(55,170)[l]{\small $G^\pm,H$}
\Text(35,120)[]{\bf (c5)}

\DashArrowLine(150,170)(170,170){4}\Photon(170,170)(200,190){2}{4}
\DashArrowLine(200,190)(200,150){4}\DashArrowLine(170,170)(200,150){4}
\Photon(200,190)(220,190){3}{2.5}\Photon(200,150)(220,150){3}{2.5}
\Text(195,195)[r]{\small $W^\pm,Z$}
\Text(195,145)[r]{\small $G^\mp,G^0$}
\Text(205,170)[l]{\small $G^\pm,H$}
\Text(185,120)[]{\bf (c6)}

\DashArrowLine(300,170)(320,170){4}\DashArrowLine(320,170)(350,190){4}
\Photon(350,190)(350,150){2}{4}\DashArrowLine(320,170)(350,150){4}
\Photon(350,190)(370,190){3}{2.5}\Photon(350,150)(370,150){3}{2.5}
\Text(350,195)[r]{\small $G^\pm,H$}
\Text(350,145)[r]{\small $G^\mp,H$}
\Text(355,170)[l]{\small $W^\pm,Z$}
\Text(335,120)[]{\bf (c7)}

\DashArrowLine(0,70)(20,70){4}\DashArrowLine(20,70)(50,90){4}
\DashArrowLine(50,90)(50,50){4}\DashArrowLine(20,70)(50,50){4}
\Photon(50,90)(70,90){3}{2.5}\Photon(50,50)(70,50){3}{2.5}
\Text(50,95)[r]{\small $G^\pm,G^0,H$}
\Text(50,45)[r]{\small $G^\mp,G^0,H$}
\Text(55,70)[l]{\small $G^\pm,H,G^0$}
\Text(35,20)[]{\bf (c8)}

\DashArrowLine(150,70)(170,70){4}\DashArrowLine(170,70)(200,90){1.5}
\DashArrowLine(200,90)(200,50){1.5}\DashArrowLine(200,50)(170,70){1.5}
\Photon(200,90)(220,90){3}{2.5}\Photon(200,50)(220,50){3}{2.5}
\Text(195,95)[r]{\small $c^\pm$}
\Text(195,45)[r]{\small $c^\pm$}
\Text(205,70)[l]{\small $c^\pm$}
\Text(185,20)[]{\bf (c9)}

\DashArrowLine(300,70)(320,70){4}\DashArrowLine(350,90)(320,70){1.5}
\DashArrowLine(350,50)(350,90){1.5}\DashArrowLine(320,70)(350,50){1.5}
\Photon(350,90)(370,90){3}{2.5}\Photon(350,50)(370,50){3}{2.5}
\Text(345,95)[r]{\small $c^\pm$}
\Text(345,45)[r]{\small $c^\pm$}
\Text(355,70)[l]{\small $c^\pm$}
\Text(335,20)[]{\bf (c10)}
\end{picture}\\
{\small {\bf Fig.\ 8:} ~Diagrams contributing to the absorptive part of
the $HZZ$ coupling in the BFG.}

\end{center}


\begin{thebibliography}{99}
\bibitem{Velt2}  It is worth emphasizing  that in Veltman's pioneering
  work \cite{Velt} these  issues do not arise, since  the toy model he
  considered      was    based    on     $(\phi^{3})_{4}$    which  is
  super-renormalizable,  {\em  i.e.},   the renormalization group   is
  trivial.
\end{thebibliography}
\end{document}